\documentclass[british,noreviewchanges,noreferee,nodrafting,latin9,usenames,dvipsnames,svgnames,table]{aa}
\pdfoutput=1
\usepackage[LGR,T1]{fontenc}
\usepackage[latin9]{inputenc}
\usepackage{textcomp}
\usepackage{amstext}

\makeatletter

\DeclareRobustCommand{\greektext}{%
  \fontencoding{LGR}\selectfont\def\encodingdefault{LGR}}
\DeclareRobustCommand{\textgreek}[1]{\leavevmode{\greektext #1}}
\DeclareFontEncoding{LGR}{}{}
\DeclareTextSymbol{\~}{LGR}{126}
\newcommand{\lyxmathsym}[1]{\ifmmode\begingroup\def\b@ld{bold}
  \text{\ifx\math@version\b@ld\bfseries\fi#1}\endgroup\else#1\fi}

\usepackage{natbib}
\bibpunct{(}{)}{;}{a}{}{,}
\usepackage{rotating}
\usepackage{drafting}

\makeatother

\usepackage{babel}

\usepackage[T1]{fontenc}
\usepackage{geometry}
\usepackage{tikz}
\usepackage{longtable}
\usepackage[nomessages]{fp}
\usepackage{subfig}
\usepackage{graphicx}
\newcommand{\lyxdot}{.}
\usepackage{xcolor}
\usepackage{wasysym}
\usepackage{url}
\usepackage{esint}
\usepackage{booktabs}
\providecommand{\tabularnewline}{\\}
\begin{document}

\date{-Received date / Accepted date}

\title{X-ray spectral modelling of the AGN obscuring region in the CDFS:
Bayesian model selection and catalogue}

\author{J.~Buchner\inst{\ref{MPE}}
\and A.~Georgakakis\inst{\ref{MPE}}
\and K.~Nandra\inst{\ref{MPE}}
\and L.~Hsu\inst{\ref{MPE}}
\and C.~Rangel\inst{\ref{IMP}}
\and M.~Brightman\inst{\ref{MPE}}
\and A.~Merloni\inst{\ref{MPE}}
\and M.~Salvato\inst{\ref{MPE}}
\and J.~Donley\inst{\ref{Alamos}}
\and D.~Kocevski\inst{\ref{Kentucky}}
}
\titlerunning{Absorption and reflection model comparison of AGN in the CDFS}
\authorrunning{Buchner et al.}
\institute{Max Planck Institut f\"ur Extraterrestrische Physik Giessenbachstrasse, 85748 Garching Germany\label{MPE}
\and Astrophysics Group, Imperial College London, Blackett Laboratory, Prince Consort Road, London SW7 2AZ\label{IMP}
\and Los Alamos National Laboratory, Los Alamos, NM  USA\label{Alamos}
\and Department of Physics and Astronomy, University of Kentucky, 505 Rose Street, Lexington, Kentucky 40506, USA\label{Kentucky}
}
\abstract {
} {
\longchange{The geometry of the obscuring ``torus'' in AGN is known
to be complex in local sources. At $z=0.5-3$, where most accretion
occurs in obscured sources, simple models of line-of-sight (LOS) obscuration
are usually assumed. A range of important analyses, such as luminosity
functions, accretion density, co-evolution of AGN with host galaxies,
rely on results based on these models. We probe the geometry of the
torus to find the best model for high-redshift ($z>0.5$) AGN.}{Active
Galactic Nuclei are known to have complex X-ray spectra that depend
on both the properties of the accreting supermassive black hole (e.g.
mass, accretion rate) and the distribution of obscuring material in
its vicinity (i.e. the ``torus''). Often however, simple and even
unphysical models are adopted to represent the X-ray spectra of AGN,
which do not capture the complexity and diversity of the observations.
In the case of blank field surveys in particular, this should have
an impact on e.g. the determination of the AGN luminosity function,
the inferred accretion history of the Universe and also on our understanding
of the relation between AGN and their host galaxies.} }{
\longchange{We develop a Bayesian framework for model comparison
and parameter estimation of X-ray spectra. We also propagate the uncertainty
associated with photometric redshifts estimates. Ten physically motivated
models of torus geometries are compared using the deepest X-ray field
to date, the Chandra Deep Field South (CDFS) of 4Ms exposure time.}{We
develop a Bayesian framework for model comparison and parameter estimation
of X-ray spectra. We take into account uncertainties associated with
both the Poisson nature of X-ray data and the determination of source
redshift using photometric methods. We also demonstrate how Bayesian
model comparison can be used to select among ten different physically
motivated X-ray spectral models the one that provides a better representation
of the observations. This methodology is applied to X-ray AGN in the
4~Ms Chandra Deep Field South. } }{
\longchange{For the $\sim350$ AGN spectra considered, our analysis
identifies four important components to represent the variety of AGN
spectra: (1) An intrinsic power law, (2) an obscurer which reprocesses
the radiation due to photo-electric absorption, Compton scattering
and Fe-K fluorescence, (3) an unabsorbed power law associated with
Thomson scattering off ionised clouds, and (4) Compton reflection,
most noticeable from a stronger-than-expected Fe-K$\alpha$ line.
Simpler models, such as a photo-electrically absorbed power law with
a Thomson scattering component, are ruled out with decisive evidence
($Z>100$). We also note that not considering the Thomson scattering
component underestimates the inferred column density $N_{H}$. Regarding
the geometry of the obscurer, we find strong evidence against both
a completely closed, or entirely open, toroidal geometry in favour
of an intermediate case.}{For the $\sim350$ AGN in that field, our
analysis identifies four components needed to represent the diversity
of the observed X-ray spectra: (1) an intrinsic power law, (2) a cold
obscurer which reprocesses the radiation due to photo-electric absorption,
Compton scattering and Fe-K fluorescence, (3) an unabsorbed power
law associated with Thomson scattering off ionised clouds, and (4)
Compton reflection, most noticeable from a stronger-than-expected
Fe-K line. Simpler models, such as a photo-electrically absorbed power
law with a Thomson scattering component, are ruled out with decisive
evidence ($B>100$). We also find that ignoring the Thomson scattering
component results in underestimation of the inferred column density,
$N_{H}$, of the obscurer. Regarding the geometry of the obscurer,
there is strong evidence against both a completely closed (e.g. sphere),
or entirely open (e.g. blob of material along the line of sight),
toroidal geometry in favour of an intermediate case.} } {
\longchange{Despite the use of low-count spectra, our methodology
is able to draw strong inferences on the geometry of the torus. Simpler
models are ruled out in favour of a geometrically extended structure
with significant Compton scattering. We confirm the presence of a
soft component associated with Thomson scattering off ionised clouds
in the opening angle of the torus, which scatters the unobscured intrinsic
radiation past the torus. The additional Compton reflection may be
a sign of a density gradient in the torus or reflection off the accretion
disk. Finally, we release a catalogue of AGN in the CDFS with estimated
parameters such as the accretion luminosity in the $2-10\,\text{keV}$
band and the column density $N_{H}$.}{Despite the use of low-count
spectra, our methodology is able to draw strong inferences on the
geometry of the torus. Simpler models are ruled out in favour of a
geometrically extended structure with significant Compton scattering.
We confirm the presence of a soft component, possibly associated with
Thomson scattering off ionised clouds in the opening angle of the
torus. The additional Compton reflection required by data over that
predicted by toroidal geometry models, may be a sign of a density
gradient in the torus or reflection off the accretion disk. Finally,
we release a catalogue of AGN in the CDFS with estimated parameters
such as the accretion luminosity in the $2-10\,\text{keV}$ band and
the column density, $N_{H}$, of the obscurer.}}
{}

\maketitle

\section{Introduction\label{sec:Introduction}}

\longchange{\para{Introduction}{Active Galactic Nuclei (AGN) are
thought to be powered by the accretion of material origin onto a super-massive
black hole (SMBH) at the centre of galaxies. This process is powerful
-- the energy released exceeds the binding energy of the host galaxy
-- and ubiquitous -- virtually every massive galaxy hosts a SMBH.
Based on this statement, the interplay between AGN and galaxy evolution
is likely, and caused research in this field to soar in the past 15
years \citep[][]{Kormendy2013}. One connection is the shared galactic
gas supply fueling both star formation and growth of the SMBH \citep[][]{Alexander2012}.
Indeed, both the star formation and the accretion density peak between
redshift $0.5$ and $3$, when Large Scale Structure formation makes
gas available to the galaxies \citep[][]{Lilly1996,Madau1996,Boyle1998,Hasinger2005,Hopkins2005}.

The majority of AGN spectra show obscuration by cold material of column
density $N_{H}>10^{22}\text{cm}^{-2}$ \citep[e.g. ][]{Turner1997,Risaliti1999,Ueda2003,Hasinger2005,LaFranca2005,Tozzi2006,Akylas2006}.
The obscurer is thought to be a small-scale, geometrically thick reservoir
called ``the torus''. The study of the obscurer is not only interesting
in its own right, especially if it is an intermediate accumulation
of gas before final accretion. To wit, comparing only the unobscured
AGN with local SMBH masses shows that accretion must predominantly
occur during these conceiled, obscured phases \citep[originally \cite{Soltan1982},
for a review see ][]{Fabian2004}. Thus obscured AGN in the redshift
range $z=0.5-3$ are particularly interesting objects. Their intrinsic
luminosity can be used to study the violent accretion process itself,
and its relation to the host galaxy.

The first step to understanding the complete accretion density is
to understand basic AGN population properties, such as the number
density of AGN, their luminosity, redshift and obscuration distributions.
For simplicity, these studies often analyse X-ray spectra with relatively
simple models or even just hardness ratios. Whether in high redshift
sources these simple models are justified however has not been demonstrated,
while in local sources obscuration is known to be complex. We thus
assume that more realistic, physically motivated models and a methodology
that incorporates all sources of uncertainty may push the limitations
of current results, especially with regards to claimed trends with
luminosity and redshift. We compare various physically motivated models
for the geometry of the obscuring region to identify the complexity
justified for high redshift sources.

X-ray observations are our most efficient method for selecting AGN.
X-rays are comparatively mildly affected by obscuration except at
highest column densities, but are soft enough to be well measurable.
A large sample of high-redshift X-ray AGN is thus the ideal field
to study accreting SMBHs. A number of issues make this endeavour
hard: (1) Low counts in the analysis of X-ray spectra of faint sources
limit the familiar $\chi^{2}$-regression to broadly binning spectra.
(2) The difficulty and thus uncertainty of distance estimation of
faint sources through photometric redshift affects both the spectral
analysis and luminosity estimates. Finally, (3) methods used to compare
spectral models have been limited to nested problems, i.e. where a
simpler model is a special case of a more complex one.}

\para{Motivation}{To overcome these issues, we introduce a novel
framework for un-binned X-ray spectral analysis based on Bayesian
inference, which propagates all the uncertainties from redshift and
photon counts. We apply this method to CDFS data to infer structural
information about the ``torus'' of AGN from X-ray spectra by comparing
various physically motivated obscurer models.}

}{

\para{Introduction}{Active Galactic Nuclei (AGN) are thought to
represent accretion events onto super-massive black holes (SMBHs)
at the centres of galaxies. The demographics of this population therefore
provide important insights into the formation history of the black
holes we observe in nearly all massive bulges in the nearby Universe
\citep[][]{Richstone1998,Shankar2009}. Moreover an increasing body of
observational evidence \citep[e.g. ][ and references therein]{Kormendy2013}
as well as theoretical calculations \citep[e.g. ][]{Silk1998,Fabian1999,King2005,King2010}
suggest that AGN are also important for understanding the formation
and evolution of galaxies. The evidence above motivated efforts to
constrain the cosmological evolution of AGN \citep[e.g. ][]{Aird2010}
and determine the statistical properties of their host galaxies \citep[e.g.
][]{Alexander2012}. Despite significant progress in these fields
in the last decade important details are still missing. There are
for example, open questions relating to the contribution of obscured
accretion to the black hole mass budget \citep[e.g. ][]{Shankar2004,Akylas2012}
as well as the overall impact of the energy released by AGN on their
immediate environment on kpc and Mpc scales \citep[][]{Kormendy2013}.}

\para{How to}{One approach for addressing the points above is via
studies of the population properties of AGN averaged over cosmological
volumes. The first step in this direction is the characterisation
of the basic properties of individual AGN in a statistical sample,
e.g. their accretion luminosities, level of line-of-sight obscuration
and if possible the basic geometrical properties of the obscuring
material in the vicinity of the SMBH. The importance of X-ray observations
for constraining these properties are twofold. Firstly, selection
at high energies yields AGN samples that are least biased in terms
of either line-of-sight obscuration or dilution by stellar light from
the host galaxy \citep[][]{Comastri2002,Severgnini2003,Mushotzky2004,Georgantopoulos2005}.
Secondly, because the X-ray flux emitted by AGN is believed to originate
at small scales close to the SMBH, spectroscopy at high energies is
a unique diagnostic of the accretion properties and the geometry of
the material in the vicinity of the central engine.}

\para{What do we bring to the table}{In this paper we develop a
novel framework based on the Bayesian inference to analyse the X-ray
spectra of AGN. This method is applied to the problem of the characterisation
of the X-ray spectral properties of AGN detected in blank field surveys.
Such datasets are extensively used to constrain the statistical properties
of AGN and their hosts at moderate and high redshift ($z\approx0.5-3$),
when the bulk of the SMBH we observe in local spheroids were put in
place. This particular application is challenging because (i) typical
deep field X-ray sources only have a small number of photon counts,
well into the Poisson regime, where the familiar $\chi^{2}$-regression
approach is often not applicable, (ii) a large fraction of the deep
field X-ray sources have redshifts measured via photometric methods
\citep[e.g. ][]{Salvato2009,Salvato2011} and therefore suffer uncertainties
that affect the inferred X-ray spectral parameters, (iii) it is often
difficult for traditional spectral fitting methods to select among
different physically plausible spectral models the one that provides
a better representation of the observations. A direct result of the
latter point is that deep field X-ray spectra are typically fit with
simple models \citep[but see for example ][]{Brightman2012}, because
it is unclear whether more complex, but also more realistic, model
families are justified by the observations.

The methodology presented in this paper overcomes the issues above
by propagating into the analysis and the parameter estimation all
the uncertainties related to both Poisson noise and redshift measurement
errors. Bayesian model comparison and selection is also used to infer
from deep survey observations structural information on the distribution
of material close to the SMBH both for individual sources and for
the overall population. The field of choice for the application of
this methodology is the 4~Ms Chandra Deep Field South (CDFS). This
is motivated by the X-ray depth and the availability of multi-wavelength
data in that field.}

}

\para{Paper structure}{We begin in Section \ref{sec:Spectral-components}
by reviewing \longchange{the}{our} current understanding of the
structure of AGN\longchange{,}{} and the physical processes relevant
to X-ray emission in the $0.5-10\,\mathrm{keV}$ band. In Section
\ref{sec:Model-definitions}, the modelling of the spectral components
is defined in detail. Section \ref{sub:Observational-Data} describes
the data used. Section \ref{sub:Statistical-analysis-methods} presents
our approach for comparing models and estimating model parameters.
\longchange{Section \ref{sec:Results} presents the results obtained,
which are }{Finally, the results are presented in Section \ref{sec:Results}
and }discussed in Section \ref{sec:Discussion}.}


We adopt a cosmology of $H_{0}=70.4\,\text{km}\,\text{s}^{\lyxmathsym{\textminus}1}\text{Mpc}^{\text{\textminus}1}$
, $\Omega_{M}=0.272$, and $\lyxmathsym{\textgreek{W}}_{\Lambda}=0.728$
\citep[][]{Komatsu2011cosmology}. Solar abundances are assumed. The
galactic photo-electric absorption along the line of sight direction
to the CDFS is modelled with $N_{H}\approx8.8\times10^{19}\text{cm}^{-2}$
\citep[][]{Stark1992GalHImap}.

\section{Spectral components\label{sec:Spectral-components}}

\para{Intro paragraph}{\change{}{This section describes the main
spectral compoments identified by both observational data and theoretical
work }which constitute the X-ray spectrum of AGN, although not all
of them may be present simultaneously in each source.}

\para{Power law component}{\stmt{X-ray observations show that
all AGN share an intrinsic power law spectrum of the form $E^{-\Gamma}$
with photon index $\Gamma$ and a steep cut-off at higher energies
\just{\citep[][]{Zdziarski2000}}.} \stmt{Observed values for the
photon index range between $1.4$ and $2.8$, and its distribution
in local AGN can be approximated by a Gaussian of \change{$\Gamma=1.95\pm0.15$}{mean
$1.95$ and standard deviation $0.15$.}} \just{\citep[][]{Nandra1994}}.
\stmt{This spectrum is thought to be the result of thermal comptonisation
of accretion disk UV radiation by a hot electron corona \just{\citep[][]{Zdziarski2000,Sunyaev1980}}.
}\just{A hotter plasma can up-scatter photons to higher energies},
thereby \stmt{producing a power law with a lower photon index}.
Up-scattering cools the corona which needs to be repopulated with
high-velocity electrons to stay in equilibrium. \just{The ultimate
source of power is accretion onto the black hole}, and \stmt{thus
$\Gamma$ is thought to be related to physical AGN properties such
as the Eddington rate} \citep[see ][ for a recent discussion]{Brightman2013}.}

\para{Absorption models}{The intrinsic power law is often obscured
by cold material in the line of sight \citep[see e.g. ][]{Turner1997,Risaliti1999,Brightman2011a}.
The most important effects of this obscuring screen for X-rays are
photo-electric absorption, Compton scattering and Fe K-shell fluorescence.
Under the unification scheme of AGN \citep[][]{Antonucci1982,AntonucciUnification1985,Antonucci1993},
which postulates that the viewing angle is the main cause for the
variety of AGN spectra, the simplest common obscuring geometry is
a toroidal structure usually referred to as ``the torus''. X-ray
observations do not resolve AGN and spectral models can not distinguish
the scales at which e.g. photo-electric absorption occurs. However,
because the circum-nuclear obscuring material is cold and molecular,
it must be distant or self-shielding \citep[see e.g. ][]{Gaskell2008}.
This picture is further motivated by high-resolution optical space-based
observations \citep[][]{Ferrarese1996,vanderMarel1998}, which show such
a torus-like structures in $\sim100-1000\text{pc}$ scale, or more
recently, mid-infrared interferometry \citep[][]{Burtscher2013}.}

\begin{figure}
\begin{centering}
\resizebox{\hsize}{!}{\includegraphics{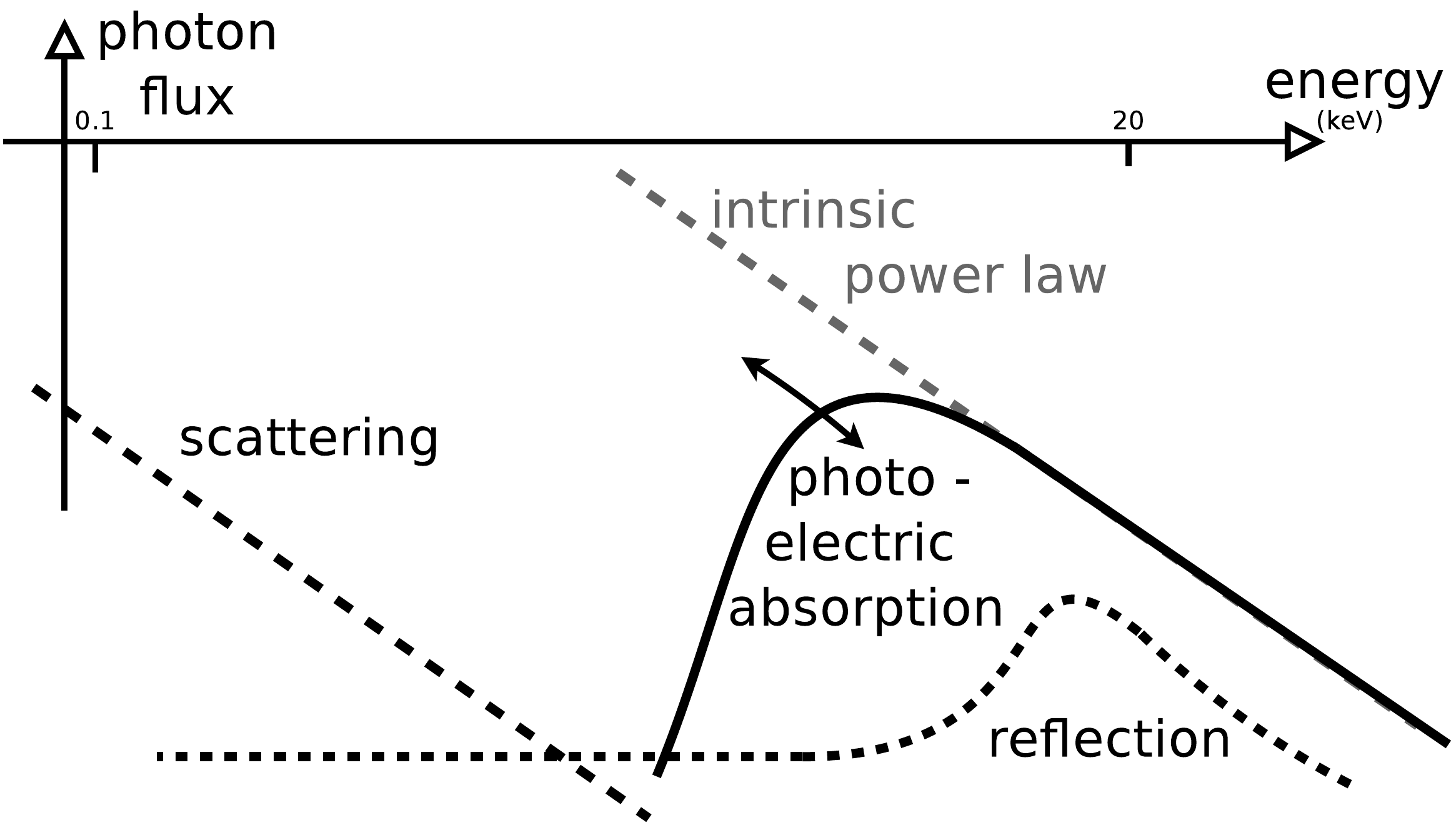}}
\par\end{centering}

\caption{\label{fig:spectralcomponents}Illustration of the typical shapes
of the discussed spectral features. Emission lines and absorption
edges have been omitted for simplicity.}
\end{figure}

\para{Reflection models; angle and geometry dependence}{In addition
to photo-electric absorption, the obscuring material in the vicinity
of the SMBH can also Compton-scatter photons in or out of the line
of sight (LOS). \stmt{Compton scattering, unlike photo-electric absorption,
is anisotropic. From a single, unresolved viewpoint, the geometry
of the obscurer thus influences the obscured spectrum. For example,
compared to a sphere, a torus geometry with a certain opening angle
produces a smaller Compton reflection hump between $\sim10-30\,\mathrm{keV}$
(see Figure \ref{fig:spectralcomponents}) , proportional to the solid
angle illuminated by the source and the surface area exposed to the
observer} \just{\citep[][]{MurphyYaqoobMyTorus2009,Brightman2011a}}.
At neutral hydrogen equivalent column densities $N_{H}\apprge10^{23}\textrm{cm}^{-2}$
Compton scattering becomes important compared to photo-electric absorption
(Compton-regime).}

\para{Soft scattering component}{\stmt{In obscured AGN, it is
believed that a fraction of the intrinsic radiation leaks through
encountering no Compton scattering or photo-electric absorption}
\just{\citep[][]{Ueda2007}}. \stmt{Explanations for this component
include Thomson scattering off ionised material within the torus opening
angle, which reflects the intrinsic spectrum into the line of sight}
\just{\citep[][]{Krolik1987,Turner1997,Guainazzi2007}}. This component
is referred to as ``scattering'' in this work, as opposed to ``Compton
scattering'' or ``reflection'' in the cold obscurer. }

\para{Soft excess}{\stmt{In unobscured AGN, extrapolating the
$2-10\,\text{keV}$ power law to softer energies shows an excess of
soft X-rays in some sources} \just{\citep[][]{Turner1989,Gondhalekar1997}}.
Processes that have been proposed to explain the soft component include
thermal disk emission by the torus at $kT\approx0.2\,\text{keV}$
\just{\citep[][]{Gierlinski2004}}, or a relativistically blurred,
photo-ionized reflection spectrum by the accretion disk \just{\citep[][]{Ross1993}}.
In medium-to-high redshift objects, which constitute the bulk of the
objects of interest in this work, the soft excess component lies outside
the observed energy range and thus we choose to neglect it.}


\section{Model definitions\label{sec:Model-definitions}}

The focus of this work is a study of the AGN population to constrain
the geometry of the X-ray obscuring material in the vicinity of SMBHs.
To identify which physical mechanisms and geometries produce the observed
spectra we identify and compare 10 physically motivated models with
different levels of complexity. For the model comparison, we develop
a Bayesian methodology (discussed in Section \ref{sub:Statistical-analysis-methods})
that propagates the uncertainties from X-ray observations and redshifts
correctly. The data and extraction method is described in Section
\ref{sub:Observational-Data}. The spectral components considered
are defined below.

\begin{figure*}
\begin{centering}
\includegraphics[bb=0bp 88bp 1293bp 655bp,clip,width=17cm]{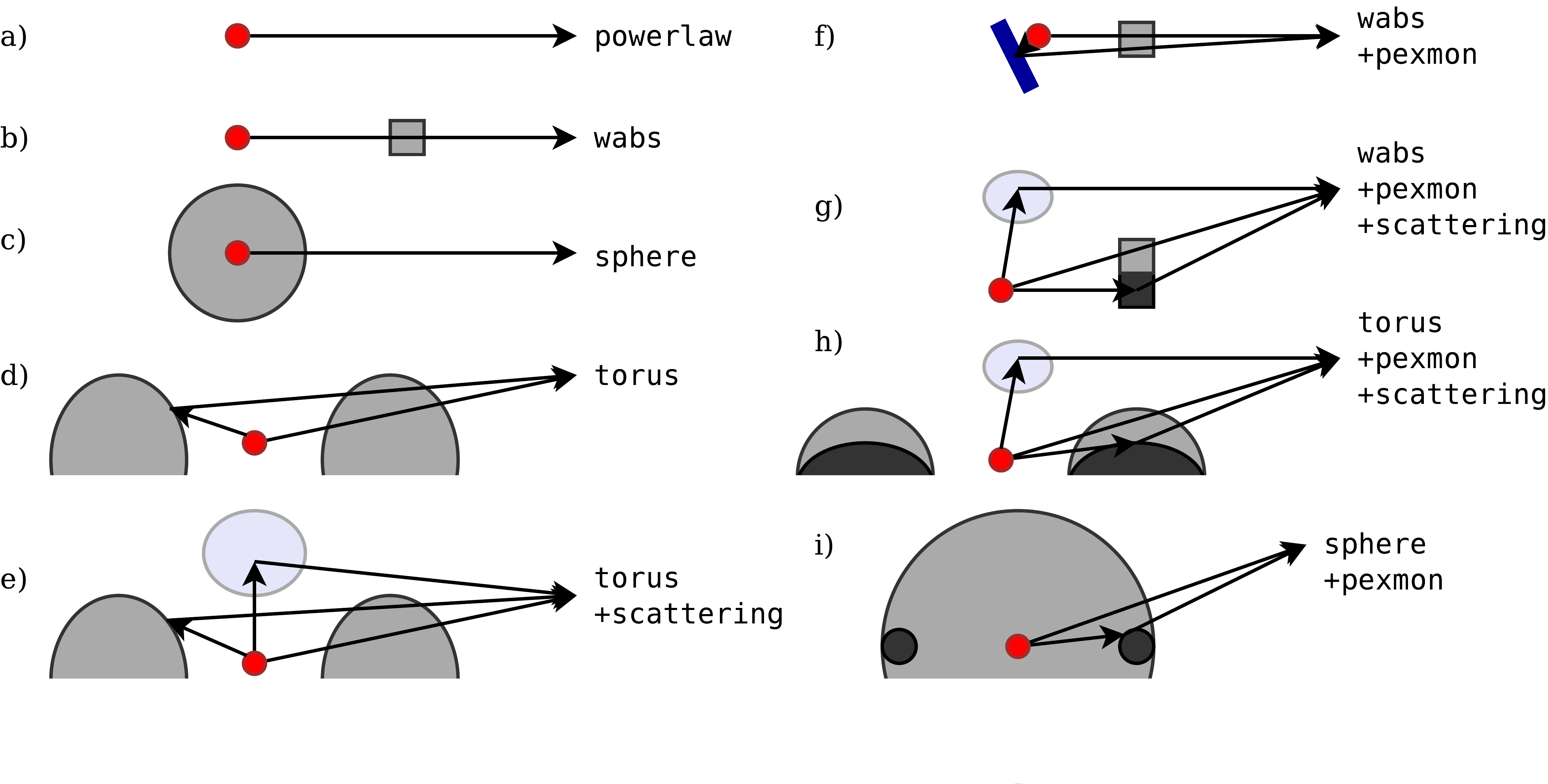}
\par\end{centering}

\caption[Geometry of models]{\label{fig:model-geometry}Cartoon illustrations of the geometries
associated with each model. The \mo{wabs} model (b) represents an
absorbing slab in the line of sight, and can also be interpreted as
the case of a torus with extreme opening angle. While \mo{torus}
(d) uses a intermediate opening angle, the \mo{sphere} (c) represents
the other extreme, a vanishing opening angle. The \mo{+scattering}
extension (e) of the named models correspond to Thomson scattering
from outside the line of sight, which does not experience any energy-dependent
effects. Finally, the reflection component (\mo{+pexmon}) corresponds
to either disk reflection (f) or additional reflection if the torus
is not viewed through the same column density as the reflection (g,
h, i). For the \mo{sphere} it should be noted that scattering is
not physically possible, as no unabsorbed radiation can escape.}
\end{figure*}

\para{Powerlaw model}{The simplest model considered is a power
law, referred to as \mo{powerlaw}. It represents the intrinsic spectrum
of AGN without any obscurer, as shown in Figure \ref{fig:model-geometry}a).
\stmt{The high-energy cutoff observed in the energy range $80-300\,\mathrm{keV}$
\citep[][]{Perola2002} lies well above the energy range used in this
work ($<10\,\text{keV}$) even at moderate redshifts. We therefore
neglect the cutoff. The parameters of \mo{powerlaw} are the normalisation
at $1\,\mathrm{keV}$ and the power law index, $\Gamma$.}}

\para{obscurer effects}{The most commonly used model to describe
obscuration is to apply photo-electric absorption to the intrinsic
power law. Absorption has the largest cross-section among the interaction
processes, and thus is a good first order approximation for the X-ray
spectra of AGN detected in deep fields. However, at higher column
densities ($N_{H}\apprge10^{24}cm^{-2}$), matter becomes optically
thick to Compton scattering, and re-emission in lines due to X-ray
fluorescence is prevalent. In contrast to photo-electric absorption,
which is opaque at low energies, Compton scattering introduces an
energy loss at each interaction, thus low-energy photons can be received
by the observer. Furthermore, Compton scattering induces a non-uniform
scattering angle distribution making the spectrum dependent on the
assumed geometry. Influenced by the solid angle illuminated by the
source and the surface area exposed to the observer, the contribution
by Compton scattering varies, as radiation is scattered into the line
of sight. For instance, a torus geometry with a certain opening angle
produces a smaller Compton reflection hump between $\sim10-30\,\mathrm{keV}$
(see Figure \ref{fig:spectralcomponents}) than a completely closed,
spherical geometry \just{\citep[][]{MurphyYaqoobMyTorus2009,Brightman2011a}}.
\stmt{The opening angle of the torus, and the viewing angle to a
minor degree, thus influences the strength of the reflection hump.}
\stmt{This allows in principle to determine the viewing/opening angle
and $N_{H}$ independently, possibly probing the density gradient
of the obscurer. However, because the effects on the spectrum are
small, particularly in the case of low photon count spectra, this
has not yet been successfully applied to discriminate these parameters
in individual obscured sources. In this work, we also assume a limited
range of geometries (e.g. $45^{\circ}$ opening angle, edge-on view)
as we do not attempt to constrain the viewing and opening angle.}
This geometry-dependence of spectra also makes modelling challenging.
\stmt{Multiple inter-dependent interactions of several elements can
realistically only be done by X-ray radiative transfer simulations
for a fixed geometry} \just{\citep[see e.g. ][]{Nandra1994a,MurphyYaqoobMyTorus2009,Brightman2011a}}.}

\para{preamble obscurers}{In this work, three obscuration models
are considered, \mo{wabs}, \mo{sphere} and \mo{torus}, which correspond
to different geometries:}

\para{wabs model}{The ``\mo{wabs}'' model applies only photo-electric
absorption on a power law with the cross-sections and polynomial approximations
computed by \cite{wabs}. This model does not include any Compton
scattering. It can be considered as an infinitely small blob (see
Figure \ref{fig:model-geometry}b) in the line of sight (LOS): No
radiation is scattered into the LOS, and all Compton scattering leaves
the LOS. However, the loss of high-energy photons due to Compton scattering
at high column densities is not modelled.}

\para{sphere+torus models}{We also consider an absorber with spherical
geometry (\mo{sphere}) illustrated in Figure \ref{fig:model-geometry}c.
This model, computed by \cite{Brightman2011a}, consistently models
photoelectric absorption, Compton scattering and K-shell fluorescence
of a power-law spectrum source at the centre of a cold, neutral medium.
Similarly, a toroidal geometry (\mo{torus}) is used, which was simulated
in the same way as \mo{sphere}, but with a bi-conical cut-out. We
restrain the torus model, using a $45\text{\textdegree}$ opening
angle viewed edge on. The physical scenario represented by this model
is shown in Figure \ref{fig:model-geometry}d, which indicates that
Compton scattering into the LOS is possible.}

\para{obscurer summary}{The three obscuration models used mark
the extreme cases of torus geometries: \mo{wabs} for a almost $90\text{\textdegree}$
half-opening angle, \mo{sphere} for a vanishing opening angle and
\mo{torus} represents an intermediate case where 30\% of the incident
radiation encounters the obscurer. Figure \ref{fig:model-geometry}b-d
illustrates these differences. In comparison to the \mo{powerlaw}
model, \mo{wabs}, \mo{sphere} and \mo{torus} have the additional
parameter $N_{H}$, the neutral hydrogen equivalent column density
in the LOS.}

\para{scattering}{Observations of some obscured AGN show that a
fraction of the intrinsic radiation escapes without encountering any
obscuration. This is understood to be Thomson scattering off ionised
material inside the opening of the obscurer, as illustrated in Figure
\ref{fig:model-geometry}e for the torus geometry. We model this component,
referred to as scattering (\mo{+scattering}) by adding a simple,
unabsorbed power law component with the same $\Gamma$ as the incident
radiation. The normalisation of this component, $f_{scat}$, is modelled
relative to the intrinsic power-law component, and may vary between
$10^{-10}$ and $10\%$. In this fashion, \mo{torus} and \mo{wabs}
are extended to \mo{torus+scattering} and \mo{wabs+scattering}.
For the sphere model, it should be noted that no unabsorbed radiation
can escape, and thus adding scattering is unphysical. However, for
completeness, we also consider \mo{sphere+scattering}\change{}{,
which mimics the case of a torus with a very large covering fraction}.}

\para{reflection}{The observed $N_{H}$ distribution requires a
stronger gradient than a constant-density torus geometry can provide
\citep[see e.g. ][]{Treister2004}. We thus consider an additional density
gradient contribution by adding a denser slab outside the LOS inside
the obscurer. The $N_{H}$ -- measured largely through photo-electric
absorption -- indicates only the ``effective'' column density along
the LOS, and thus remains unaffected. However, a denser region outside
the LOS can scatter Compton reflection into the LOS, as illustrated
in Figure \ref{fig:model-geometry}g-i. Alternatively, reflection
off the accretion disk may also be part of the LOS spectrum entering
the obscurer \citep[][]{Fabian1989,pexmonlines}. This is illustrated
in Figure \ref{fig:model-geometry}f for \mo{wabs}. We thus add a
Compton reflection component \mo{+pexmon} to the obscured spectrum.
We assume that this component passes through the obscuring column,
so it is introduced photon-absorbed with the same column density as
the LOS obscuring material. The used model, \sw{PEXMON} \citep[][]{pexmon}
combines Compton reflection off a neutral dense semi-infinite slab
geometry \citep[\sw{PEXRAV}, ][]{pexrav} consistently with $Fe\, K\alpha$,
$Fe\, K\beta$ and $Ni\, K\alpha$ emission lines and the $Fe\, K\alpha$
Compton shoulder \citep[][]{pexmonlines,pexmonshoulder}. We use the
same incident spectrum (photon index $\Gamma$ and no cutoff) for
the reflection, and follow \cite{pexmon} in assuming a fixed inclination
of $60\degr$. The normalisation of this component is modelled relative
to the \mo{wabs} component ($R$ parameter), and may vary between
$0.1$ and $100$. Figure \ref{fig:model-geometry} illustrates various
causes for an additional reflection spectrum (f-i). In the case of
\mo{wabs} model in particular, the \mo{+pexmon} component can compensate
for the lack of forward-scattering inside the obscurer.}

\begin{figure}
\begin{centering}
\resizebox{\hsize}{!}{\includegraphics{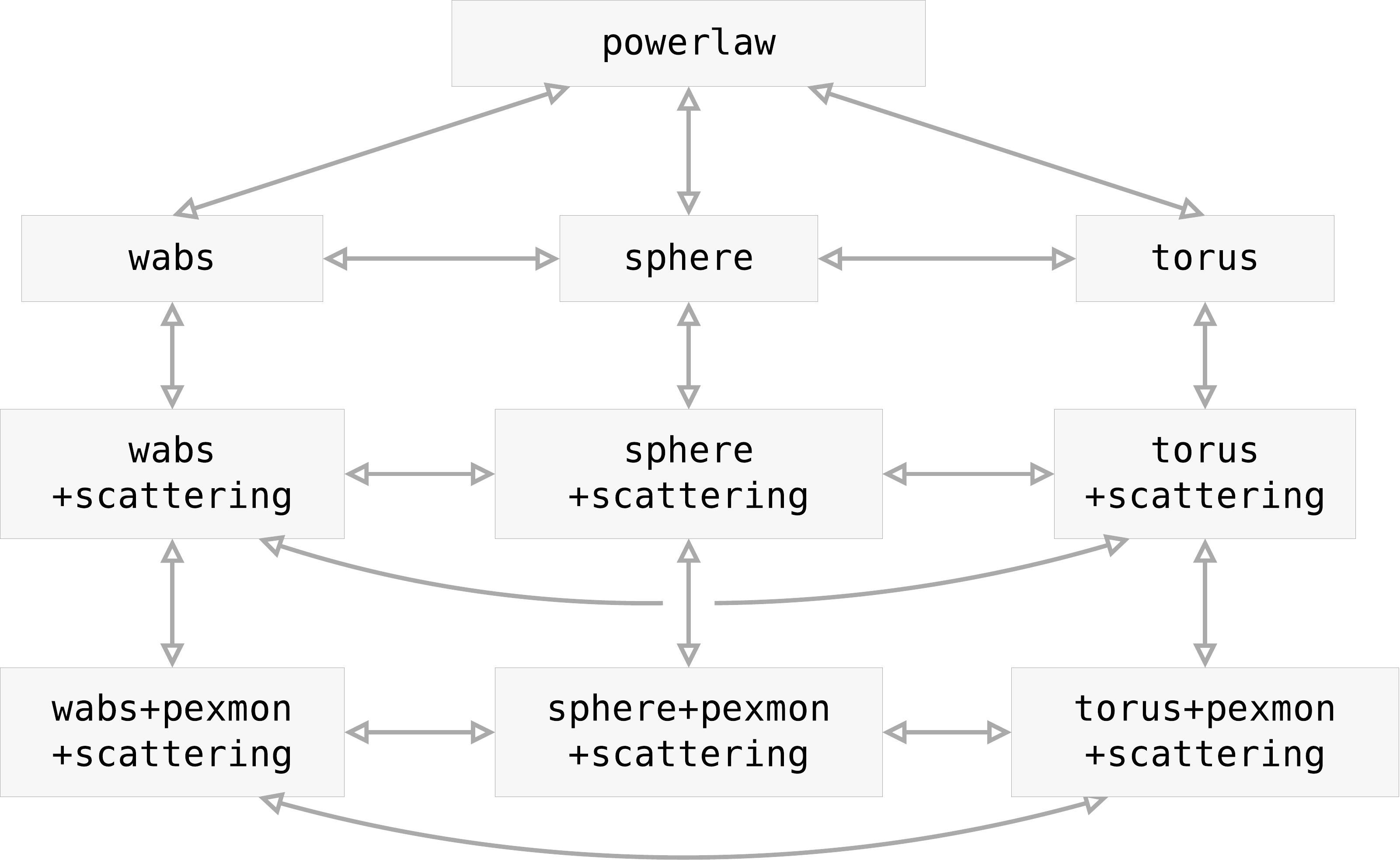}}
\par\end{centering}

\caption[Model hierarchy]{\label{fig:model-hierarchy}Models considered. For model comparison,
we compute the evidence (see Section \ref{sub:Model-selection}) for
each source and model. We then start by assuming a power law (\mo{powerlaw})
and move along the arrows to more complex models if model comparison
justifies the preference. \protect \\
The three obscurer models, \mo{wabs}, \mo{sphere} and \mo{torus},
are compared against each other, as well as the introduction of additional
features (\mo{+scattering}, \mo{+pexmon}). See text for details.}
\end{figure}
\para{final summary}{To summarise, we consider 10 models identified
by \mo{typewriter} font style. The statistical method used for model
comparison is described in Section \ref{sub:Statistical-analysis-methods}.
Figure \ref{fig:model-hierarchy} illustrates the comparisons we are
interested in with arrows. Firstly, obscurer geometries are compared
(\mo{powerlaw}: no obscurer, Figure \ref{fig:model-geometry}a; \mo{wabs}:
bullet-like blob, \ref{fig:model-geometry}b; \mo{sphere}: complete
covering of the source, \ref{fig:model-geometry}c; \mo{torus}: intermediate
case, \ref{fig:model-geometry}d). We test for the existence of a
soft scattering component which is represented by \mo{wabs+scattering},
\mo{sphere+scattering} and \mo{torus+scattering} (Figure \ref{fig:model-geometry}e).
Finally, we explore the need for additional Compton reflection as
shown in Figure \ref{fig:model-geometry} f,g for \mo{wabs+pexmon+scattering},
\ref{fig:model-geometry}i for \mo{sphere+pexmon+scattering} and
\ref{fig:model-geometry}h for \mo{torus+pexmon+scattering}. The
following section describes the data to which we apply our methodology.}

\section{Observational Data\label{sub:Observational-Data}}

\para{Intro}{We intend to study the obscuring material in the vicinity
of active SMBHs by analysing X-ray spectra with a variety of physically
motivated models. \change{Most obscured AGN are found at higher redshifts,
where most of the accretion takes place}{The typical SMBH grows through
active accretion at redshift $0.5-3$ \citep[see e.g. ][]{Aird2010}}.
To obtain a sample of \change{}{such }high-redshift AGN, long exposure
times are needed. The deepest X-ray survey to date is the \telescope{Chandra}
Deep Field South (CDFS) campaign. This survey region has been observed
extensively providing excellent multi-wavelength coverage, which we
utilise for redshift estimation.}

\para{Source detection}{The CDFS survey consists of 51 observations
giving a total exposure time of $4\,\text{Ms}$ on an area of $464.5\,\text{arcmin}^{2}$.
Data reduction and source detection followed \cite{Laird2009} and
is described in detail in \cite{Rangel2013,Rangel2013b}. Briefly,
hot pixels, cosmic ray afterglows and times of anomalously high backgrounds
were removed to produce clean level-2 event files. These were then
aligned using bright sources and subsequently merged. Images and exposure
maps in four energy bands ($0.5\lyxmathsym{\textendash}7$, $0.5\lyxmathsym{\textendash}2$,
$2\lyxmathsym{\textendash}7$ and $4\text{\textendash}7\,\text{keV}$)
were computed. A candidate source list was created using \sw{wavdetect}
with a low significance threshold ($10^{-4}$). Source and background
counts were extracted on the found positions using the Chandra point
spread function tables of \cite{Laird2009}. The source region is
constructed to contain $70\%$ encircled energy fraction (EEF) of
the point spread function (PSF). The background region is a surrounding
annulus between $1.5\times90\%$ EEF PSF and an additional 100 pixel
radius. Other candidate sources are masked from the background extraction
region. For each candidate source position, the Poisson probability
that the observed counts are a background fluctuation was computed,
and the source accepted if the significance exceeds $4\times10^{-6}$
\citep[][]{NandraSrcDet2005}. This yields a final catalogue of 569 sources.}

\para{Data extraction}{The \sw{ACIS EXTRACT} (AE) software package
\citep[][]{BroosACISEXTRACT2010} was used to extract spectra for each
source following \cite{Brightman2012}. Initially, each source and
each pointing is dealt with separately. AE simulates the PSFs at each
source position. Regions enclosing $90\%$ PSF at 1.5keV were used
to extract source spectra. The background regions are constructed
around the sources such that they contain at least 100 counts, with
other sources masked out. AE also constructs local response matrix
files (RMF) and auxiliary matrix files (ARF) using ray-tracing. As
a final step, AE merges the extracted spectra so that each source
has a single source spectrum, a single local background spectrum,
ARF and RMF. It was possible to extract 567 spectra.}

\para{Background model}{For consistent analysis using Poisson statistics,
a model has to be compared to the observed counts. The background
contribution can not be subtracted away because unlike in Gaussian
distributions, the subtraction of two Poisson distributions does not
yield a analytic distribution. It is common practise to use per-bin
background estimates. However, this yields unstable results in bins
with few counts. We thus choose to model the background in a parametric
way using a Gaussian mixture model (see Appendix \ref{sec:Appendix-I:-Goodness-1}
for details). This may not be physical but provides a good approximation
to the background which is very similar to the On-orbit background
measurements in \cite{ACISMemo162} (compare the left panel of Figure
2 therein with Figure \ref{fig:gof-318} in the Appendix). In particular
we model a number of instrumental emission lines by introducing Gaussians
at mean energies of 1.486 ($Al\, K\alpha$), 1.739 ($Si\, K\alpha$),
2.142 ($Au\, M\alpha,\beta$), 7.478 ($Ni\, K\alpha$), 9.713 \citep[$Au\, L\alpha$;
all in keV, ][]{thompson2001x}. We allow the centres of these
lines to vary within $0.1\,\text{keV}$. A feature at $\sim8.3\,\text{keV}$
(possibly $Ni$/$Au$ lines) is described by three additional Gaussians,
while the overall continuum shape is modelled of using a flat continuum
level and two Gaussians at softer energies (see Appendix \ref{sec:Appendix-I:-Goodness-1}
for details). All background model parameters (means, widths, heights)
are then fitted to the background spectrum of each source. The goodness
of the fit is evaluated using Q\textendash{}Q plots (see Appendix
\ref{sec:Appendix-I:-Goodness-1} for details). The found parameters
then remained fixed for the subsequent X-ray spectral analysis of
the source spectrum. In general, there may be cases where simultaneous
analysis of background and source model parameters provides better
results. However, because the larger background region captures many
more background photons than the source region, the background model
is well constrained by the background data alone.}

\newcommand{\selectionall}{526}
\newcommand{\selectionzphot}{476}
\newcommand{\selectionzspec}{301}
\newcommand{\selectionzphotonly}{185}
\newcommand{\selectionzwogal}{352}
\newcommand{\selectionzwoostars}{346}
\newcommand{\selectionzwooounfit}{334}
\newcommand{\selectionunfit}{13}

\para{Redshifts}{X-ray spectra extracted from survey data have
to be associated with multi-wavelength data to obtain a measure of
the distance of each source, i.e. redshifts. For identification of
counterparts, the J/K/H-band positions from the CANDELS \citep[][]{Koekemoer2011,Grogin2012}
CDFS field \citep[][]{Guo2013,Hsieh2012} are used in combination with
mid-infrared positions from IRAC \citep[][]{Damen2011,Ashby2013}. A
Bayesian method was developed that matches X-ray positions to mid-
and near-infrared counterparts simultaneously \citep[][Salvato
et al. in prep]{Budavari2008}. For 528 sources optical counterparts have been found
(see Hsu et al., in prep). For \selectionzspec\  sources, reliable
spectroscopic redshifts are available (Hathi et al. in prep); for
\selectionzphotonly\  of the remaining sources, photometric redshifts
were successfully computed, giving a total of \selectionzphot\  sources.

Optical, IR and UV photometry was used from the TFIT 10-epoch multi-wavelength
catalogue \citep[][]{Guo2013} within the CANDELS GOODS South Survey,
32-band data from the MUSYC survey \citep[][]{Cardamone2010} and Far/Near-UV
from GALEX. Outside the CANDELS region, TENIS J/K band data \citep[][]{Hsieh2012}
was used. The computation of photometric redshift by Hsu et al. (in
prep) follows the same SED fitting procedures described in \cite{Salvato2011},
using SED templates from \cite{Ilbert2009} and \cite{Salvato2009}
and software developed for \cite{Arnouts1999} and \cite{Ilbert2006}.

While it is common practise to merely use the best fit redshifts,
here we marginalise the probabilities to obtain photometric redshift
probability distributions (``photo-z PDFs''). The benefit is that
all solutions are considered in subsequent analysis. The benefits
of propagating the uncertainty of the redshift determination is demonstrated
in Appendix \ref{sec:pdz-test}, by comparing to using a point-estimator.
By an instructive example, it is demonstrated there that Maximum Likelihood
fitting methods can significantly underestimate the uncertainty in
the derived parameters. A future improvement could be to incorporate
systematic errors into these PDFs.}

\para{Contamination}{Star-burst galaxies can contaminate the sample
and skew the inferences we try to make about AGN, especially with
regards to the soft energy ranges. We adopt the criteria for inclusion
of AGN from \cite{Xue2011}: Sources with $L_{X,\,2-10\text{keV}}<3\times10^{42}$,
effective power law index $\Gamma_{eff}>1$ and $\log\, L_{X}/L_{opt}<-1$
are not used. The criterion selects weak X-ray sources that are much
brighter in the optical, and additionally emit mostly soft X-rays.
This selection is designed to exclude sources with non-negligable
host contribution. Thus, AGNs in star-burst galaxies and low-luminosity
AGN \change{}{are }not studied. Furthermore, objects classified as
stars or galaxies (i.e. host-dominated sources in the X-ray) by \cite{Xue2011}
are removed. \selectionzwoostars\ AGN remain.

Model selection assumes that one model is the correct one (see Section
\ref{sub:Model-verification} below). For determining whether the
model could produce the data at hand, we adaptively bin the spectrum
counts so each bin contains 10 counts. Then we compute the $\chi^{2}$-Statistic
for the best fit model parameters. \change{if}{If} $\chi^{2}/n>2$,
where $n$ is the number of bins, the object is not used for model
comparison. For low count spectrum, this criterion is relaxed further
(2.3 if less than 500 counts, 3 if less than 100 counts, 5 if less
than 50 counts) due to the stronger Poisson variance. These limits
were obtained by simulating a flat Poisson spectrum across bins, so
that they exclude the true value in fewer than 1\% of the simulations.
The \selectionunfit\ affected sources were visually inspected in
the X-ray and optical and at least 5 of them can be clearly explained
by outliers in the photometric redshift due to contaminated photometry
or incorrect association. With an outlier fraction of $\eta=4-6\%$,
the expected number of outliers is $\sim20$. \selectionzwooounfit\ AGN
remain.

\begin{figure}
\begin{centering}
\resizebox{\hsize}{!}{\includegraphics{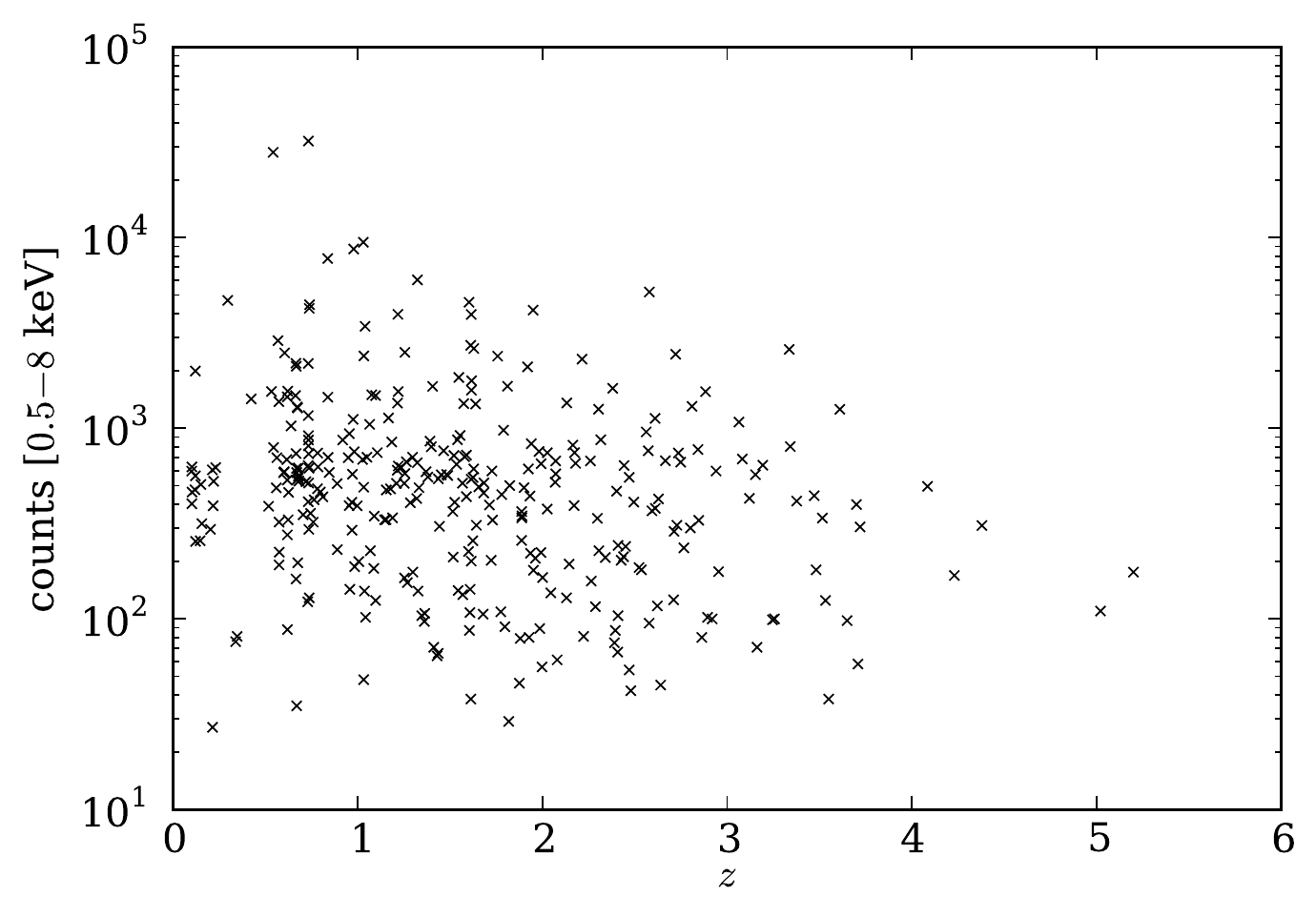}}
\par\end{centering}

\caption{\label{fig:counts-z}The redshift and number of counts in the $0.5-8\text{ keV}$
band of the AGN sample\replace{ (red crosses)}{}. For the redshift,
the spectral redshift is shown where available and otherwise the median
on the photometric redshift distribution is used.}
\end{figure}

It is worth noting that in this work the selection criteria are of
minor importance, as the unification argument is not applied. Instead
of assuming a common torus that is viewed from a variety of angles,
we only require that all sources considered are built using (a subset
of) the same physical processes.

}

\section{Statistical analysis methods\label{sub:Statistical-analysis-methods}}

\subsection{Parameter estimation}

\para{Statistics: $\chi^{2}$ is evil}{In X-ray spectral analysis
it is common practice to vary the parameters of the spectral components
until a certain statistic is optimized with regard to the observed
X-ray spectrum. 

Because X-ray sources in deep extragalactic surveys are typically
faint, their spectral bins are filled by counts according to a Poisson
process. A $\chi^{2}$ statistic is not applicable. Re-binning the
data to use a $\chi^{2}$-statistic is a common practise, however
this results in loss of information in the energy resolution. Additionally,
the reliability of the $\chi^{2}$-estimator is unknown if the incorrect
model ($\chi^{2}\neq n$) is used, and thus its confidence intervals
are problematic.

The maximum likelihood C-statistic $C=-2\times\ln\mathcal{L}_{Poisson}+\mathrm{const}$
\citep[][]{Cash1979}, based on the Poisson likelihood\change{}{ $\mathcal{L}_{Poisson}$},
does not suffer from these issues. However it is not useful as a measure
of goodness-of-fit. Both the $C$-statistic and $\chi^{2}$ denote
the $-2\times\ln$ of a likelihood, and the statements made for $C$
are also applicable to $\chi^{2}$. While $C$ is not $\chi^{2}$-distributed,
$\Delta\chi_{C}^{2}$ is, according to Wilks' theorem \citep[][]{Cash1979,wilks1938large}.}

\para{Optimising is difficult.}{For optimising the parameters,
one can not simply step through the parameter space due to the curse
of dimensionality. Typically, local optimisation algorithms like the
Levenberg-Marquardt algorithm \citep[][]{levenberg:1944,Marquardt1963}
are employed to iteratively explore the space from a starting point.
The final point is taken as the most likely parameter vector. If there
are multiple, separate, adequate solutions, i.e. local probability
maxima, in the parameter space, these algorithms can not identify
them or jump from one local maximum to the other. }

\para{Error estimation is even more difficult.}{For error estimation,
there are three commonly used options: (1) Assume independence of
parameters and Gaussian errors, and estimate the error for each parameter
in turn by finding the value where the statistic drops by a level
corresponding to $1\sigma$. (2) assume Gaussian errors, estimate
the local derivatives and invert the Fisher matrix, (3) compute contours
of some $\Delta C$ value motivated by the $\chi^{2}$ distribution.
The last is more computationally costly than optimisation and can
at most be done in 2 dimensions at a time. These three error estimates
are approximations that are only sufficient in the case where the
probability distributions either approach the shape of a single Gaussian,
or the interdependency between parameters is negligible. A general
maximum likelihood approach that does not have these shortcomings
is to estimate the error using simulated data to find a confidence
interval ($\Delta C$) that contains the used parameter e.g. $95\%$
of the time. This is the most costly option, as it fits to generated
data many times.}

\para{Bayesian approach}{Alternatively, a Bayesian approach can
be adopted, e.g. \cite{Dyk2001} introduces Bayesian X-ray spectral
analysis in detail. As before, the Poisson likelihood is used to explore
a parameter space. However, there are some important differences.
The space is now weighted or deformed by priors. The goal in Bayesian
parameter estimation is to identify sub-volumes which constitute the
bulk of the probability integral over the parameter space. This approach
does optimisation and error estimation simultaneously, but requires
an integration and exploration technique that does not suffer from
the curse of dimensionality.}

\para{MCMC}{Markov Chain Monte Carlo is a commonly employed integration
method for Bayesian parameter estimation. This algorithm tests a starting
point against a new random point chosen from the prior distribution
with a local bias. With a probability proportional to the ratio of
the two points, the algorithm picks the new point as the new starting
point. With each iteration producing a point, a chain is created with
the interesting property that parameter vectors are represented proportional
to their probability. This full representation of the parameter space
can be marginalized to look separately at each parameters distribution.
Furthermore, error propagation in derived quantities can be done correctly
without assuming a underlying normal distribution.}

\para{MCMC problems}{MCMC has its own share of problems however.
First, convergence, i.e. if a chain has become a proper representation
of the parameter space, can not be tested for. Only non-convergence
may be detected. For example, a chain can stagnate for a while but
at a later point suddenly discover a parameter sub-space of high probability.
When or if this would happen however, can not be determined. As a
local algorithm, MCMC has great difficulty finding and jumping between
well-separated maxima. MCMC is a powerful algorithm also applicable
to high dimensionality, and thus many extensions and variants have
been developed for MCMC to mitigate these issues.}

\para{Our approach}{Our approach for circumventing the problems
of MCMC in exploration and integration of parameter spaces is to use
a different Monte Carlo algorithm, Nested Sampling \citep[][]{Skilling2004}.
Nested Sampling keeps a set of fixed size of parameter vectors (``live
points'') sorted by their likelihood. These points are always randomly
drawn from the prior distribution. The algorithm then removes the
least likely point by drawing points until one is found with a higher
likelihood. Effectively, this approach ``scans'' the parameter space
vertically from the least probable zones to the most probable. For
each removed point, the volume this point is representative for is
computed, and the according probability mass added to the integration.
When the remaining parameter volume is negligible, the algorithm terminates.
A difficulty of this algorithm is to avoid the curse of dimensionality
when drawing from the prior distribution to get higher valued points.}

\para{MultiNest}{MultiNest \citep[][]{Feroz2009} solves this problem
efficiently by clustering the live points into multi-dimensional ellipses,
and drawing from these subspaces under the assumption that higher-valued
points can only be found in proximity of already drawn points. Because
of the clustering, MultiNest can follow distinct local maxima without
difficulty. MultiNest is applicable to the low-dimensional problems
of X-ray spectral modeling. It can compute points of equal weighting
akin to a Markov Chain, provide values and error estimates for each
local maximum as well as marginal probability distributions for each
parameter. The integral over the parameter space is computed globally
and for each local maximum. }

\subsection{Model comparison\label{sub:Model-selection}}

\para{Classic model comparison}{For classic model comparison, Wilks'
theorem can be used, which states that the difference in the best
fit $C$-statistic is $\chi^{2}$-distributed with the degrees of
freedom equal to the difference in number of parameters \citep[][]{wilks1938large}.
This asymptotic result is only valid for nested models, i.e. M1 is
a special case of M2.}

\para{Bayesian model comparison }{Bayesian model comparison is
done by comparing the integrals over parameter space, called the evidence
$Z=\int\,\pi(\overrightarrow{\theta})\,\exp\left[-\frac{1}{2}C(\overrightarrow{\theta})\right]\, d\overrightarrow{\theta}$,
where\change{}{$\overrightarrow{\theta}$ is the parameter vector
and } $\pi(\overrightarrow{\theta})$ represents the weighting or
the deformation of the parameter space by the prior. Often, the logarithm
$\log\, Z$ is computed. The Bayes factor $B_{12}=Z_{1}/Z_{2}$ is
then compared to the \textit{a priori} expectation. This method does
not require models to be nested nor does it make assumptions about
the parameter space or the data. Without altering the Bayes factor,
the used statistic can be modified by arbitrary constants as long
as they are independent of the parameter\change{ vector $\overrightarrow{\theta}$}{s}.}

\para{Information-based model comparison}{Alternatively, information
criteria can be used. These are approximations in the limit using
certain assumptions, which compare the difference in the best fit
$C$-statistic between the models to an over-fitting penalisation
term. Typically, the more complex model is punished by the additional
number of parameters. Computation is thus very similar to Wilks' theorem,
although the background is very different.

The Bayesian information criterion \citep[BIC, ][]{Schwarz1978} is an
approximation to Bayesian model comparison. Unlike in the Bayesian
evidence integration, only the maximum likelihood is needed, as the
posterior is assumed to be strongly single-peaked, making the prior
negligible. Because the BIC uses the Laplace integral approximation,
its results are in principle unreliable at the boundaries of the parameter
space (e.g. checking whether a non-negative parameter is zero). The
BIC decides, just like Bayesian model selection, which model is more
probable to have produced the data. The model with the smallest $BIC=C-m\times\ln\, n$
should be preferred, where $n$ is the number of observations (data
points) and $m$ donates the number of free parameters of the model.

The Akaike Information Criterion \citep[AIC, ][]{AIC} is not rooted
in Bayesian inference, but information theory. The AIC measures the
information loss by using a specific model. Thus it can be said to
consider the case of having multiple data sets and predicting the
next. Technically, the Kullback-Leibler divergence, sometimes referred
to as the ``surprise'' or ``information gain'' is minimised. The
model with the smallest $AIC=C-2\times m$ should be preferred.}

\para{The false selection rate and effectiveness of model comparison
methods is tested with simulations in Appendix \ref{sec:False-discovery-rate}.}

\subsection{Model verification\label{sub:Model-verification}}

Any model comparison, or more generally any inference problem, assumes
that one of the stated hypotheses is the true one. If no model is
correct, the least bad model will be preferred. However, this assumption
also requires examination.

Traditionally, this is quantitatively addressed by Goodness of Fit
(GoF) measures such as the reduced $\chi^{2}$ for normal distributions,
with posterior-based approaches such as posterior predictive tests
recently being developed \citep[see e.g. ][]{Bayarri2008}. No consensus
has been reached on these methods. Asymptotic results make assumptions
not applicable in practice \citep[e.g. the Likelihood-ratio and F-test are
invalid at boundaries, making them unsuitable for feature-detection
problems, see ][]{Protassov2002}. Posterior-based approaches are
diverse, and can be overly conservative \citep[see e.g. ][]{Sinharay2003209}.
In the strict sense, there is no probabilistic framework available,
frequentist or Bayesian, for testing whether the fitted model is correct.

We can, however, relax the question to ask whether the model ''explains''
the data by looking at the range covered in the observable. Monte
Carlo simulations (parametric bootstrap) can be performed with the
best-fit parameters to check whether the spectrum can actually be
produced by the given model. This can be computationally expensive
however. It is easier to use a measure of ``distance'' between model
and data (e.g. the Kolmogorov-Smirnov statistic) to discover potentially
problematic cases and to visually examine the data.

Q\textendash{}Q plots \citep[][\change{}{, see Appendix \ref{sec:Appendix-I:-Goodness-1}}]{1968}
provide a generic tool to visualise the goodness of fit, and are independent
of the underlying distribution. The quantiles of the integrated observed
counts are plotted against the integrated expected counts from the
model. A good fit shows a straight line ($y=x$). This method is shown
in Appendix \ref{sec:Appendix-I:-Goodness-1}: The Q\textendash{}Q
plot is used for model discovery for improved fits, while model comparison
using the AIC tests the significance of the model alterations.

\subsection{Experiment design and predictions}

When a result turns out to be inconclusive, it is often interesting
to find out whether a future observation can improve the discriminatory
power, i.e. how well parameters can be constrained or whether two
models can be distinguished. Classically, simulations are employed
with a specific setup. A large number of data sets is generated and
fitted. The distribution of best fit values then shows the uncertainty.

Alternatively, the Bayesian analysis can be employed. Here we only
consider a simple approach, while Bayesian Experimental Design theory
-- a theory based on maximising the information gain -- is not covered.
Since the diversity of possible spectra is less interesting than determining
the discriminatory power regarding parameters or models, it is sufficient
to generate and analyse a single spectrum for each problem as the
Poisson likelihood is already informative of the uncertainty. This
was done in \cite{Georgakakis2013} for Athena+/WFI predictions to
detect and measure outflows. For a grid of problems ($L,\, z$) the
ability to distinguish a warm absorbers from a cold absorber were
tested using the same Bayesian spectral analysis methodology laid
out here, and the uncertainty of parameters recorded.

\subsection{Priors\label{sub:Priors}}

\para{why priors}{A model definition in the Bayesian framework
would be incomplete without describing the priors to the models defined
above. These encode an \textit{a priori} weighting within the parameter
space. Equivalently, a transformation of the unit space to the parameter
space is sought. The simplest case is a uniform prior, giving equal
weight in the parameter space. However, a uniform prior means something
different whether the parameter or its logarithm is used. The parameters
are very similar across the models used, so we describe them all together.}

\para{Which Priors used here}{In the problem at hand, we assume
the prior distributions to be independent. We use log-uniform priors
for scale variables such as normalisations and $N_{H}$. The normalisation
of spectral models is allowed to vary between $10^{-30}$ and $1$.
This conservatively contains physically possible photon fluxes. However,
since all considered models contain this parameter, the limits are
irrelevant in practise. The normalisation of individual spectral components
is then modelled relative to this model normalisation. $N_{H}$ is
defined from $10^{20}$ to $10^{26}\,\mathrm{cm}^{-2}$. \stmt{\replace{}{The
photon index }$\Gamma$ is modelled after the local sample analysed
in \cite{Nandra1994}, as a normal distribution \replace{of $\Gamma=1.95\pm0.15$}{with
mean $1.95$ and a standard deviation of $0.15$}.} \just{This
encodes the assumption that distant AGNs behave like local AGNs with
regard to the\replace{}{ir} intrinsic spectrum}. There is a degeneracy
between $N_{H}$ and $\Gamma$ in that a steeper power law can be
flattened by absorption. Hence, \replace{the}{placing a} prior
on $\Gamma$ plays an important role in constraining $N_{H}$. \replace{}{Nevertheless,
we found that our results (e.g. the $N_{H}$ distribution of the analysed
sample) are insensitive to other choices \citep[for instance $\Gamma=1.68\pm0.3$
from][]{deRosa2012}, as the data drive the result in the majority
of cases. }While redshift and $\Gamma$ use informed priors, all
remaining parameters, unless otherwise stated, are assumed to be location
parameters and thus have a uniform prior.} 

\para{Are priors dominant}{Of course we are not interested in our
prior beliefs, but in how the data strengthens or weakens hypotheses
and re-weights parameter space. If the data has no discriminatory
power, there is no information gain and the prior and posterior probability
distributions will look the same. The difference between the prior
and posterior can be measured using the Kullback-Leibler divergence
\citep[][]{kullback1951information} as $\text{KL}=\int\, posterior(x)\,\log\frac{posterior(x)}{prior(x)}\,{\rm d}x$,
essentially a integral difference across the parameter space. The
KL divergence is measured in $\text{ban}$, a unit of information
or entropy. A considerable information gain is e.g. $KL>0.13\,\text{bans}$
which corresponds to halving the standard error of a Gaussian. We
will restrict ourselves to computing the information gain only for
the $N_{H}$ parameter.}

\para{comparing models}{For model comparison, we furthermore need
to specify our \textit{a priori} preference of models. We consider
two approaches: a) pair-wise comparisons from low to high complexity
(e.g. \mo{torus} vs. \mo{torus+scattering}, see Figure \ref{fig:model-hierarchy}),
and between models of the same complexity. We adopt the scale of \cite{jeffreys1961theory}:
A Bayes factor above 100 is ``decisive'', 30 ``very strong evidence'',
10-30 ``strong evidence'', 3-10 ``substantial evidence''. In $\log\, Z$,
this corresponds to differences of 2, 1.5, 1 and 0.5 respectively.
In case of a Bayes factor below 10 we remain with the simpler model.
b) comparing all models simultaneously to find the model with the
highest evidence. In both cases we consider the models \textit{a priori}
equally probable.}

\subsubsection{Implementation}

\newcommand{\swname}[0]{BXA}

\para{Software written}{For the spectral analysis in the framework
of Bayesian analysis, the \sw{MultiNest} library was used and two
software packages were created: \sw{PyMultiNest} is a generic package
for connecting Python probability functions with the \sw{MultiNest}
library, as well as model comparison and parameter estimation analysis
of the \sw{MultiNest} output. \sw{\swname} is a package that connects
the \sw{Sherpa} X-ray analysis framework \citep[][]{Freeman2001} with
our Bayesian methodology. \sw{(Py)MultiNest} repeatedly suggests
parameters on a unit hypercube which are transformed by \sw{\swname}
into model parameters using the prior definitions. \sw{\swname} then
computes a probability using \sw{Sherpa}s C-stat implementation,
which is passed back to \sw{(Py)MultiNest}. \sw{(Py)MultiNest}
can then be used to compute Bayes factors, create one or two-dimensional
marginalised posterior probability distributions (PDFs) and output
summarising Gaussian approximations.}

\para{Software versions}{\sw{\swname} and \sw{PyMultiNest} are
publicly available on \url{http://github.com/JohannesBuchner}. In
this work, MultiNest v2.17 is being used by \sw{\swname} on \sw{Sherpa}
version 4.4v2 with 400 live points and a log-evidence accuracy of
$0.1$. For further analyses, we made extensive use of the Numpy/Scipy,
Matplotlib and Cosmolopy packages \citep[][\url{http://roban.github.com/CosmoloPy/}]{scipy,matplotlib}
packages.}


\section{Results \label{sec:Results}}

\begin{figure*}[t]
\begin{centering}
\includegraphics[width=8.5cm]{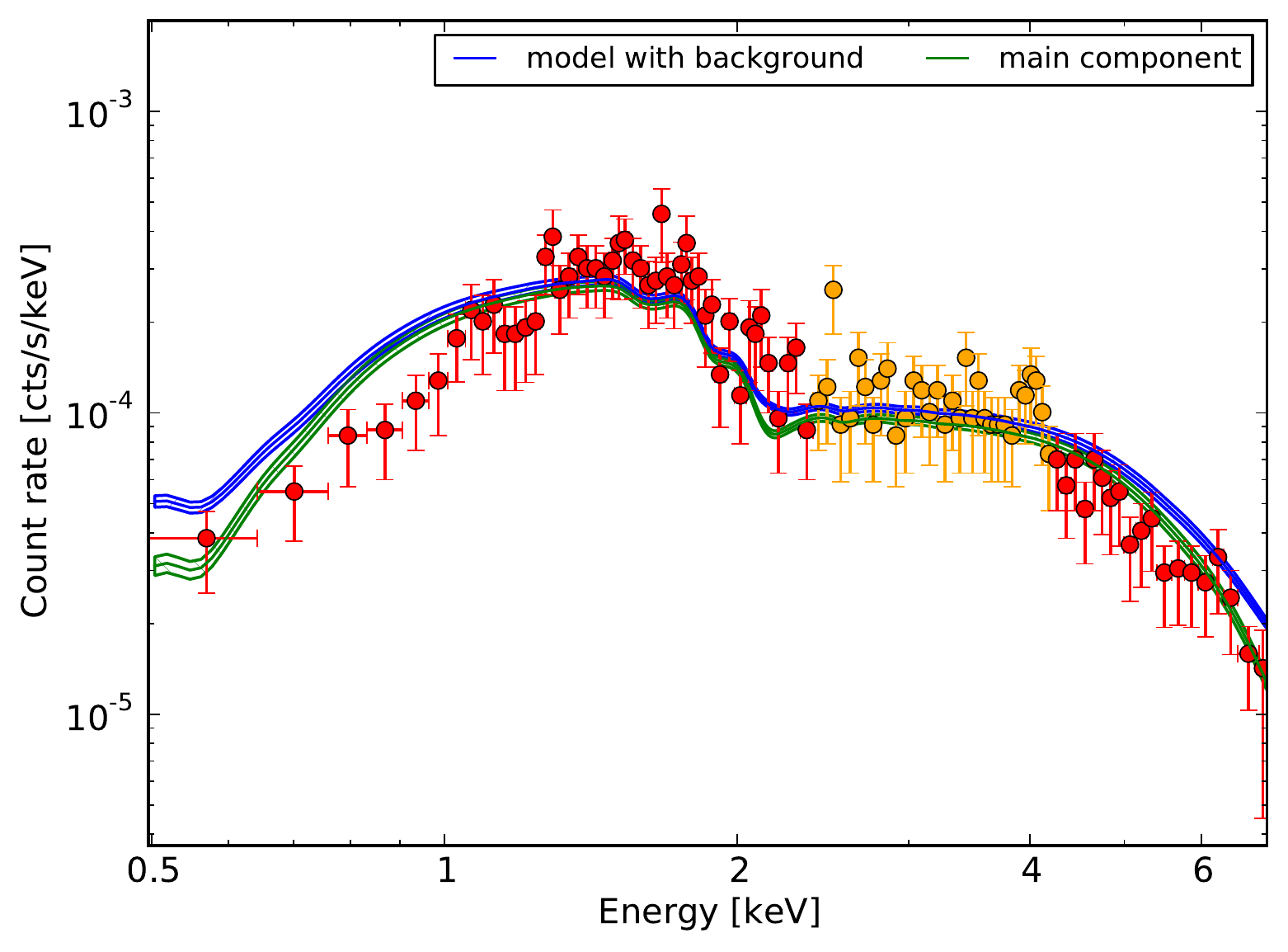}\includegraphics[width=8.5cm]{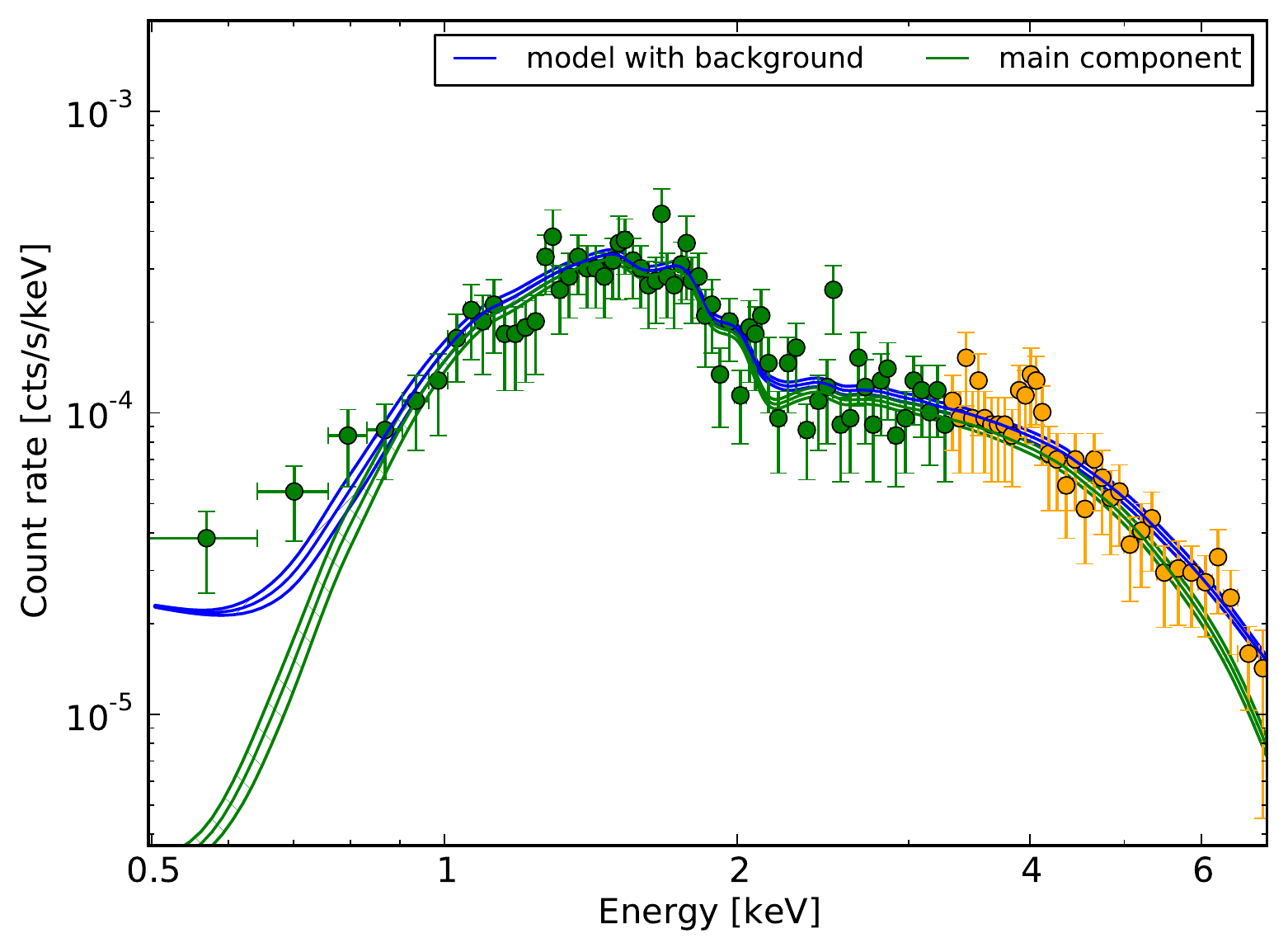}
\par\end{centering}

\begin{centering}
\includegraphics[width=8.5cm]{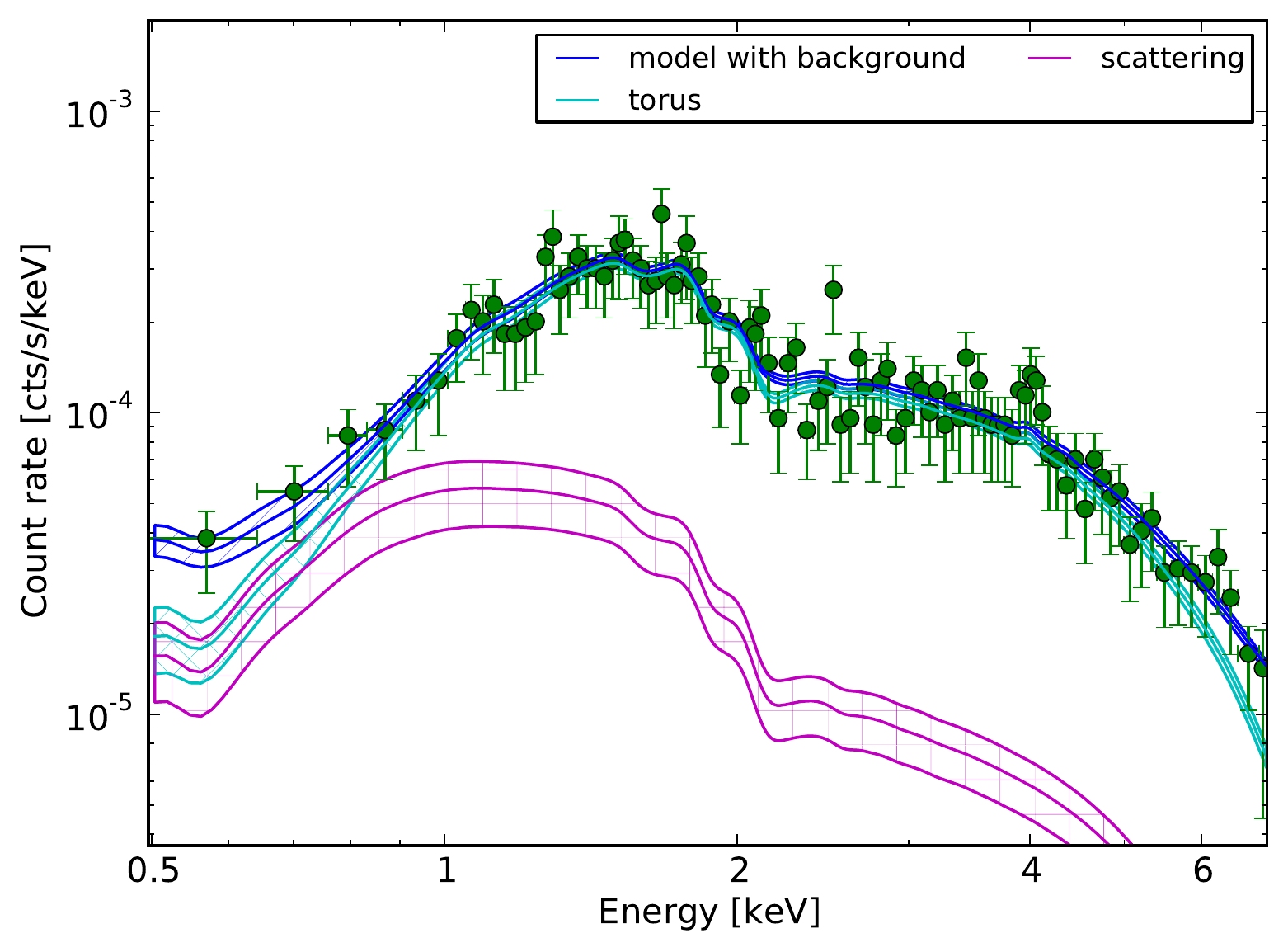}\includegraphics[width=8.5cm]{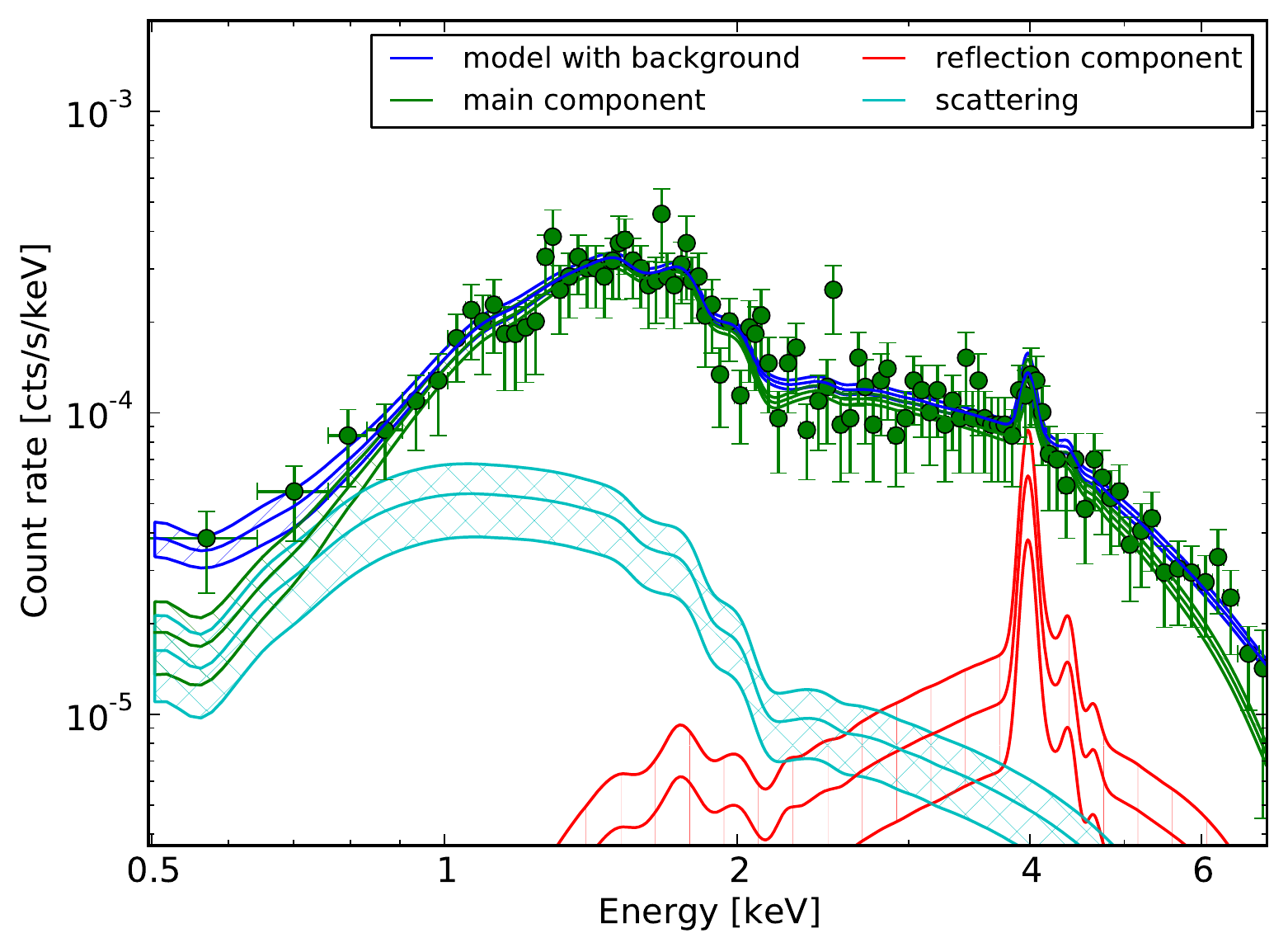}
\par\end{centering}

\caption{\label{fig:refl-example-spec} Observed (convolved) spectrum of object
179, binned for plotting to 10 counts per bin. Shown are analyses
using various models and their individual components: \mo{powerlaw}
(upper left), \mo{wabs} (upper right), \mo{torus+scattering} (lower
left) and \mo{wabs+pexmon+scattering} (lower right). The posterior
of the parameters are used to compute the median and 10\%-quantiles
of each model component.}
\end{figure*}

\begin{figure*}[p]
\includegraphics[height=3.5cm]{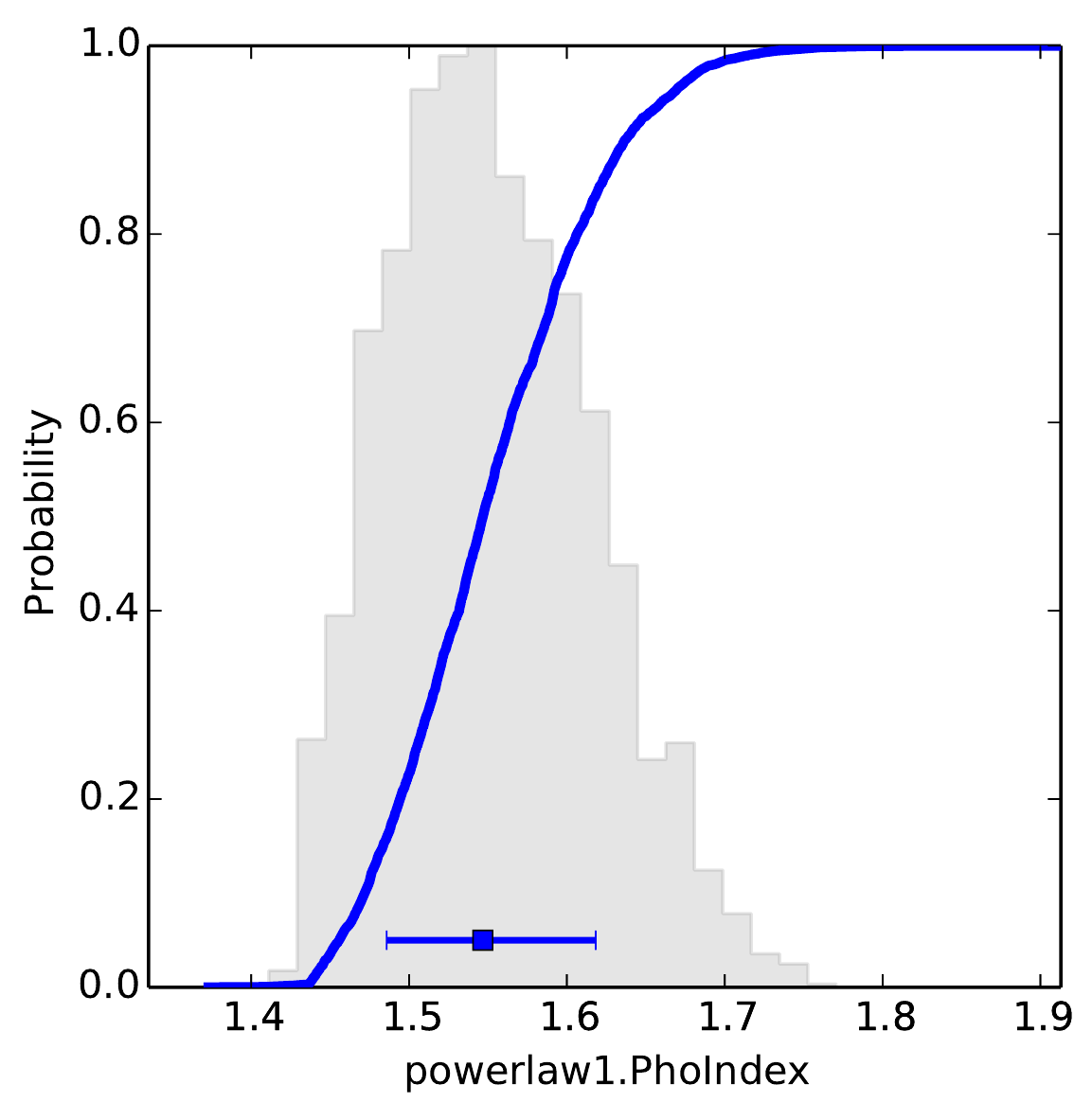}\includegraphics[height=3.5cm]{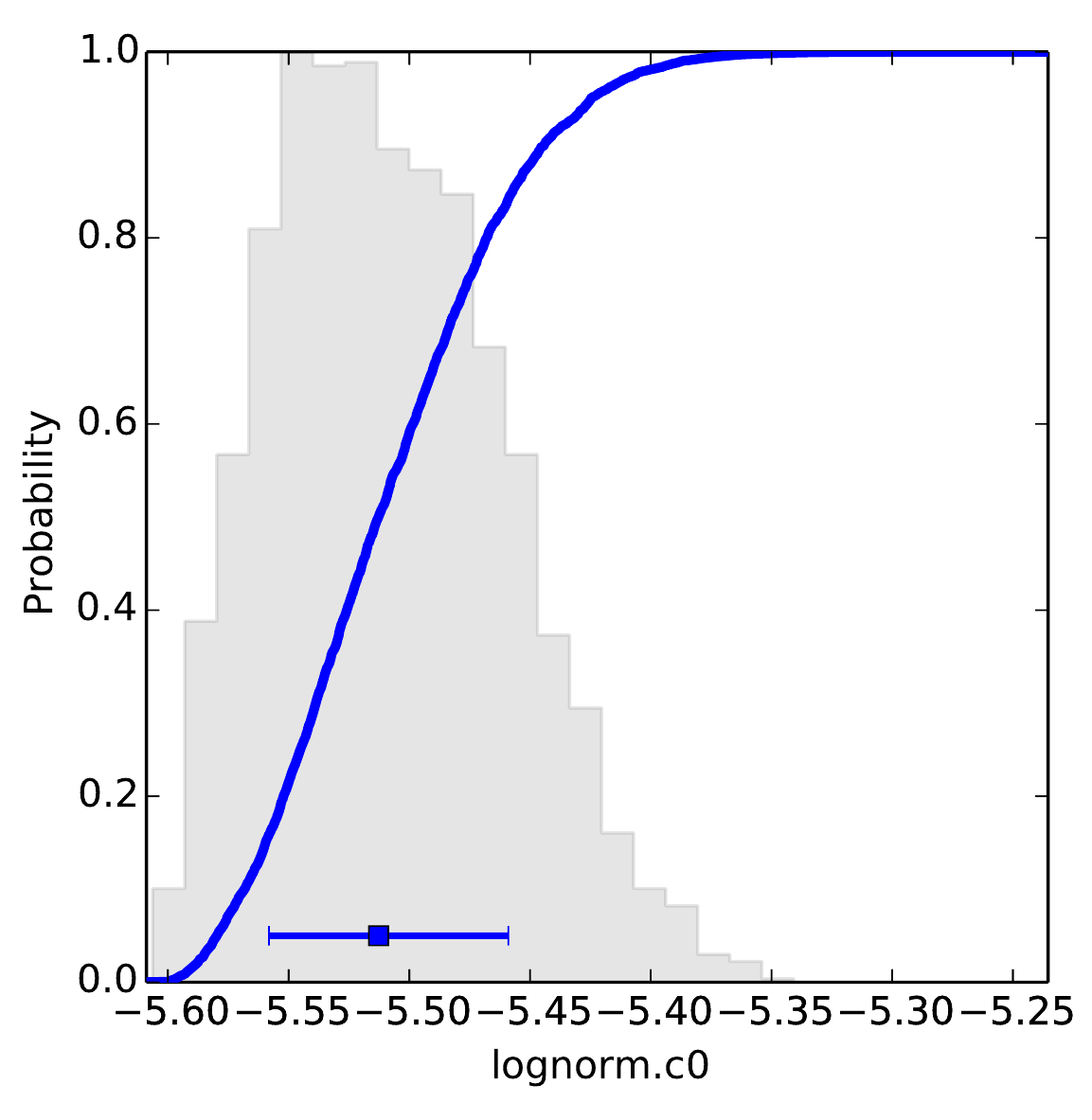}\includegraphics[height=3.5cm]{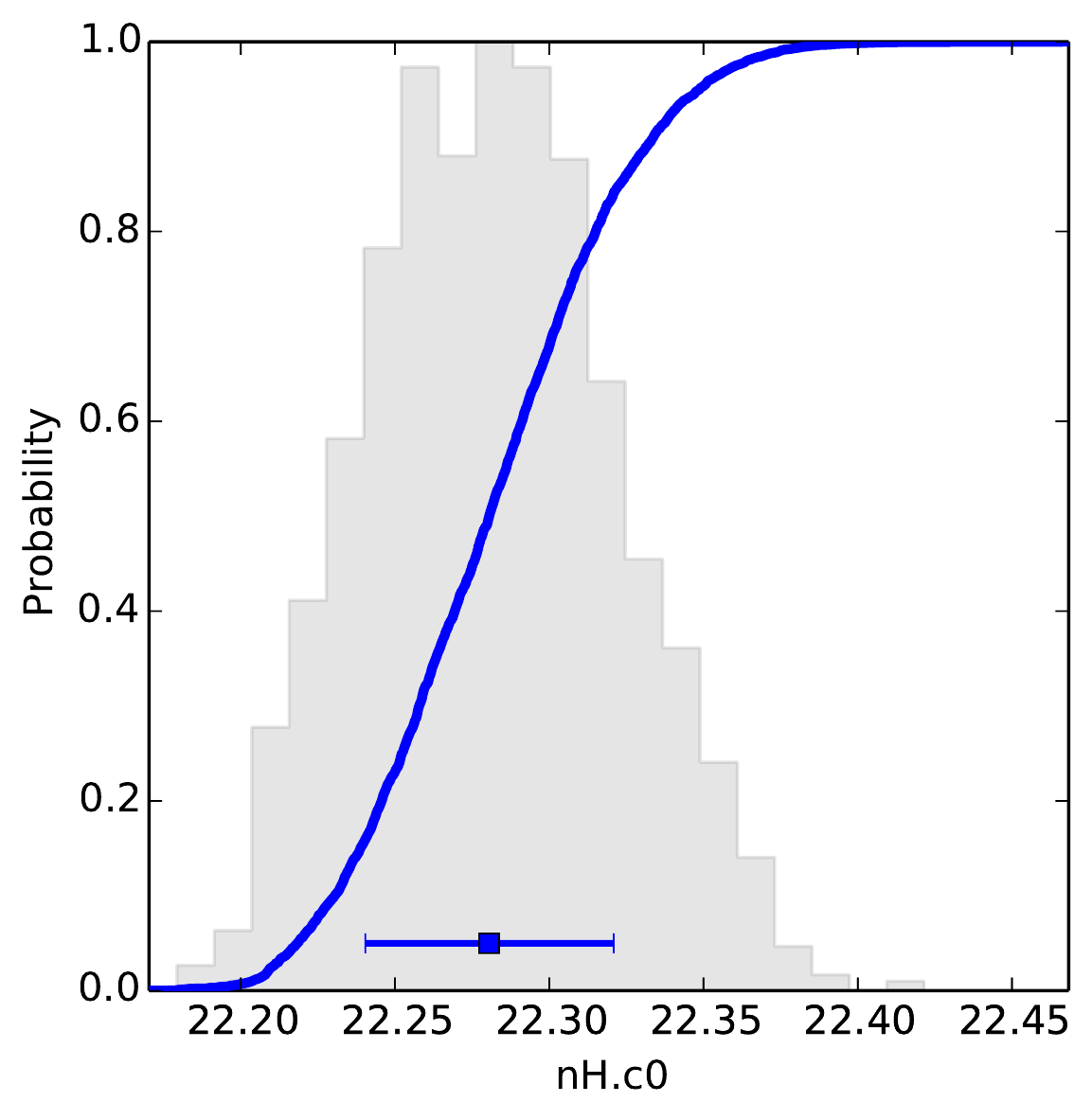}

\includegraphics[height=3.5cm]{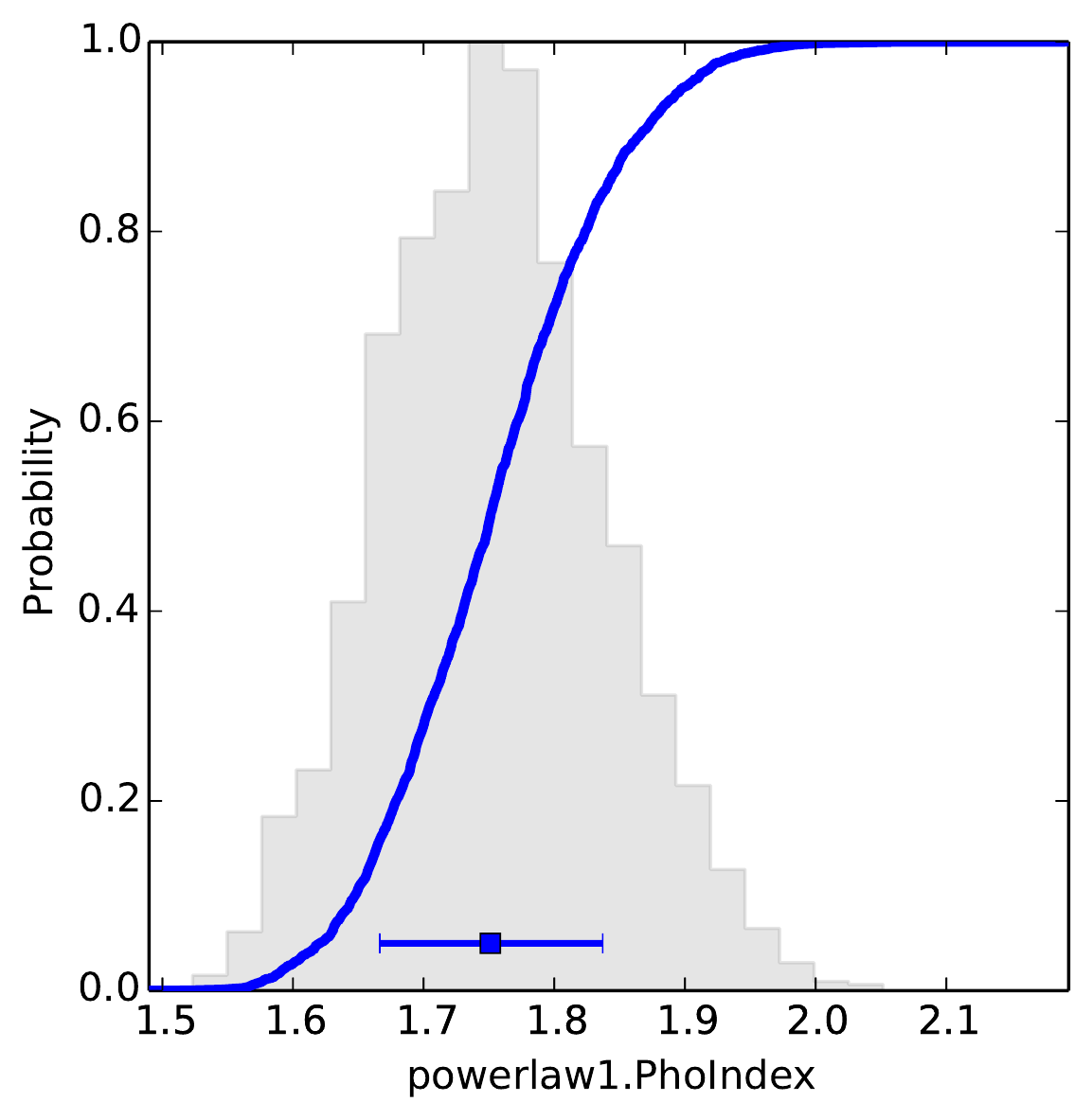}\includegraphics[height=3.5cm]{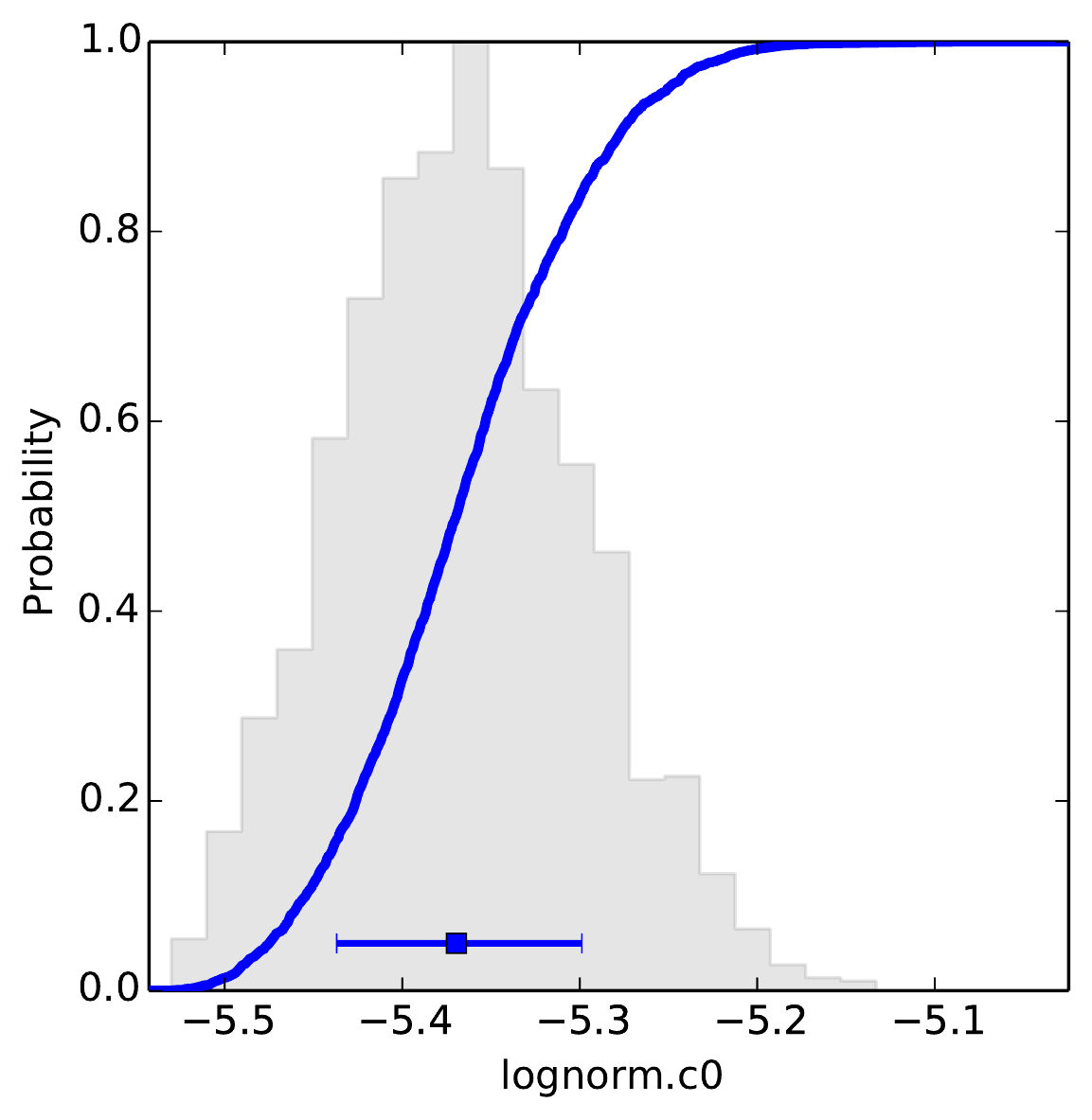}\includegraphics[height=3.5cm]{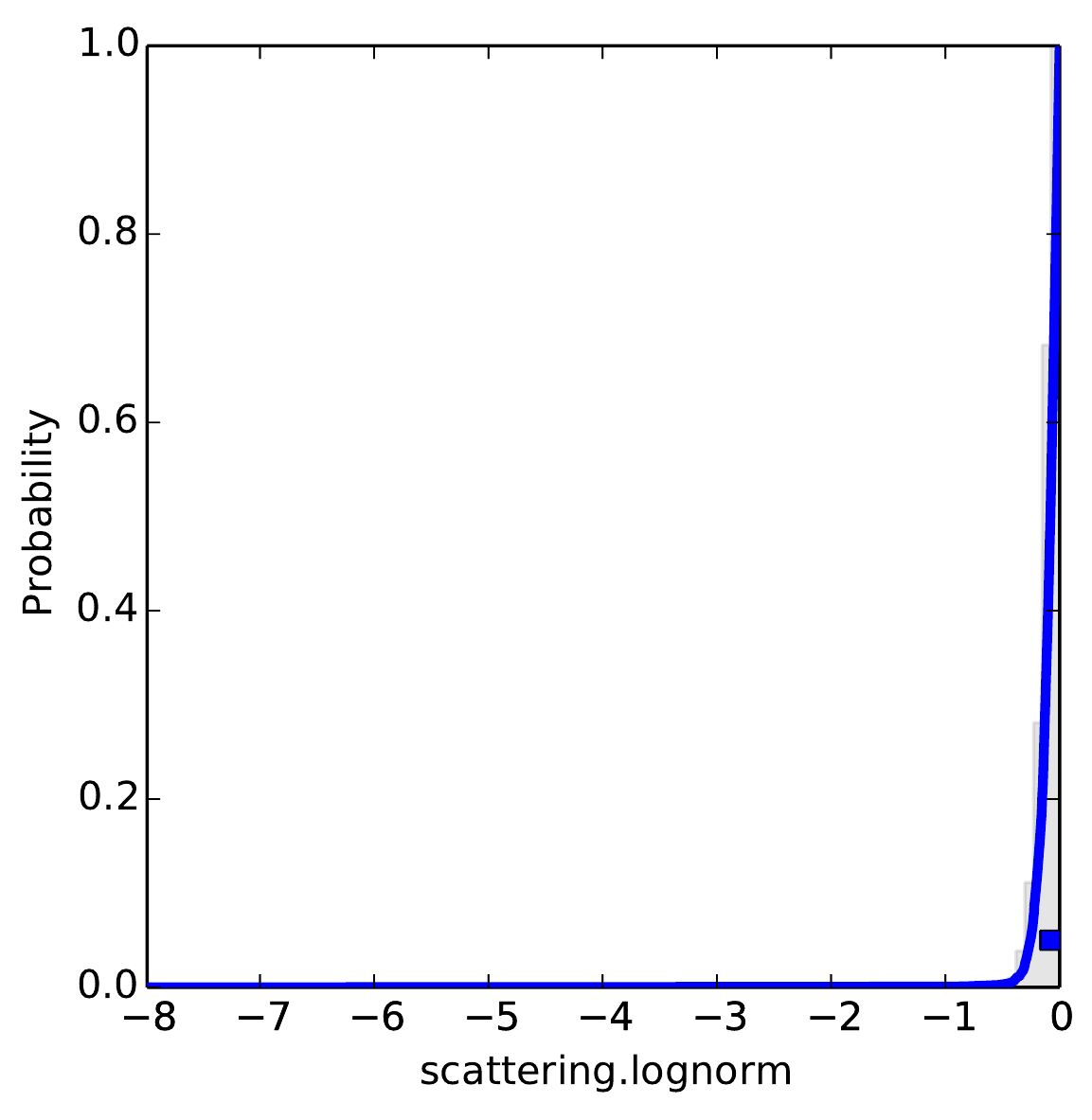}\includegraphics[height=3.5cm]{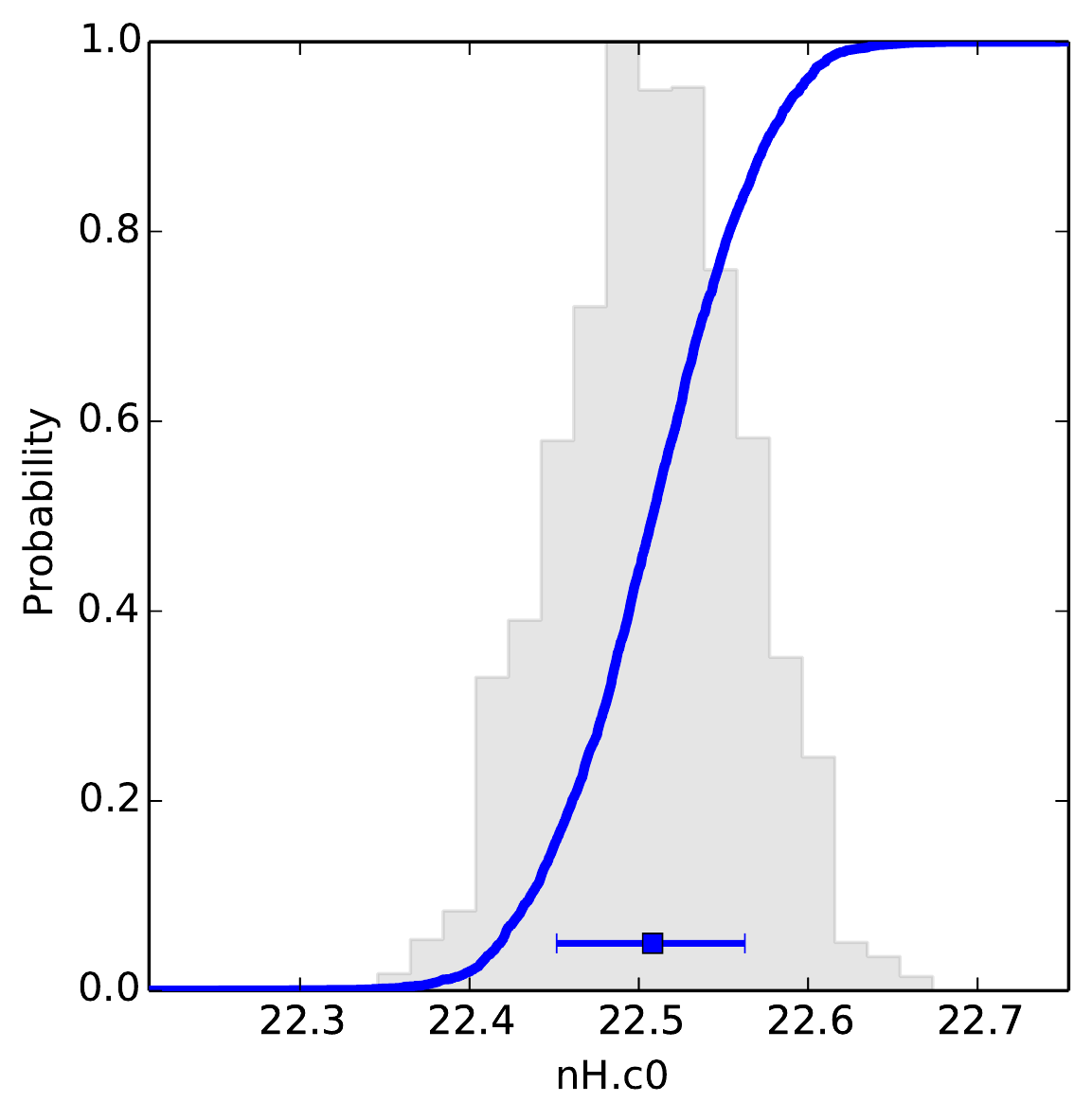}

\includegraphics[height=3.5cm]{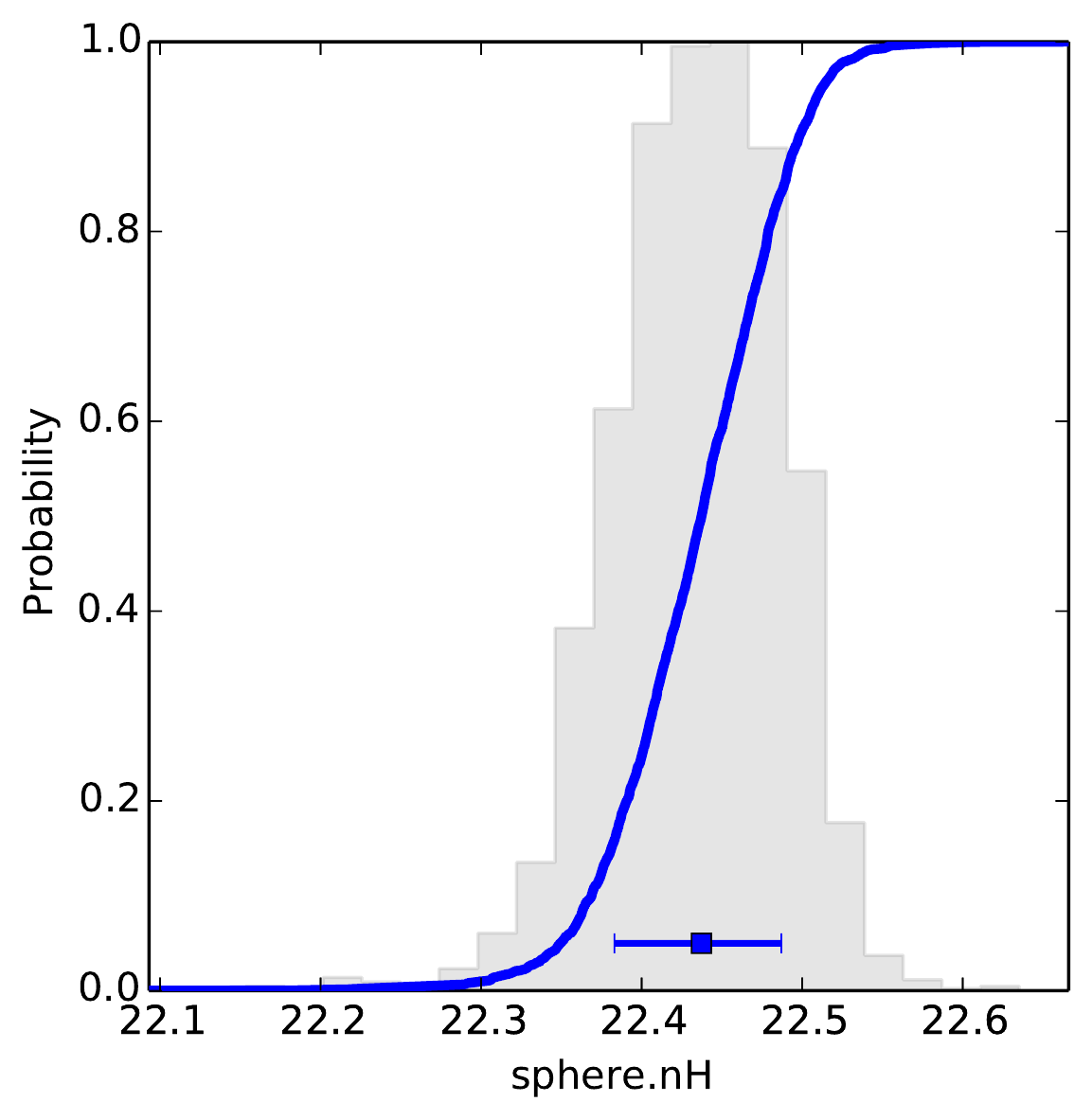}\includegraphics[height=3.5cm]{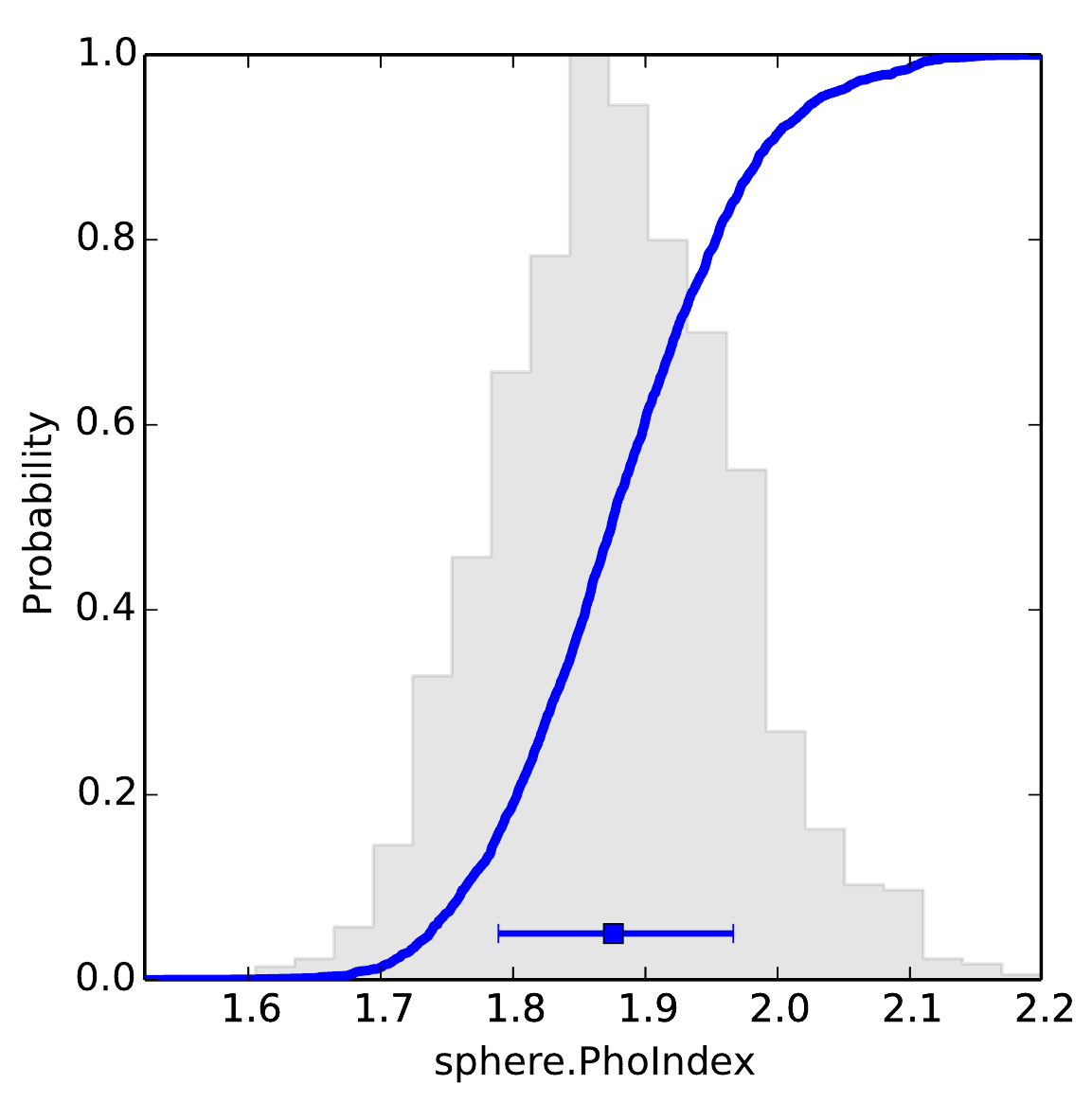}\includegraphics[height=3.5cm]{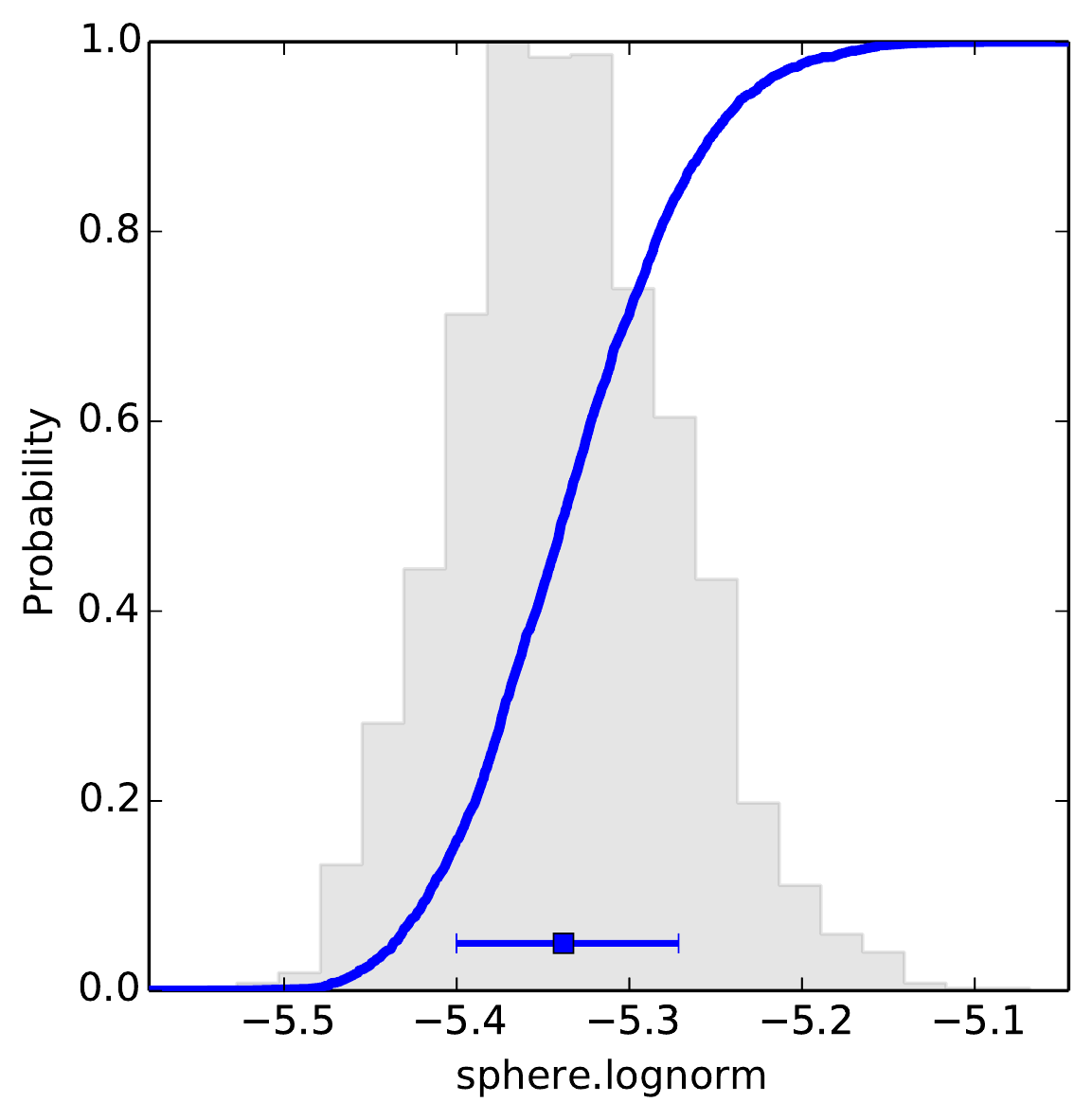}\includegraphics[height=3.5cm]{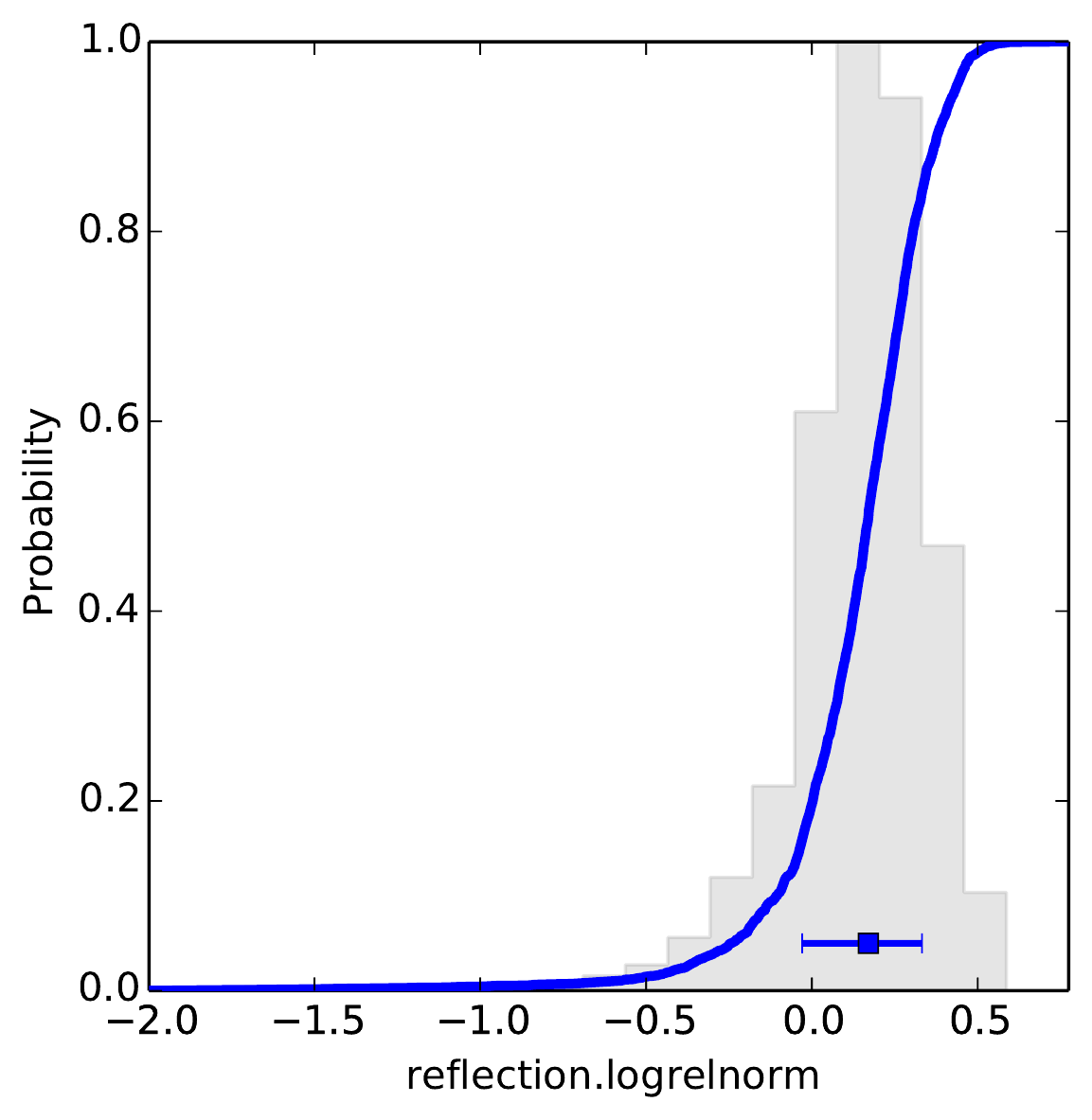}\includegraphics[height=3.5cm]{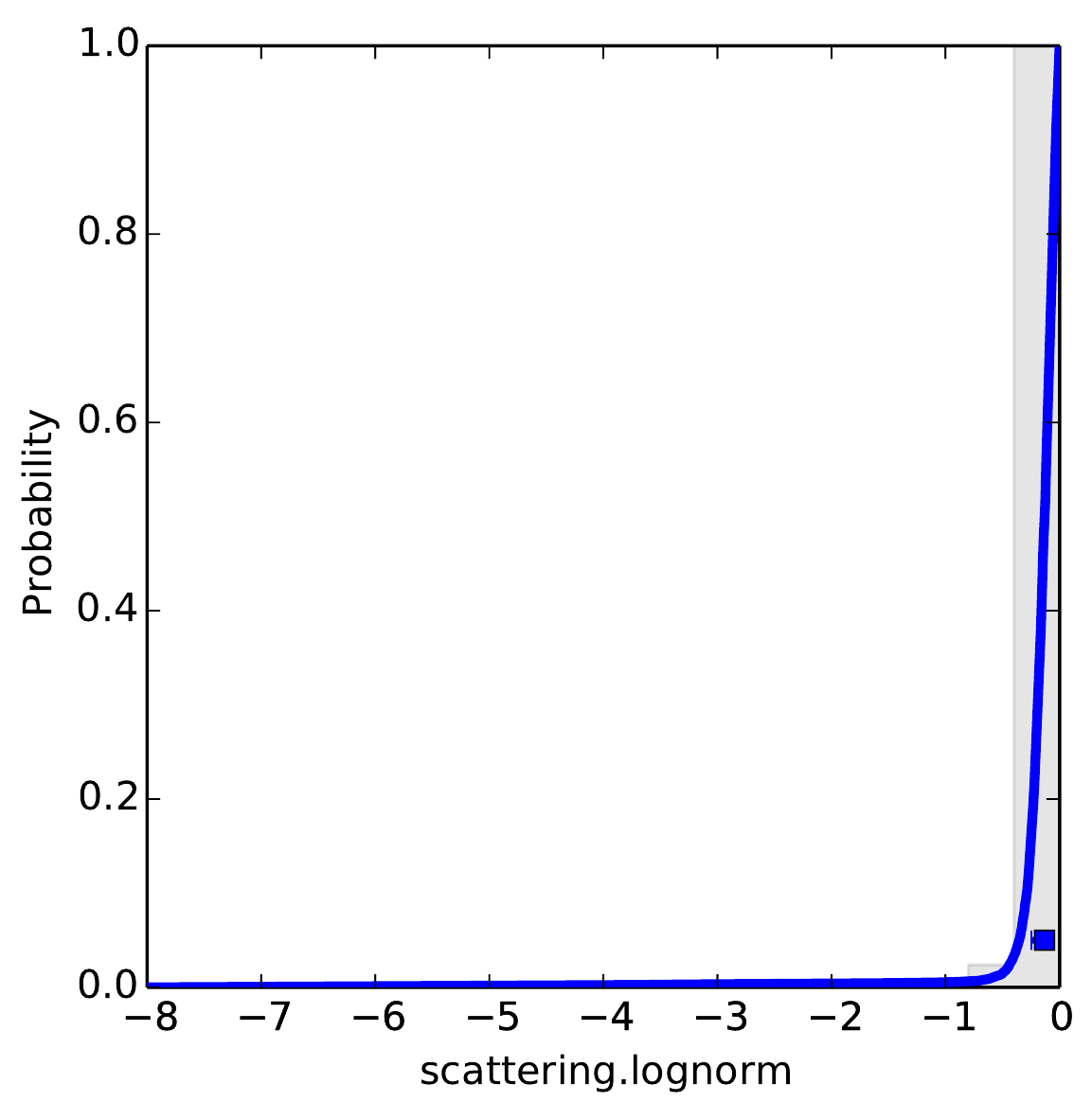}
\includegraphics[height=3.5cm]{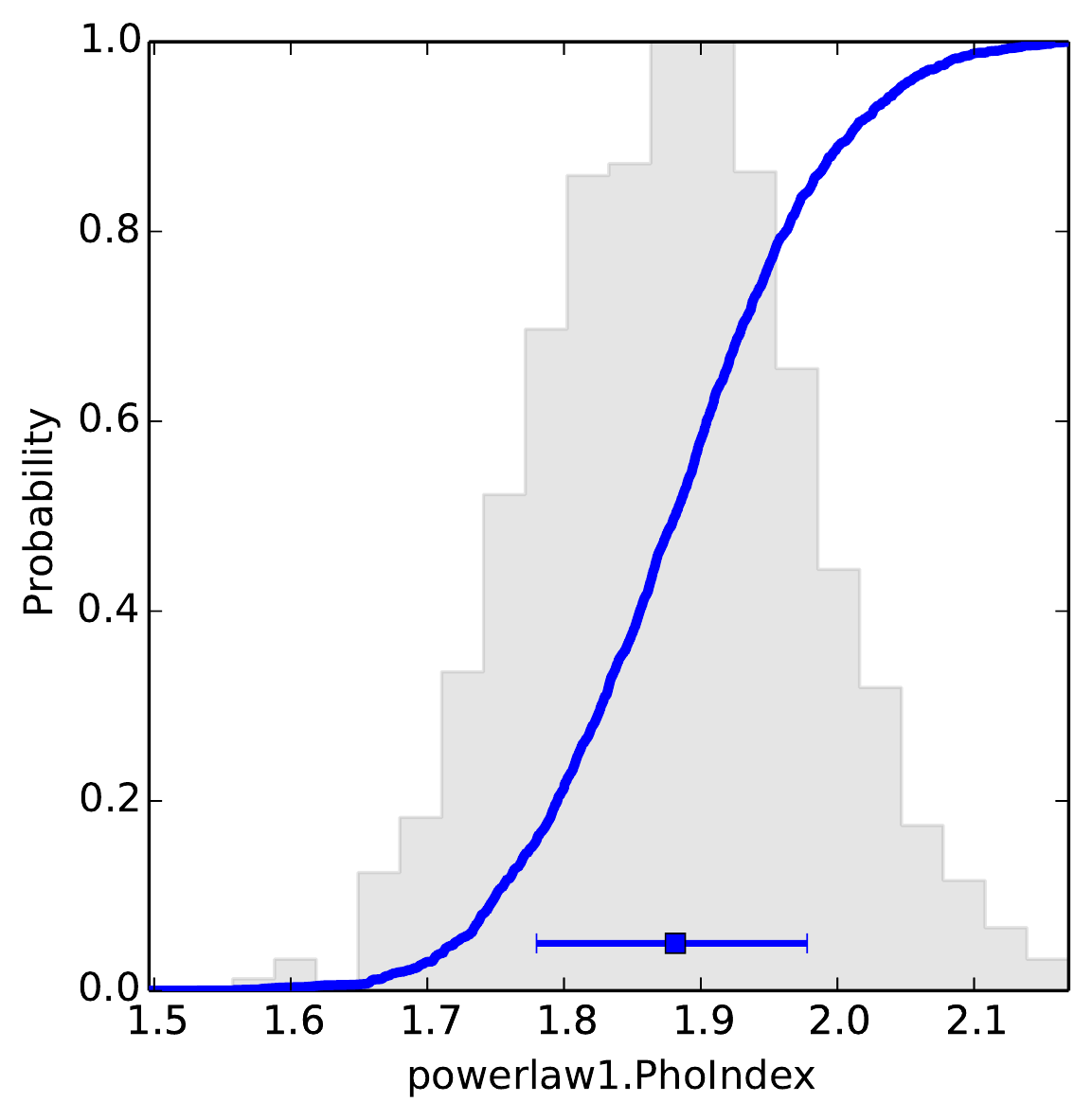}\includegraphics[height=3.5cm]{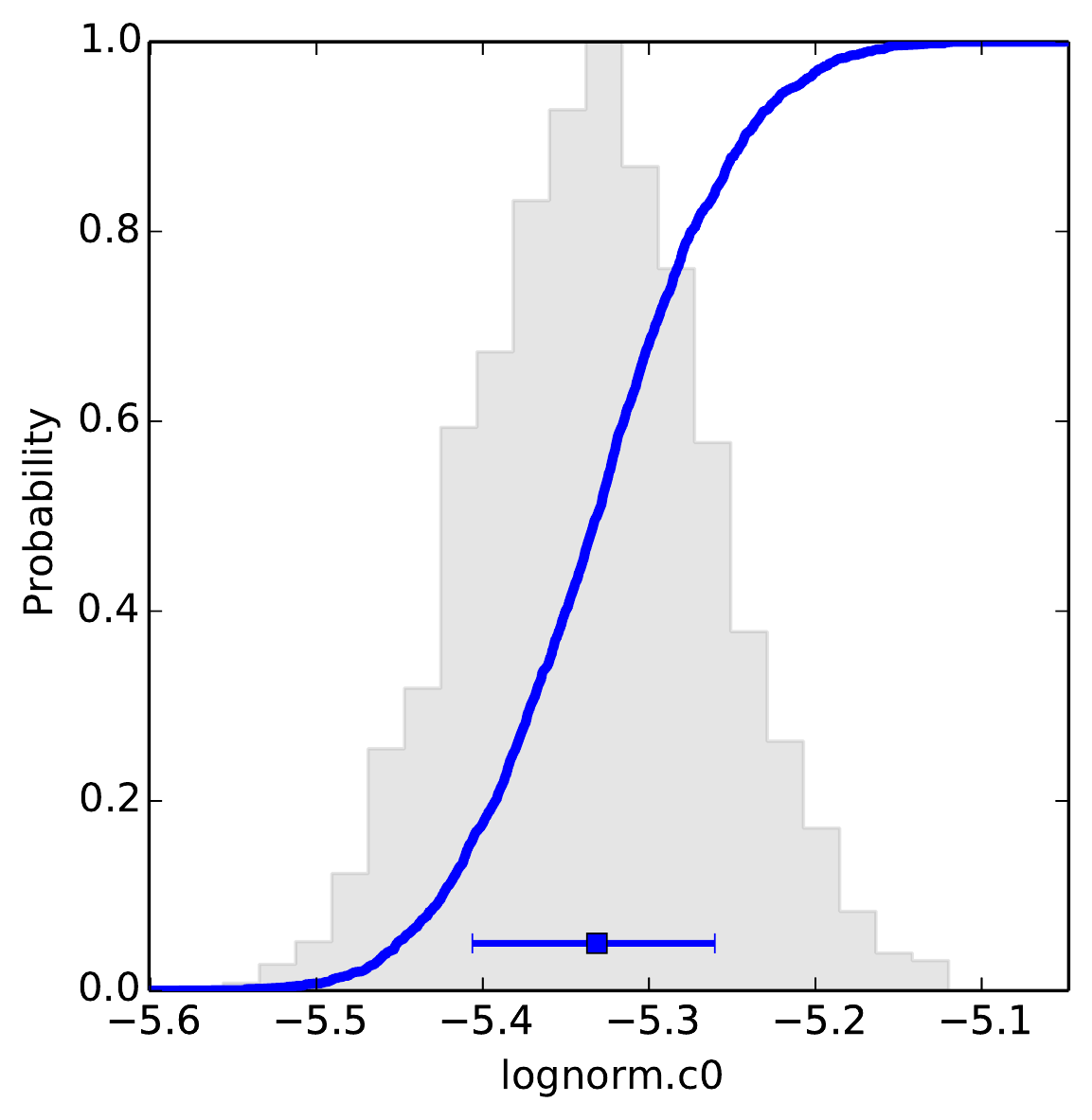}\includegraphics[height=3.5cm]{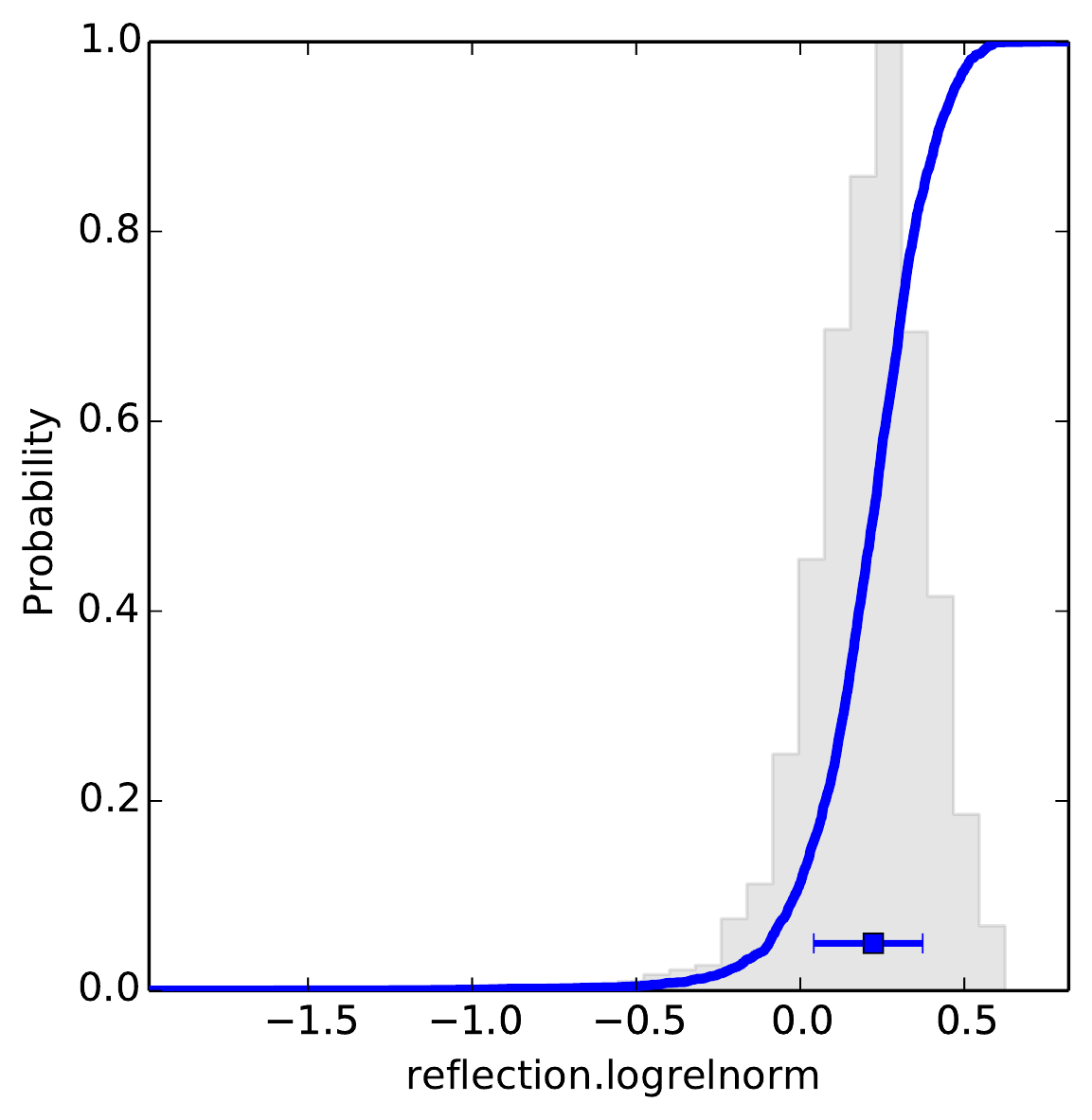}\includegraphics[height=3.5cm]{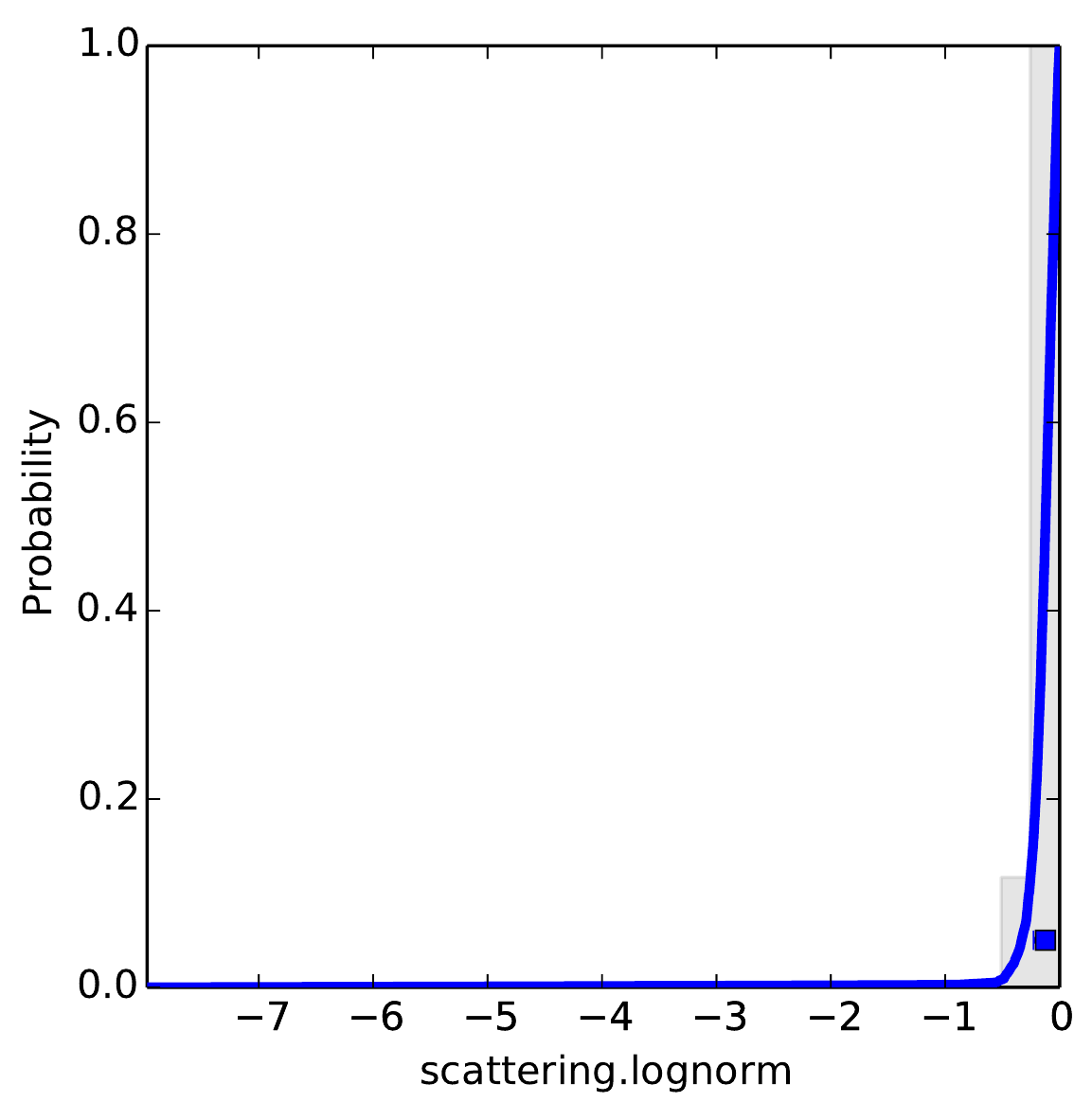}\includegraphics[height=3.5cm]{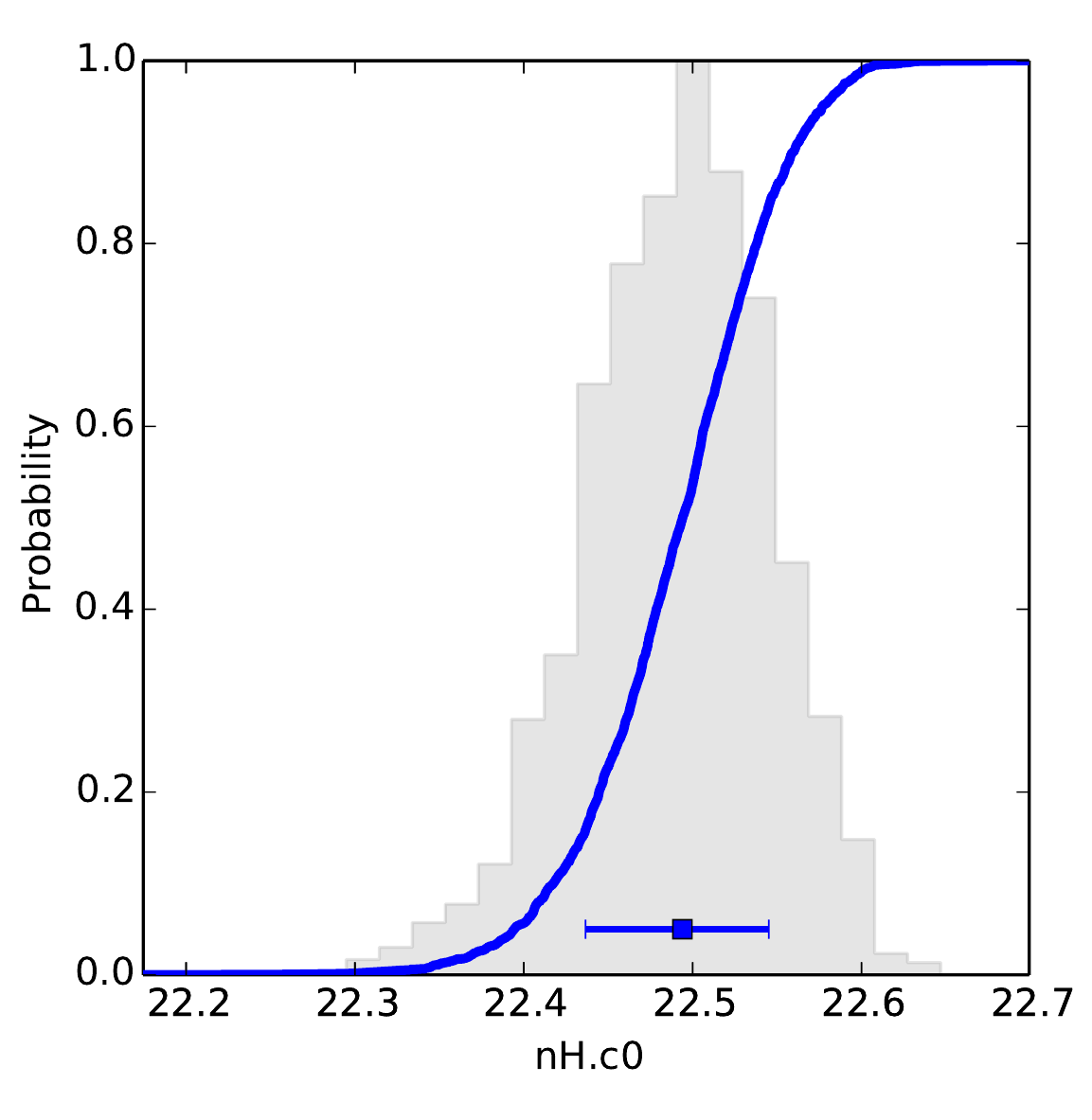}

\caption{\label{fig:refl-example-refl-marg}\label{fig:refl-example-torus-marg}
Marginalised parameters of the \mo{wabs} (top), \mo{wabs+scattering},
\mo{sphere+pexmon+scattering} and \mo{wabs+pexmon+scattering} model
(bottom) for source 179. The posterior probability density distribution,
normalised to the maximum, is shown by grey bars. The blue line indicates
the cumulative posterior distribution. For summary of the error, the
median and 10/90\% quantiles can be used, or as the blue error bar
indicates, the 1 standard-deviation equivalent probabilities.}
\end{figure*}

\begin{table*}[ht]
\caption{Model selection results for object 179. \change{Derived luminosity in erg/s (2) and column density (3, in ${cm}^{-2}$) are shown with $1\sigma$-equivalent quantile errors (see Figure \ref{fig:refl-example-torus-marg} for the posterior distributions). The third column shows the information gain in column density in ban (see text). The forth column shows the intrinsic photon index.}{Derived luminosity (logarithmic, in erg/s) (2), photon index (3), and column density (4, logarithmic, in $\text{cm}^{-2}$) are shown with $1\sigma$-equivalent quantile errors (see Figure \ref{fig:refl-example-torus-marg} for the posterior distributions). The fifth column shows the information gain in column density in ban (see text).} The last four columns show the model comparison results based on log-evidence, posterior assuming all models have same prior probability, BIC and AIC.}
\label{table:refl-modelselection}
\centering
\begin{tabular}{l | r r r r |r r r r}
\hline\hline
Model & $L_{2-10\text{keV}}$ & $\Gamma$ & $N_H$ & $\text{KL}|_{N_H}$ & $\log Z$ & $p(M|D)$ & BIC & AIC \\ 
 (1) & (2) & (3) & (4) & (5) & (6) & (7) & (8)\\
\hline
\mo{torus+pexmon+scattering} & $43.35_{-0.01}^{+0.01}$ & $1.87_{-0.1}^{+0.1}$ & $22.44_{-0.06}^{+0.05}$ & 1.43 & 0.0 & 40.0\% & -109.40 & -122.59\\

\mo{sphere+pexmon+scattering} & $43.35_{-0.01}^{+0.01}$ & $1.88_{-0.1}^{+0.1}$ & $22.44_{-0.05}^{+0.05}$ & 1.46 & -0.1 & 30.2\% & -109.40 & -122.59\\

\mo{wabs+pexmon+scattering} & $43.34_{-0.01}^{+0.01}$ & $1.88_{-0.1}^{+0.1}$ & $22.49_{-0.06}^{+0.05}$ & 1.44 & -0.1 & 29.0\% & -109.43 & -122.62\\

\mo{torus+scattering} & $43.39_{-0.01}^{+0.01}$ & $1.76_{-0.1}^{+0.1}$ & $22.45_{-0.06}^{+0.05}$ & 1.42 & -2.0 & 0.4\% & -105.18 & -113.97\\

\mo{sphere+scattering} & $43.39_{-0.01}^{+0.01}$ & $1.77_{-0.1}^{+0.1}$ & $22.46_{-0.06}^{+0.05}$ & 1.45 & -2.1 & 0.3\% & -105.87 & -114.67\\

\mo{wabs+scattering} & $43.39_{-0.01}^{+0.01}$ & $1.75_{-0.1}^{+0.1}$ & $22.51_{-0.06}^{+0.05}$ & 1.44 & -2.7 & 0.1\% & -103.08 & -111.88\\

\mo{sphere} & $43.35_{-0.01}^{+0.01}$ & $1.57_{-0.1}^{+0.1}$ & $22.23_{-0.04}^{+0.05}$ & 1.52 & -4.7 & 0.0\% & -98.07 & -102.47\\

\mo{torus} & $43.35_{-0.01}^{+0.01}$ & $1.56_{-0.1}^{+0.1}$ & $22.22_{-0.04}^{+0.05}$ & 1.54 & -4.7 & 0.0\% & -97.63 & -102.02\\

\mo{wabs} & $43.35_{-0.01}^{+0.01}$ & $1.55_{-0.1}^{+0.1}$ & $22.28_{-0.04}^{+0.04}$ & 1.58 & -5.5 & 0.0\% & -95.13 & -99.53\\

\mo{powerlaw} & $43.35_{-0.01}^{+0.01}$ & $0.80_{-0.0}^{+0.0}$ &  &  & -27.9 & 0.0\% & 0.00 & 0.00\\

\hline
\end{tabular}
\end{table*}

The Bayesian methodology of parameter estimation and model comparison
presented in Section \ref{sub:Statistical-analysis-methods} is applied
using the models introduced in Section \ref{sec:Model-definitions}
to all sample spectra (see Section \ref{sub:Observational-Data}).
To demonstrate the methodology, we first discuss a single source.
Then we apply model comparison across the full sample.

\subsection{Source 179}

Source 179 (spectroscopic $z = 0.605$,  \object{GOODS-MUSIC 15626}) was detected with 2485 counts in the $0.5-10 \mathrm{keV}$ band at RA/Dec=(3:32:13.23, -27:42:41.02).
This source was chosen because it illustrates the model selection
well, showing several features, namely the Fe-K$\alpha$ line, scattering
and absorption. In the next few paragraphs we present the source's
spectrum and how well different models reproduce them. Figure \ref{fig:refl-example-spec}
overlays different models to the observed data. For brevity, only
a subset of models are included in this presentation, namely \mo{powerlaw},
\mo{wabs}, \mo{torus+scattering} and \mo{wabs+pexmon+scattering}.
We then show the results of model-selection for this object in Table
\ref{table:refl-modelselection}, where the $\log Z$-column (normalised
to highest) shows the computed evidence. Finally, the derived posterior
parameters are shown.

A power law model (\mo{powerlaw}) does not provide a good fit. This
can be seen in the deviations between model and data points in the
upper left panel, and also in the fact that this model has the lowest
evidence of all models Table \ref{table:refl-modelselection} (column
5). Furthermore, the derived photon index, $\Gamma=0.8\pm0.05$ is
unlikely and would constitute a $7\sigma$ outlier.

Obscuration is expected in some sources, and the \mo{wabs} model
indeed improves the fit. The model follows the spectrum much closer
(upper right panel in Figure \ref{fig:refl-example-spec}) with a
line-of-sight absorption of $N_{H}=22.5\pm0.1$. The evidence for
this model is significantly higher, and rules out the \mo{powerlaw}
model. Here we consider a difference of $\log\, Z_{1}-\log\, Z_{2}>\log\,10=1$
as a significant preference (see Section \ref{sub:Priors}). The same
is true for \mo{sphere} and \mo{torus} models not shown here.

Comparing the data with the model prediction in the \mo{wabs} spectrum,
a line is visible at $\sim4\text{keV}$ as well as an excess of soft
energy counts. The former coincides with the Fe-K$\alpha$ line, while
we expect to model the latter with the \mo{+scattering} component.
Considering \mo{torus+scattering} and \mo{wabs+pexmon+scattering}
(lower panels in Figure \ref{fig:refl-example-spec}), they both model
the observed counts well. However, the Fe-K$\alpha$ line is clearly
visible in the data. Compared to the simple absorption models, the
addition of \mo{+scattering} and \mo{+pexmon} increases the evidence
(see Table \ref{table:refl-modelselection}), hence these models are
preferred, while the other models, e.g. the widely used \mo{wabs+scattering},
are ruled out for this source.

When comparing models, an important aspect is the model-dependence
of derived physical parameters. In Table \ref{table:refl-modelselection},
we show the intrinsic luminosity\change{}{, photon index} and line-of-sight
column density. These are computed by the posterior values and summarised
using $1\sigma$-equivalent quantiles. The marginal posterior distributions
are shown in Figure \ref{fig:refl-example-torus-marg}. \change{Adding
a \mo{+pexmon} component decreases the intrinsic luminosity slightly,
and adding the \mo{+scattering} component increases $N_{H}$ by a
small amount. However, between the}{Simple absorption model like
\mo{wabs} try to compensate the soft X-rays by using a flatter spectrum
with less absorption. Both the \mo{+scattering} and \mo{+pexmon}
component increase the photon index. As these additional components
only take up a fraction of the intrinsic power law component, the
changes in intrinsic luminosity are small. Despite these changes,
in between} obscurer geometries (i.e. \mo{wabs}, \mo{torus}, \mo{sphere}),
the values are consistent.

Furthermore, it is important to check whether the results on the derived
physical parameters are strongly influenced by the prior. For example,
for a weak source with data of no discriminatory power, the posterior
of $N_{H}$ would look like the prior (log-uniform). We use the KL-divergence
(see Section \ref{sub:Priors}), also known as the information gain
or knowledge update, to measures the ``distance'' between prior
and posterior of $N_{H}$. For this particular source, the values
are shown in column 4 of Table \ref{table:refl-modelselection}. For
reference, a considerable information gain is e.g. $KL>0.13\,\text{bans}$
which corresponds to halving the standard error of a Gaussian. Because
$N_{H}$ is well-constrained in this source compared to the prior,
the $KL|_{N_{H}}$ values are high, and the posterior is not dominated
by the prior.

The BIC and AIC values (column 7 and 8 of Table \ref{table:refl-modelselection},
lower values are preferred) show the same preferences as the Bayesian
evidence computations. They are an approximate method of model selection
based on the likelihood ratio and parameter penalisation. There are
however important differences to the Bayesian model selection (see
Section \ref{sub:Model-selection} for details). The evidence and
its approximation, the BIC, can be used to express how much more probable
a model is compared to the others based on the data (column 6). The
AIC, in contrast, measures the information loss by using a specific
model. In this single source, the model selection prefers models with
absorption, scattering and an additional reflection component (\mo{pexmon}).
However, no preference is found between the geometries of \mo{wabs},
\mo{torus} and \mo{sphere}. To improve the discriminatory power,
we combine the evidence of the full sample.


\subsection{Model selection on the full sample}

\ifx\du\undefined
  \newlength{\du}
\fi
\setlength{\du}{12\unitlength}
\usetikzlibrary{positioning,arrows}
\usetikzlibrary{decorations.markings}
\usetikzlibrary{shapes.arrows}

\newcommand{\modeldiagramstart}[0]{
\begin{tikzpicture}[thick,scale=0.8, every node/.style={scale=0.8}, every edge/.style={thick,line width=0.5mm},
model/.style={rectangle,draw,draw=gray!50,fill=lightgray!20,text width=2.5cm,align=center, 
},
node distance=15em, on grid, >=latex
] 

}

\newcommand{\modeldiagramnodes}[0]{

\node (powerlaw) [model] {\mo{powerlaw}};
\node (sphere) [below=3\du of powerlaw,model] {\mo{sphere}};
\node (wabs) [left=7\du of sphere,model] {\mo{wabs}};
\node (torus) [right=7\du of sphere,model] {\mo{torus}};

\node (wabs+scattering) [below=3\du of wabs,model] {\mo{wabs\\+scattering}};
\node (sphere+scattering) [right=7\du of wabs+scattering,model] {\mo{sphere\\+scattering}};
\node (torus+scattering) [right=7\du of sphere+scattering,model] {\mo{torus\\+scattering}};

\node (wabs+pexmon+scattering) [below=4\du of wabs+scattering,model] {\mo{wabs\\+pexmon\\+scattering}};
\node (sphere+pexmon+scattering) [right=7\du of wabs+pexmon+scattering,model] {\mo{sphere\\+pexmon\\+scattering}};
\node (torus+pexmon+scattering) [right=7\du of sphere+pexmon+scattering,model] {\mo{torus\\+pexmon\\+scattering}};

}

\newcommand{\powerlawmodeldiagramnodes}[0]{

\node (powerlaw) [model] {\mo{powerlaw}};
\node (powerlaw+flatgamma) [below left=2\du and 7\du of powerlaw,model] {\mo{powerlaw\\+flatgamma}};
\node (powerlaw+pexmon) [below right=2\du and 7\du of powerlaw,model] {\mo{powerlaw\\+pexmon}};
\node (wabs) [below=4\du of powerlaw,model] {\mo{wabs}};

}

\newcommand{\reflmodeldiagramnodes}[0]{

\node (torus+scattering) [model] {\mo{torus\\+scattering}};
\node (torus+pexmon+scattering) [right=7\du of torus+scattering,model] {\mo{torus\\+pexmon\\+scattering}};
\node (vartorus+scattering) [below=4\du of torus+scattering,model] {\mo{vartorus\\+scattering}};
\node (vartorus+pexmon+scattering) [right=7\du of vartorus+scattering,model] {\mo{vartorus\\+pexmon\\+scattering}};
\node (incvartorus+scattering) [below=4\du of vartorus+scattering,model] {\mo{incvartorus\\+scattering}};
\node (incvartorus+pexmon+scattering) [below=4\du of vartorus+pexmon+scattering,model] {\mo{incvartorus\\+pexmon\\+scattering}};

}

\newcommand{\modeldiagramend}[0]{
\end{tikzpicture}

}

\newcommand{\prefcountsXspherescatteringXnothingstar}{208}
\newcommand{\prefcountsXspherescatteringXspherereflectionscattering}{1}
\newcommand{\prefcountsXspherescatteringXwabsreflectionscattering}{3}
\newcommand{\prefcountsXspherescatteringXwarmreflectionscattering}{5}
\newcommand{\prefcountsXspherescatteringXtorusBdreflectionscattering}{3}
\newcommand{\prefcountsXspherescatteringXpowerlawkalpha}{173}
\newcommand{\prefcountsXspherescatteringXwabs}{33}
\newcommand{\prefcountsXspherescatteringXpowerlawreflection}{173}
\newcommand{\prefcountsXspherescatteringXsphere}{23}
\newcommand{\prefcountsXspherescatteringXpowerlawflatgamma}{125}
\newcommand{\prefcountsXspherescatteringXpowerlaw}{173}
\newcommand{\prefcountsXspherescatteringXtorusdecreflectionscattering}{3}
\newcommand{\prefcountsXspherescatteringXwabsscattering}{17}
\newcommand{\prefcountsXspherescatteringXtorusscattering}{4}
\newcommand{\prefcountsXspherescatteringXpowerlawflatgammaplasmahmxb}{167}
\newcommand{\prefcountsXspherescatteringXtorusincreflectionscattering}{2}
\newcommand{\prefcountsXspherescatteringXtorus}{17}
\newcommand{\prefcountsXspherescatteringXtorusreflectionscattering}{3}
\newcommand{\prefcountsXnothingstarXspherescattering}{7}
\newcommand{\prefcountsXnothingstarXspherereflectionscattering}{7}
\newcommand{\prefcountsXnothingstarXwabsreflectionscattering}{7}
\newcommand{\prefcountsXnothingstarXwarmreflectionscattering}{6}
\newcommand{\prefcountsXnothingstarXtorusBdreflectionscattering}{8}
\newcommand{\prefcountsXnothingstarXpowerlawkalpha}{13}
\newcommand{\prefcountsXnothingstarXwabs}{8}
\newcommand{\prefcountsXnothingstarXpowerlawreflection}{12}
\newcommand{\prefcountsXnothingstarXsphere}{8}
\newcommand{\prefcountsXnothingstarXpowerlawflatgamma}{8}
\newcommand{\prefcountsXnothingstarXpowerlaw}{12}
\newcommand{\prefcountsXnothingstarXtorusdecreflectionscattering}{8}
\newcommand{\prefcountsXnothingstarXwabsscattering}{7}
\newcommand{\prefcountsXnothingstarXtorusscattering}{8}
\newcommand{\prefcountsXnothingstarXpowerlawflatgammaplasmahmxb}{49}
\newcommand{\prefcountsXnothingstarXtorusincreflectionscattering}{8}
\newcommand{\prefcountsXnothingstarXtorus}{8}
\newcommand{\prefcountsXnothingstarXtorusreflectionscattering}{8}
\newcommand{\prefcountsXspherereflectionscatteringXspherescattering}{9}
\newcommand{\prefcountsXspherereflectionscatteringXnothingstar}{211}
\newcommand{\prefcountsXspherereflectionscatteringXwabsreflectionscattering}{2}
\newcommand{\prefcountsXspherereflectionscatteringXwarmreflectionscattering}{4}
\newcommand{\prefcountsXspherereflectionscatteringXtorusBdreflectionscattering}{1}
\newcommand{\prefcountsXspherereflectionscatteringXpowerlawkalpha}{179}
\newcommand{\prefcountsXspherereflectionscatteringXwabs}{45}
\newcommand{\prefcountsXspherereflectionscatteringXpowerlawreflection}{179}
\newcommand{\prefcountsXspherereflectionscatteringXsphere}{31}
\newcommand{\prefcountsXspherereflectionscatteringXpowerlawflatgamma}{134}
\newcommand{\prefcountsXspherereflectionscatteringXpowerlaw}{180}
\newcommand{\prefcountsXspherereflectionscatteringXtorusdecreflectionscattering}{1}
\newcommand{\prefcountsXspherereflectionscatteringXwabsscattering}{26}
\newcommand{\prefcountsXspherereflectionscatteringXtorusscattering}{9}
\newcommand{\prefcountsXspherereflectionscatteringXpowerlawflatgammaplasmahmxb}{172}
\newcommand{\prefcountsXspherereflectionscatteringXtorusincreflectionscattering}{0}
\newcommand{\prefcountsXspherereflectionscatteringXtorus}{25}
\newcommand{\prefcountsXspherereflectionscatteringXtorusreflectionscattering}{0}
\newcommand{\prefcountsXwabsreflectionscatteringXspherescattering}{14}
\newcommand{\prefcountsXwabsreflectionscatteringXnothingstar}{213}
\newcommand{\prefcountsXwabsreflectionscatteringXspherereflectionscattering}{5}
\newcommand{\prefcountsXwabsreflectionscatteringXwarmreflectionscattering}{1}
\newcommand{\prefcountsXwabsreflectionscatteringXtorusBdreflectionscattering}{4}
\newcommand{\prefcountsXwabsreflectionscatteringXpowerlawkalpha}{176}
\newcommand{\prefcountsXwabsreflectionscatteringXwabs}{45}
\newcommand{\prefcountsXwabsreflectionscatteringXpowerlawreflection}{178}
\newcommand{\prefcountsXwabsreflectionscatteringXsphere}{40}
\newcommand{\prefcountsXwabsreflectionscatteringXpowerlawflatgamma}{134}
\newcommand{\prefcountsXwabsreflectionscatteringXpowerlaw}{178}
\newcommand{\prefcountsXwabsreflectionscatteringXtorusdecreflectionscattering}{1}
\newcommand{\prefcountsXwabsreflectionscatteringXwabsscattering}{23}
\newcommand{\prefcountsXwabsreflectionscatteringXtorusscattering}{11}
\newcommand{\prefcountsXwabsreflectionscatteringXpowerlawflatgammaplasmahmxb}{175}
\newcommand{\prefcountsXwabsreflectionscatteringXtorusincreflectionscattering}{2}
\newcommand{\prefcountsXwabsreflectionscatteringXtorus}{29}
\newcommand{\prefcountsXwabsreflectionscatteringXtorusreflectionscattering}{1}
\newcommand{\prefcountsXwarmreflectionscatteringXspherescattering}{29}
\newcommand{\prefcountsXwarmreflectionscatteringXnothingstar}{218}
\newcommand{\prefcountsXwarmreflectionscatteringXspherereflectionscattering}{12}
\newcommand{\prefcountsXwarmreflectionscatteringXwabsreflectionscattering}{7}
\newcommand{\prefcountsXwarmreflectionscatteringXtorusBdreflectionscattering}{9}
\newcommand{\prefcountsXwarmreflectionscatteringXpowerlawkalpha}{180}
\newcommand{\prefcountsXwarmreflectionscatteringXwabs}{60}
\newcommand{\prefcountsXwarmreflectionscatteringXpowerlawreflection}{184}
\newcommand{\prefcountsXwarmreflectionscatteringXsphere}{51}
\newcommand{\prefcountsXwarmreflectionscatteringXpowerlawflatgamma}{138}
\newcommand{\prefcountsXwarmreflectionscatteringXpowerlaw}{182}
\newcommand{\prefcountsXwarmreflectionscatteringXtorusdecreflectionscattering}{9}
\newcommand{\prefcountsXwarmreflectionscatteringXwabsscattering}{36}
\newcommand{\prefcountsXwarmreflectionscatteringXtorusscattering}{25}
\newcommand{\prefcountsXwarmreflectionscatteringXpowerlawflatgammaplasmahmxb}{180}
\newcommand{\prefcountsXwarmreflectionscatteringXtorusincreflectionscattering}{11}
\newcommand{\prefcountsXwarmreflectionscatteringXtorus}{39}
\newcommand{\prefcountsXwarmreflectionscatteringXtorusreflectionscattering}{9}
\newcommand{\prefcountsXtorusBdreflectionscatteringXspherescattering}{12}
\newcommand{\prefcountsXtorusBdreflectionscatteringXnothingstar}{214}
\newcommand{\prefcountsXtorusBdreflectionscatteringXspherereflectionscattering}{2}
\newcommand{\prefcountsXtorusBdreflectionscatteringXwabsreflectionscattering}{6}
\newcommand{\prefcountsXtorusBdreflectionscatteringXwarmreflectionscattering}{9}
\newcommand{\prefcountsXtorusBdreflectionscatteringXpowerlawkalpha}{181}
\newcommand{\prefcountsXtorusBdreflectionscatteringXwabs}{46}
\newcommand{\prefcountsXtorusBdreflectionscatteringXpowerlawreflection}{180}
\newcommand{\prefcountsXtorusBdreflectionscatteringXsphere}{36}
\newcommand{\prefcountsXtorusBdreflectionscatteringXpowerlawflatgamma}{136}
\newcommand{\prefcountsXtorusBdreflectionscatteringXpowerlaw}{180}
\newcommand{\prefcountsXtorusBdreflectionscatteringXtorusdecreflectionscattering}{0}
\newcommand{\prefcountsXtorusBdreflectionscatteringXwabsscattering}{29}
\newcommand{\prefcountsXtorusBdreflectionscatteringXtorusscattering}{7}
\newcommand{\prefcountsXtorusBdreflectionscatteringXpowerlawflatgammaplasmahmxb}{171}
\newcommand{\prefcountsXtorusBdreflectionscatteringXtorusincreflectionscattering}{0}
\newcommand{\prefcountsXtorusBdreflectionscatteringXtorus}{25}
\newcommand{\prefcountsXtorusBdreflectionscatteringXtorusreflectionscattering}{0}
\newcommand{\prefcountsXpowerlawkalphaXspherescattering}{2}
\newcommand{\prefcountsXpowerlawkalphaXnothingstar}{139}
\newcommand{\prefcountsXpowerlawkalphaXspherereflectionscattering}{2}
\newcommand{\prefcountsXpowerlawkalphaXwabsreflectionscattering}{2}
\newcommand{\prefcountsXpowerlawkalphaXwarmreflectionscattering}{2}
\newcommand{\prefcountsXpowerlawkalphaXtorusBdreflectionscattering}{5}
\newcommand{\prefcountsXpowerlawkalphaXwabs}{6}
\newcommand{\prefcountsXpowerlawkalphaXpowerlawreflection}{10}
\newcommand{\prefcountsXpowerlawkalphaXsphere}{9}
\newcommand{\prefcountsXpowerlawkalphaXpowerlawflatgamma}{4}
\newcommand{\prefcountsXpowerlawkalphaXpowerlaw}{3}
\newcommand{\prefcountsXpowerlawkalphaXtorusdecreflectionscattering}{6}
\newcommand{\prefcountsXpowerlawkalphaXwabsscattering}{1}
\newcommand{\prefcountsXpowerlawkalphaXtorusscattering}{5}
\newcommand{\prefcountsXpowerlawkalphaXpowerlawflatgammaplasmahmxb}{85}
\newcommand{\prefcountsXpowerlawkalphaXtorusincreflectionscattering}{6}
\newcommand{\prefcountsXpowerlawkalphaXtorus}{7}
\newcommand{\prefcountsXpowerlawkalphaXtorusreflectionscattering}{7}
\newcommand{\prefcountsXwabsXspherescattering}{1}
\newcommand{\prefcountsXwabsXnothingstar}{204}
\newcommand{\prefcountsXwabsXspherereflectionscattering}{4}
\newcommand{\prefcountsXwabsXwabsreflectionscattering}{0}
\newcommand{\prefcountsXwabsXwarmreflectionscattering}{0}
\newcommand{\prefcountsXwabsXtorusBdreflectionscattering}{2}
\newcommand{\prefcountsXwabsXpowerlawkalpha}{166}
\newcommand{\prefcountsXwabsXpowerlawreflection}{167}
\newcommand{\prefcountsXwabsXsphere}{3}
\newcommand{\prefcountsXwabsXpowerlawflatgamma}{113}
\newcommand{\prefcountsXwabsXpowerlaw}{165}
\newcommand{\prefcountsXwabsXtorusdecreflectionscattering}{1}
\newcommand{\prefcountsXwabsXwabsscattering}{0}
\newcommand{\prefcountsXwabsXtorusscattering}{0}
\newcommand{\prefcountsXwabsXpowerlawflatgammaplasmahmxb}{157}
\newcommand{\prefcountsXwabsXtorusincreflectionscattering}{2}
\newcommand{\prefcountsXwabsXtorus}{0}
\newcommand{\prefcountsXwabsXtorusreflectionscattering}{2}
\newcommand{\prefcountsXpowerlawreflectionXspherescattering}{2}
\newcommand{\prefcountsXpowerlawreflectionXnothingstar}{140}
\newcommand{\prefcountsXpowerlawreflectionXspherereflectionscattering}{2}
\newcommand{\prefcountsXpowerlawreflectionXwabsreflectionscattering}{3}
\newcommand{\prefcountsXpowerlawreflectionXwarmreflectionscattering}{1}
\newcommand{\prefcountsXpowerlawreflectionXtorusBdreflectionscattering}{6}
\newcommand{\prefcountsXpowerlawreflectionXpowerlawkalpha}{18}
\newcommand{\prefcountsXpowerlawreflectionXwabs}{7}
\newcommand{\prefcountsXpowerlawreflectionXsphere}{9}
\newcommand{\prefcountsXpowerlawreflectionXpowerlawflatgamma}{3}
\newcommand{\prefcountsXpowerlawreflectionXpowerlaw}{17}
\newcommand{\prefcountsXpowerlawreflectionXtorusdecreflectionscattering}{6}
\newcommand{\prefcountsXpowerlawreflectionXwabsscattering}{4}
\newcommand{\prefcountsXpowerlawreflectionXtorusscattering}{5}
\newcommand{\prefcountsXpowerlawreflectionXpowerlawflatgammaplasmahmxb}{81}
\newcommand{\prefcountsXpowerlawreflectionXtorusincreflectionscattering}{5}
\newcommand{\prefcountsXpowerlawreflectionXtorus}{7}
\newcommand{\prefcountsXpowerlawreflectionXtorusreflectionscattering}{5}
\newcommand{\prefcountsXsphereXspherescattering}{0}
\newcommand{\prefcountsXsphereXnothingstar}{204}
\newcommand{\prefcountsXsphereXspherereflectionscattering}{0}
\newcommand{\prefcountsXsphereXwabsreflectionscattering}{1}
\newcommand{\prefcountsXsphereXwarmreflectionscattering}{2}
\newcommand{\prefcountsXsphereXtorusBdreflectionscattering}{0}
\newcommand{\prefcountsXsphereXpowerlawkalpha}{168}
\newcommand{\prefcountsXsphereXwabs}{16}
\newcommand{\prefcountsXsphereXpowerlawreflection}{169}
\newcommand{\prefcountsXsphereXpowerlawflatgamma}{116}
\newcommand{\prefcountsXsphereXpowerlaw}{167}
\newcommand{\prefcountsXsphereXtorusdecreflectionscattering}{1}
\newcommand{\prefcountsXsphereXwabsscattering}{6}
\newcommand{\prefcountsXsphereXtorusscattering}{0}
\newcommand{\prefcountsXsphereXpowerlawflatgammaplasmahmxb}{158}
\newcommand{\prefcountsXsphereXtorusincreflectionscattering}{1}
\newcommand{\prefcountsXsphereXtorus}{0}
\newcommand{\prefcountsXsphereXtorusreflectionscattering}{1}
\newcommand{\prefcountsXpowerlawflatgammaXspherescattering}{10}
\newcommand{\prefcountsXpowerlawflatgammaXnothingstar}{194}
\newcommand{\prefcountsXpowerlawflatgammaXspherereflectionscattering}{9}
\newcommand{\prefcountsXpowerlawflatgammaXwabsreflectionscattering}{9}
\newcommand{\prefcountsXpowerlawflatgammaXwarmreflectionscattering}{6}
\newcommand{\prefcountsXpowerlawflatgammaXtorusBdreflectionscattering}{10}
\newcommand{\prefcountsXpowerlawflatgammaXpowerlawkalpha}{160}
\newcommand{\prefcountsXpowerlawflatgammaXwabs}{20}
\newcommand{\prefcountsXpowerlawflatgammaXpowerlawreflection}{160}
\newcommand{\prefcountsXpowerlawflatgammaXsphere}{19}
\newcommand{\prefcountsXpowerlawflatgammaXpowerlaw}{162}
\newcommand{\prefcountsXpowerlawflatgammaXtorusdecreflectionscattering}{9}
\newcommand{\prefcountsXpowerlawflatgammaXwabsscattering}{12}
\newcommand{\prefcountsXpowerlawflatgammaXtorusscattering}{12}
\newcommand{\prefcountsXpowerlawflatgammaXpowerlawflatgammaplasmahmxb}{72}
\newcommand{\prefcountsXpowerlawflatgammaXtorusincreflectionscattering}{10}
\newcommand{\prefcountsXpowerlawflatgammaXtorus}{15}
\newcommand{\prefcountsXpowerlawflatgammaXtorusreflectionscattering}{10}
\newcommand{\prefcountsXpowerlawXspherescattering}{2}
\newcommand{\prefcountsXpowerlawXnothingstar}{139}
\newcommand{\prefcountsXpowerlawXspherereflectionscattering}{2}
\newcommand{\prefcountsXpowerlawXwabsreflectionscattering}{2}
\newcommand{\prefcountsXpowerlawXwarmreflectionscattering}{2}
\newcommand{\prefcountsXpowerlawXtorusBdreflectionscattering}{6}
\newcommand{\prefcountsXpowerlawXpowerlawkalpha}{13}
\newcommand{\prefcountsXpowerlawXwabs}{6}
\newcommand{\prefcountsXpowerlawXpowerlawreflection}{18}
\newcommand{\prefcountsXpowerlawXsphere}{7}
\newcommand{\prefcountsXpowerlawXpowerlawflatgamma}{3}
\newcommand{\prefcountsXpowerlawXtorusdecreflectionscattering}{5}
\newcommand{\prefcountsXpowerlawXwabsscattering}{2}
\newcommand{\prefcountsXpowerlawXtorusscattering}{5}
\newcommand{\prefcountsXpowerlawXpowerlawflatgammaplasmahmxb}{83}
\newcommand{\prefcountsXpowerlawXtorusincreflectionscattering}{6}
\newcommand{\prefcountsXpowerlawXtorus}{6}
\newcommand{\prefcountsXpowerlawXtorusreflectionscattering}{6}
\newcommand{\prefcountsXtorusdecreflectionscatteringXspherescattering}{11}
\newcommand{\prefcountsXtorusdecreflectionscatteringXnothingstar}{215}
\newcommand{\prefcountsXtorusdecreflectionscatteringXspherereflectionscattering}{2}
\newcommand{\prefcountsXtorusdecreflectionscatteringXwabsreflectionscattering}{6}
\newcommand{\prefcountsXtorusdecreflectionscatteringXwarmreflectionscattering}{11}
\newcommand{\prefcountsXtorusdecreflectionscatteringXtorusBdreflectionscattering}{0}
\newcommand{\prefcountsXtorusdecreflectionscatteringXpowerlawkalpha}{178}
\newcommand{\prefcountsXtorusdecreflectionscatteringXwabs}{49}
\newcommand{\prefcountsXtorusdecreflectionscatteringXpowerlawreflection}{178}
\newcommand{\prefcountsXtorusdecreflectionscatteringXsphere}{36}
\newcommand{\prefcountsXtorusdecreflectionscatteringXpowerlawflatgamma}{135}
\newcommand{\prefcountsXtorusdecreflectionscatteringXpowerlaw}{179}
\newcommand{\prefcountsXtorusdecreflectionscatteringXwabsscattering}{29}
\newcommand{\prefcountsXtorusdecreflectionscatteringXtorusscattering}{7}
\newcommand{\prefcountsXtorusdecreflectionscatteringXpowerlawflatgammaplasmahmxb}{171}
\newcommand{\prefcountsXtorusdecreflectionscatteringXtorusincreflectionscattering}{0}
\newcommand{\prefcountsXtorusdecreflectionscatteringXtorus}{24}
\newcommand{\prefcountsXtorusdecreflectionscatteringXtorusreflectionscattering}{0}
\newcommand{\prefcountsXwabsscatteringXspherescattering}{3}
\newcommand{\prefcountsXwabsscatteringXnothingstar}{206}
\newcommand{\prefcountsXwabsscatteringXspherereflectionscattering}{5}
\newcommand{\prefcountsXwabsscatteringXwabsreflectionscattering}{1}
\newcommand{\prefcountsXwabsscatteringXwarmreflectionscattering}{3}
\newcommand{\prefcountsXwabsscatteringXtorusBdreflectionscattering}{7}
\newcommand{\prefcountsXwabsscatteringXpowerlawkalpha}{170}
\newcommand{\prefcountsXwabsscatteringXwabs}{22}
\newcommand{\prefcountsXwabsscatteringXpowerlawreflection}{170}
\newcommand{\prefcountsXwabsscatteringXsphere}{26}
\newcommand{\prefcountsXwabsscatteringXpowerlawflatgamma}{120}
\newcommand{\prefcountsXwabsscatteringXpowerlaw}{171}
\newcommand{\prefcountsXwabsscatteringXtorusdecreflectionscattering}{4}
\newcommand{\prefcountsXwabsscatteringXtorusscattering}{6}
\newcommand{\prefcountsXwabsscatteringXpowerlawflatgammaplasmahmxb}{160}
\newcommand{\prefcountsXwabsscatteringXtorusincreflectionscattering}{5}
\newcommand{\prefcountsXwabsscatteringXtorus}{18}
\newcommand{\prefcountsXwabsscatteringXtorusreflectionscattering}{7}
\newcommand{\prefcountsXtorusscatteringXspherescattering}{4}
\newcommand{\prefcountsXtorusscatteringXnothingstar}{210}
\newcommand{\prefcountsXtorusscatteringXspherereflectionscattering}{3}
\newcommand{\prefcountsXtorusscatteringXwabsreflectionscattering}{6}
\newcommand{\prefcountsXtorusscatteringXwarmreflectionscattering}{9}
\newcommand{\prefcountsXtorusscatteringXtorusBdreflectionscattering}{0}
\newcommand{\prefcountsXtorusscatteringXpowerlawkalpha}{174}
\newcommand{\prefcountsXtorusscatteringXwabs}{36}
\newcommand{\prefcountsXtorusscatteringXpowerlawreflection}{176}
\newcommand{\prefcountsXtorusscatteringXsphere}{27}
\newcommand{\prefcountsXtorusscatteringXpowerlawflatgamma}{127}
\newcommand{\prefcountsXtorusscatteringXpowerlaw}{177}
\newcommand{\prefcountsXtorusscatteringXtorusdecreflectionscattering}{0}
\newcommand{\prefcountsXtorusscatteringXwabsscattering}{20}
\newcommand{\prefcountsXtorusscatteringXpowerlawflatgammaplasmahmxb}{167}
\newcommand{\prefcountsXtorusscatteringXtorusincreflectionscattering}{0}
\newcommand{\prefcountsXtorusscatteringXtorus}{15}
\newcommand{\prefcountsXtorusscatteringXtorusreflectionscattering}{0}
\newcommand{\prefcountsXpowerlawflatgammaplasmahmxbXspherescattering}{32}
\newcommand{\prefcountsXpowerlawflatgammaplasmahmxbXnothingstar}{201}
\newcommand{\prefcountsXpowerlawflatgammaplasmahmxbXspherereflectionscattering}{27}
\newcommand{\prefcountsXpowerlawflatgammaplasmahmxbXwabsreflectionscattering}{23}
\newcommand{\prefcountsXpowerlawflatgammaplasmahmxbXwarmreflectionscattering}{21}
\newcommand{\prefcountsXpowerlawflatgammaplasmahmxbXtorusBdreflectionscattering}{24}
\newcommand{\prefcountsXpowerlawflatgammaplasmahmxbXpowerlawkalpha}{175}
\newcommand{\prefcountsXpowerlawflatgammaplasmahmxbXwabs}{48}
\newcommand{\prefcountsXpowerlawflatgammaplasmahmxbXpowerlawreflection}{173}
\newcommand{\prefcountsXpowerlawflatgammaplasmahmxbXsphere}{43}
\newcommand{\prefcountsXpowerlawflatgammaplasmahmxbXpowerlawflatgamma}{36}
\newcommand{\prefcountsXpowerlawflatgammaplasmahmxbXpowerlaw}{174}
\newcommand{\prefcountsXpowerlawflatgammaplasmahmxbXtorusdecreflectionscattering}{23}
\newcommand{\prefcountsXpowerlawflatgammaplasmahmxbXwabsscattering}{35}
\newcommand{\prefcountsXpowerlawflatgammaplasmahmxbXtorusscattering}{30}
\newcommand{\prefcountsXpowerlawflatgammaplasmahmxbXtorusincreflectionscattering}{24}
\newcommand{\prefcountsXpowerlawflatgammaplasmahmxbXtorus}{35}
\newcommand{\prefcountsXpowerlawflatgammaplasmahmxbXtorusreflectionscattering}{23}
\newcommand{\prefcountsXtorusincreflectionscatteringXspherescattering}{13}
\newcommand{\prefcountsXtorusincreflectionscatteringXnothingstar}{215}
\newcommand{\prefcountsXtorusincreflectionscatteringXspherereflectionscattering}{3}
\newcommand{\prefcountsXtorusincreflectionscatteringXwabsreflectionscattering}{6}
\newcommand{\prefcountsXtorusincreflectionscatteringXwarmreflectionscattering}{9}
\newcommand{\prefcountsXtorusincreflectionscatteringXtorusBdreflectionscattering}{0}
\newcommand{\prefcountsXtorusincreflectionscatteringXpowerlawkalpha}{180}
\newcommand{\prefcountsXtorusincreflectionscatteringXwabs}{46}
\newcommand{\prefcountsXtorusincreflectionscatteringXpowerlawreflection}{180}
\newcommand{\prefcountsXtorusincreflectionscatteringXsphere}{36}
\newcommand{\prefcountsXtorusincreflectionscatteringXpowerlawflatgamma}{131}
\newcommand{\prefcountsXtorusincreflectionscatteringXpowerlaw}{181}
\newcommand{\prefcountsXtorusincreflectionscatteringXtorusdecreflectionscattering}{0}
\newcommand{\prefcountsXtorusincreflectionscatteringXwabsscattering}{31}
\newcommand{\prefcountsXtorusincreflectionscatteringXtorusscattering}{7}
\newcommand{\prefcountsXtorusincreflectionscatteringXpowerlawflatgammaplasmahmxb}{172}
\newcommand{\prefcountsXtorusincreflectionscatteringXtorus}{24}
\newcommand{\prefcountsXtorusincreflectionscatteringXtorusreflectionscattering}{0}
\newcommand{\prefcountsXtorusXspherescattering}{3}
\newcommand{\prefcountsXtorusXnothingstar}{208}
\newcommand{\prefcountsXtorusXspherereflectionscattering}{3}
\newcommand{\prefcountsXtorusXwabsreflectionscattering}{4}
\newcommand{\prefcountsXtorusXwarmreflectionscattering}{5}
\newcommand{\prefcountsXtorusXtorusBdreflectionscattering}{0}
\newcommand{\prefcountsXtorusXpowerlawkalpha}{170}
\newcommand{\prefcountsXtorusXwabs}{24}
\newcommand{\prefcountsXtorusXpowerlawreflection}{171}
\newcommand{\prefcountsXtorusXsphere}{11}
\newcommand{\prefcountsXtorusXpowerlawflatgamma}{121}
\newcommand{\prefcountsXtorusXpowerlaw}{170}
\newcommand{\prefcountsXtorusXtorusdecreflectionscattering}{0}
\newcommand{\prefcountsXtorusXwabsscattering}{12}
\newcommand{\prefcountsXtorusXtorusscattering}{0}
\newcommand{\prefcountsXtorusXpowerlawflatgammaplasmahmxb}{164}
\newcommand{\prefcountsXtorusXtorusincreflectionscattering}{0}
\newcommand{\prefcountsXtorusXtorusreflectionscattering}{0}
\newcommand{\prefcountsXtorusreflectionscatteringXspherescattering}{10}
\newcommand{\prefcountsXtorusreflectionscatteringXnothingstar}{215}
\newcommand{\prefcountsXtorusreflectionscatteringXspherereflectionscattering}{3}
\newcommand{\prefcountsXtorusreflectionscatteringXwabsreflectionscattering}{6}
\newcommand{\prefcountsXtorusreflectionscatteringXwarmreflectionscattering}{11}
\newcommand{\prefcountsXtorusreflectionscatteringXtorusBdreflectionscattering}{0}
\newcommand{\prefcountsXtorusreflectionscatteringXpowerlawkalpha}{181}
\newcommand{\prefcountsXtorusreflectionscatteringXwabs}{47}
\newcommand{\prefcountsXtorusreflectionscatteringXpowerlawreflection}{180}
\newcommand{\prefcountsXtorusreflectionscatteringXsphere}{38}
\newcommand{\prefcountsXtorusreflectionscatteringXpowerlawflatgamma}{135}
\newcommand{\prefcountsXtorusreflectionscatteringXpowerlaw}{183}
\newcommand{\prefcountsXtorusreflectionscatteringXtorusdecreflectionscattering}{0}
\newcommand{\prefcountsXtorusreflectionscatteringXwabsscattering}{32}
\newcommand{\prefcountsXtorusreflectionscatteringXtorusscattering}{6}
\newcommand{\prefcountsXtorusreflectionscatteringXpowerlawflatgammaplasmahmxb}{170}
\newcommand{\prefcountsXtorusreflectionscatteringXtorusincreflectionscattering}{0}
\newcommand{\prefcountsXtorusreflectionscatteringXtorus}{26}
\begin{figure}
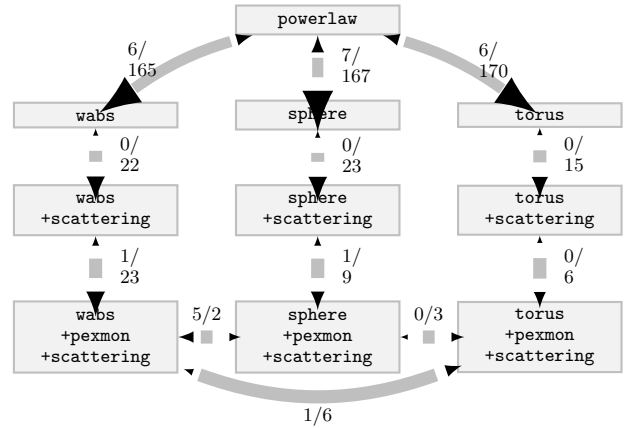

\modeldiagramstart
\modeldiagramnodes

\path (wabs) edge [draw=gray!50,
  line width=4.243126,
  decoration={markings,
  mark=at position 0.14 with {\arrow[black,line width=4.25610214007]{<}};,
  mark=at position 1 with {\arrow[black,line width=1.11650102355]{>}};,
  },
  postaction={decorate}, shorten >= 3mm, shorten <= 3mm, bend left=10,
] node[left]{\small{\begin{tabular}{l}
$6/$\\
$165$\end{tabular}}} (powerlaw);

\path (sphere) edge [draw=gray!50,
  line width=4.223365,
  decoration={markings,
  mark=at position 0.14 with {\arrow[black,line width=4.26892544024]{<}};,
  mark=at position 1 with {\arrow[black,line width=1.22693517987]{>}};,
  },
  postaction={decorate}, shorten >= 3mm, shorten <= 3mm, ,
] node[right]{\small{\begin{tabular}{l}
$7/$\\
$167$\end{tabular}}} (powerlaw);

\path (torus) edge [draw=gray!50,
  line width=4.209989,
  decoration={markings,
  mark=at position 0.14 with {\arrow[black,line width=4.28788081354]{<}};,
  mark=at position 1 with {\arrow[black,line width=1.11650102355]{>}};,
  },
  postaction={decorate}, shorten >= 3mm, shorten <= 3mm, bend right=10,
] node[right]{\small{\begin{tabular}{l}
$6/$\\
$170$\end{tabular}}} (powerlaw);

\path (wabs+scattering) edge [draw=gray!50,
  line width=4.937581,
  decoration={markings,
  mark=at position 0.14 with {\arrow[black,line width=2.20015035934]{<}};,
  mark=at position 1 with {\arrow[black,line width=0.0]{>}};,
  },
  postaction={decorate}, shorten >= 3mm, shorten <= 3mm, ,
] node[right]{\small{\begin{tabular}{l}
$0/$\\
$22$\end{tabular}}} (wabs);

\path (sphere+scattering) edge [draw=gray!50,
  line width=4.934133,
  decoration={markings,
  mark=at position 0.14 with {\arrow[black,line width=2.24217325954]{<}};,
  mark=at position 1 with {\arrow[black,line width=0.0]{>}};,
  },
  postaction={decorate}, shorten >= 3mm, shorten <= 3mm, ,
] node[right]{\small{\begin{tabular}{l}
$0/$\\
$23$\end{tabular}}} (sphere);

\path (wabs+pexmon+scattering) edge [draw=gray!50,
  line width=4.930673,
  decoration={markings,
  mark=at position 0.14 with {\arrow[black,line width=2.24217325954]{<}};,
  mark=at position 1 with {\arrow[black,line width=0.284358156752]{>}};,
  },
  postaction={decorate}, shorten >= 3mm, shorten <= 3mm, ,
] node[right]{\small{\begin{tabular}{l}
$1/$\\
$23$\end{tabular}}} (wabs+scattering);

\path (sphere+pexmon+scattering) edge [draw=gray!50,
  line width=4.978132,
  decoration={markings,
  mark=at position 0.14 with {\arrow[black,line width=1.41892173221]{<}};,
  mark=at position 1 with {\arrow[black,line width=0.284358156752]{>}};,
  },
  postaction={decorate}, shorten >= 3mm, shorten <= 3mm, ,
] node[right]{\small{\begin{tabular}{l}
$1/$\\
$9$\end{tabular}}} (sphere+scattering);

\path (sphere+pexmon+scattering.west) edge [draw=gray!50,
  line width=4.988037,
  decoration={markings,
  mark=at position 0.1 with {\arrow[black,line width=0.509486975542]{<}};,
  mark=at position 1 with {\arrow[black,line width=0.993548959565]{>}};,
  },
  postaction={decorate}, shorten >= 3mm, shorten <= 3mm, ,
] node[above]{\small{$5/2$}} (wabs+pexmon+scattering);

\path (torus+scattering) edge [draw=gray!50,
  line width=4.961419,
  decoration={markings,
  mark=at position 0.14 with {\arrow[black,line width=1.84913216131]{<}};,
  mark=at position 1 with {\arrow[black,line width=0.0]{>}};,
  },
  postaction={decorate}, shorten >= 3mm, shorten <= 3mm, ,
] node[right]{\small{\begin{tabular}{l}
$0/$\\
$15$\end{tabular}}} (torus);

\path (torus+pexmon+scattering) edge [draw=gray!50,
  line width=4.991319,
  decoration={markings,
  mark=at position 0.14 with {\arrow[black,line width=1.11650102355]{<}};,
  mark=at position 1 with {\arrow[black,line width=0.0]{>}};,
  },
  postaction={decorate}, shorten >= 3mm, shorten <= 3mm, ,
] node[right]{\small{\begin{tabular}{l}
$0/$\\
$6$\end{tabular}}} (torus+scattering);

\path (torus+pexmon+scattering.west) edge [draw=gray!50,
  line width=5.001105,
  decoration={markings,
  mark=at position 0.1 with {\arrow[black,line width=0.695856977675]{<}};,
  mark=at position 1 with {\arrow[black,line width=0.0]{>}};,
  },
  postaction={decorate}, shorten >= 3mm, shorten <= 3mm, ,
] node[above]{\small{$0/3$}} (sphere+pexmon+scattering);

\path (torus+pexmon+scattering) edge [draw=gray!50,
  line width=4.988037,
  decoration={markings,
  mark=at position 0.04 with {\arrow[black,line width=1.11650102355]{<}};,
  mark=at position 0.98 with {\arrow[black,line width=0.284358156752]{>}};,
  },
  postaction={decorate}, shorten >= 3mm, shorten <= 3mm, bend left=20,
] node[below]{\small{$1/6$}} (wabs+pexmon+scattering);

\modeldiagramend
\caption{\label{powerlawmodeldiagram}\label{modeldiagram}Model comparison for the three obscuration models. The arrow size and numbers indicate the number of sources, for which one model is strongly preferred over the other.}
\end{figure}
\begin{figure}
\end{figure}

We present the results for the full sample in two forms. Firstly,
in each source, we perform model comparison between each pair of models
(arrows in Figure \ref{fig:model-hierarchy}). We count for how many
sources a preference was found. This is shown in Figure \ref{modeldiagram},
where the two numbers indicate the two directions. For instance, in
\prefcountsXpowerlawXtorus\  sources, the (simpler) \mo{powerlaw}
model is preferred over the \mo{torus}, while in \prefcountsXtorusXpowerlaw\ 
sources, the torus is preferred. The size of the arrows visualises
the same information. There is clear preference for absorption, scattering
and reflection in $\sim170$, $\sim20$ and $>5$ cases respectively,
showing that in a substantial number of sources these components are
required. Between \mo{wabs+pexmon+scattering}, \mo{torus+pexmon+scattering}
and \mo{sphere+pexmon+scattering}, there is no clear trend, however
the torus geometry is preferred most often. This indicates some variety
in the obscurer geometry between sources.

\begin{center}\small

\begin{table*}[ht]
\caption{Sample model comparison. Considering each source in turn, the second column shows in how many sources model comparison ruled out the particular model. The third column shows the total evidence across all sources, relative to the best evidence. The following columns show the same statistics, but using bootstrapping on the sample, making the computed quantities more robust against outliers. The last column computes how often in the bootstrapping the model was ruled out based on the total evidence (10 times less likely than other models). In the lower table segments, the sample is split by $\log N_H$, estimated using the 90\% quantiles  of the \mo{torus+scattering} model posterior. Models with mean 100\% and root mean square 0\% in column 6 are not shown in the lower segments.}
\label{table:totalmodel}
\centering
\begin{tabular}{l r r |r r r}
\hline\hline
 & \multicolumn{2}{c|}{Sample results} & \multicolumn{3}{c}{Bootstrapped results} \\Model & \# rej & $\sum \log Z$ & \# rej & $\sum \log Z$ & ruled out \\ 
 (1) & (2) & (3) & (4) & (5) & (6)\\
\hline
\multicolumn{3}{l|}{\textit{All (334 sources)}} & \multicolumn{3}{l}{}  \\
{\mo{torus+pexmon+scattering}} & $11$ & $-85.6$ & $10 \pm 3.4$ & $-85.4 \pm 8.0$ & $9 \pm 28 \%$\\

{\mo{wabs+pexmon+scattering}} & $ 8$ & $-93.4$ & $ 8 \pm 2.7$ & $-93.6 \pm 7.6$ & $85 \pm 35 \%$\\

{\mo{sphere+pexmon+scattering}} & $11$ & $-95.3$ & $10 \pm 3.3$ & $-95.1 \pm 7.8$ & $100 \pm 5 \%$\\

{\mo{torus+scattering}} & $16$ & $-116.8$ & $16 \pm 4.0$ & $-117.3 \pm 11.8$ & $100 \pm 0 \%$\\

{\mo{sphere+scattering}} & $21$ & $-127.3$ & $21 \pm 4.6$ & $-127.3 \pm 10.3$ & $100 \pm 0 \%$\\

{\mo{wabs+scattering}} & $37$ & $-167.6$ & $37 \pm 5.7$ & $-168.6 \pm 14.2$ & $100 \pm 0 \%$\\

{\mo{torus}} & $39$ & $-188.2$ & $39 \pm 5.9$ & $-190.4 \pm 20.0$ & $100 \pm 0 \%$\\

{\mo{sphere}} & $55$ & $-233.1$ & $55 \pm 6.8$ & $-235.3 \pm 22.3$ & $100 \pm 0 \%$\\

{\mo{wabs}} & $59$ & $-270.9$ & $59 \pm 6.8$ & $-274.1 \pm 26.9$ & $100 \pm 0 \%$\\

{\mo{powerlaw}} & $187$ & $-2657.9$ & $186 \pm 9.6$ & $-2704.8 \pm 412.6$ & $100 \pm 0 \%$\\

\hline
\multicolumn{3}{l|}{\textit{$\log N_H=20-22$ (54 sources)}} & \multicolumn{3}{l}{} \\
{\mo{wabs+pexmon+scattering}} & $ 2$ & $-24.0$ & $ 2 \pm 1.3$ & $-24.4 \pm 5.1$ & $54 \pm 50 \%$\\

{\mo{sphere+pexmon+scattering}} & $ 2$ & $-24.4$ & $ 2 \pm 1.3$ & $-24.9 \pm 5.3$ & $63 \pm 48 \%$\\

{\mo{powerlaw}} & $ 7$ & $-24.9$ & $ 7 \pm 2.7$ & $-25.2 \pm 9.4$ & $44 \pm 50 \%$\\

\hline
\multicolumn{3}{l|}{\textit{$\log N_H=22-23$ (47 sources)}} & \multicolumn{3}{l}{} \\
{\mo{torus+pexmon+scattering}} & $ 0$ & $-5.3$ & $ 0 \pm 0.0$ & $-5.2 \pm 0.7$ & $0 \pm 0 \%$\\

{\mo{sphere+pexmon+scattering}} & $ 0$ & $-6.6$ & $ 0 \pm 0.0$ & $-6.5 \pm 0.7$ & $70 \pm 46 \%$\\

{\mo{wabs+pexmon+scattering}} & $ 0$ & $-7.5$ & $ 0 \pm 0.0$ & $-7.5 \pm 0.7$ & $98 \pm 14 \%$\\

{\mo{torus+scattering}} & $ 3$ & $-13.1$ & $ 2 \pm 1.6$ & $-12.6 \pm 3.5$ & $98 \pm 14 \%$\\

\hline
\multicolumn{3}{l|}{\textit{$\log N_H=23-24$ (51 sources)}} & \multicolumn{3}{l}{} \\
{\mo{wabs+pexmon+scattering}} & $ 0$ & $-10.1$ & $ 0 \pm 0.0$ & $-10.1 \pm 1.1$ & $6 \pm 24 \%$\\

{\mo{torus+pexmon+scattering}} & $ 4$ & $-12.6$ & $ 3 \pm 1.8$ & $-12.4 \pm 2.2$ & $67 \pm 47 \%$\\

\hline
\multicolumn{3}{l|}{\textit{$\log N_H=24-26$ (14 sources)}} & \multicolumn{3}{l}{} \\
{\mo{torus+scattering}} & $ 0$ & $-1.5$ & $ 0 \pm 0.0$ & $-1.5 \pm 0.5$ & $7 \pm 26 \%$\\

{\mo{torus+pexmon+scattering}} & $ 0$ & $-1.7$ & $ 0 \pm 0.0$ & $-1.7 \pm 0.3$ & $7 \pm 26 \%$\\

{\mo{torus}} & $ 3$ & $-9.8$ & $ 3 \pm 2.1$ & $-10.3 \pm 7.1$ & $93 \pm 26 \%$\\

\hline
\multicolumn{3}{l|}{\textit{$\log N_H=22-26$ (176 sources)}} & \multicolumn{3}{l}{} \\
{\mo{torus+pexmon+scattering}} & $ 4$ & $-30.2$ & $ 4 \pm 1.9$ & $-30.0 \pm 3.1$ & $0 \pm 0 \%$\\

{\mo{wabs+pexmon+scattering}} & $ 6$ & $-46.4$ & $ 6 \pm 2.5$ & $-46.6 \pm 5.1$ & $99 \pm 8 \%$\\

\hline
\multicolumn{3}{l|}{\textit{$z>1$ (229 sources)}} & \multicolumn{3}{l}{} \\
{\mo{torus+pexmon+scattering}} & $ 8$ & $-57.6$ & $ 8 \pm 3.0$ & $-58.9 \pm 7.7$ & $11 \pm 32 \%$\\

{\mo{wabs+pexmon+scattering}} & $ 5$ & $-64.5$ & $ 5 \pm 2.3$ & $-65.2 \pm 7.7$ & $80 \pm 40 \%$\\

{\mo{torus+scattering}} & $ 7$ & $-67.7$ & $ 7 \pm 2.5$ & $-68.5 \pm 7.7$ & $100 \pm 7 \%$\\

\hline
\multicolumn{3}{l|}{\textit{$z>2$ (96 sources)}} & \multicolumn{3}{l}{} \\
{\mo{torus+pexmon+scattering}} & $ 4$ & $-28.4$ & $ 4 \pm 2.0$ & $-28.8 \pm 6.1$ & $25 \pm 43 \%$\\

{\mo{torus+scattering}} & $ 3$ & $-30.2$ & $ 2 \pm 1.6$ & $-30.8 \pm 6.0$ & $75 \pm 43 \%$\\

{\mo{wabs+pexmon+scattering}} & $ 2$ & $-30.8$ & $ 2 \pm 1.3$ & $-31.7 \pm 5.9$ & $66 \pm 47 \%$\\

{\mo{sphere+scattering}} & $ 7$ & $-39.0$ & $ 7 \pm 2.4$ & $-39.9 \pm 5.8$ & $99 \pm 10 \%$\\

\hline
\end{tabular}
\end{table*}

\end{center}

Secondly, we show the model comparison across all sources in Table
\ref{table:totalmodel}. For each model, the evidence is stacked in
column 3 by summing the $\log\, Z$ values. This form of model comparison
assumes that one common model describes all sources. For comparison,
column 2 shows the number of sources for which the model was ruled
out by the other models. As shown with Source 179 above, individual
sources can already have strong preferences in the model selection.
To have a result descriptive of the sample that is not dominated by
outliers, we apply bootstrapping. Sources are drawn with repetition,
and the same quantities (number of rejections, summed $\log\, Z$
values) are computed, and taking all draws together the mean and root
mean square is estimated. Additionally, we compute for each draw whether
this model is ruled out ($\sum\log\, Z_{1}>\sum\log\, Z_{2}+1$).
If the model is ruled out in $100\%$ of draws (column 6), the result
is robust against bootstrapping and thus there is confidence that
this result will also hold for the parent sample and is not dominated
by outliers.

The power law model (\mo{powerlaw}) has the lowest evidence and is
ruled out by simple absorption models. These in turn are ruled out
by absorption with additional scattering (in Figure \ref{modeldiagram},
strongly preferred in 15-23 objects). Then, for 6-23 objects additional
\mo{pexmon} reflection is strongly preferred.

The remaining models are thus such that absorption, scattering and
reflection are required. The number of objects for which a model is
rejected remains comparable between \mo{wabs+pexmon+scattering},
\mo{sphere+pexmon+scattering} and \mo{torus+pexmon+scattering},
with the latter having the highest evidence. The results for the full
sample have significant large differences in $\sum\log\, Z$. But
when bootstrapping the values show large variation and overlap broadly,
showing that the difference is not robust against bootstrapping, and
sample-dependent. This indicates considerable variation in evidence,
i.e. variation in the models preferred, between sources. We investigate
this by splitting the samples by $N_{H}$ (lower segments of Table
\ref{table:totalmodel}). For this, we assign a source to a $N_{H}$
bin if the $10\%$ quantiles of the posterior values determined from
\mo{torus+scattering} fall inside. As indicated before, $N_{H}$
is comparable between absorption models. Figure \ref{fig:Individual-evidence}
shows the differences in evidence between the two best models, \mo{wabs+pexmon+scattering}
and \mo{torus+pexmon+scattering} in blue circles.

In the Compton-thick regime ($N_{H}\apprge10^{24}\text{cm}^{-2}$),
\mo{torus+pexmon+scattering} is consistently strongly preferred.
However, both the table and the figure show only mild indication that
the \mo{torus+pexmon+scattering} model is the best across the full
sample. Thus some sources must favour \mo{wabs+pexmon+scattering},
while others favour \mo{torus+pexmon+scattering} (see Figure \ref{fig:Individual-evidence}).

A further concern might be that low-redshift sources with many counts
dominate the result, ignoring the target population of our inference.
The last two segments of Table \ref{table:totalmodel} shows the result
of selecting only sources with $z>1$ and $z>2$ respectively. The
inference results in this regime are entirely consistent with the
results for the full sample.

\begin{figure}[h]
\begin{centering}
\resizebox{\hsize}{!}{\includegraphics[bb=0bp 7bp 330bp 278bp,clip]{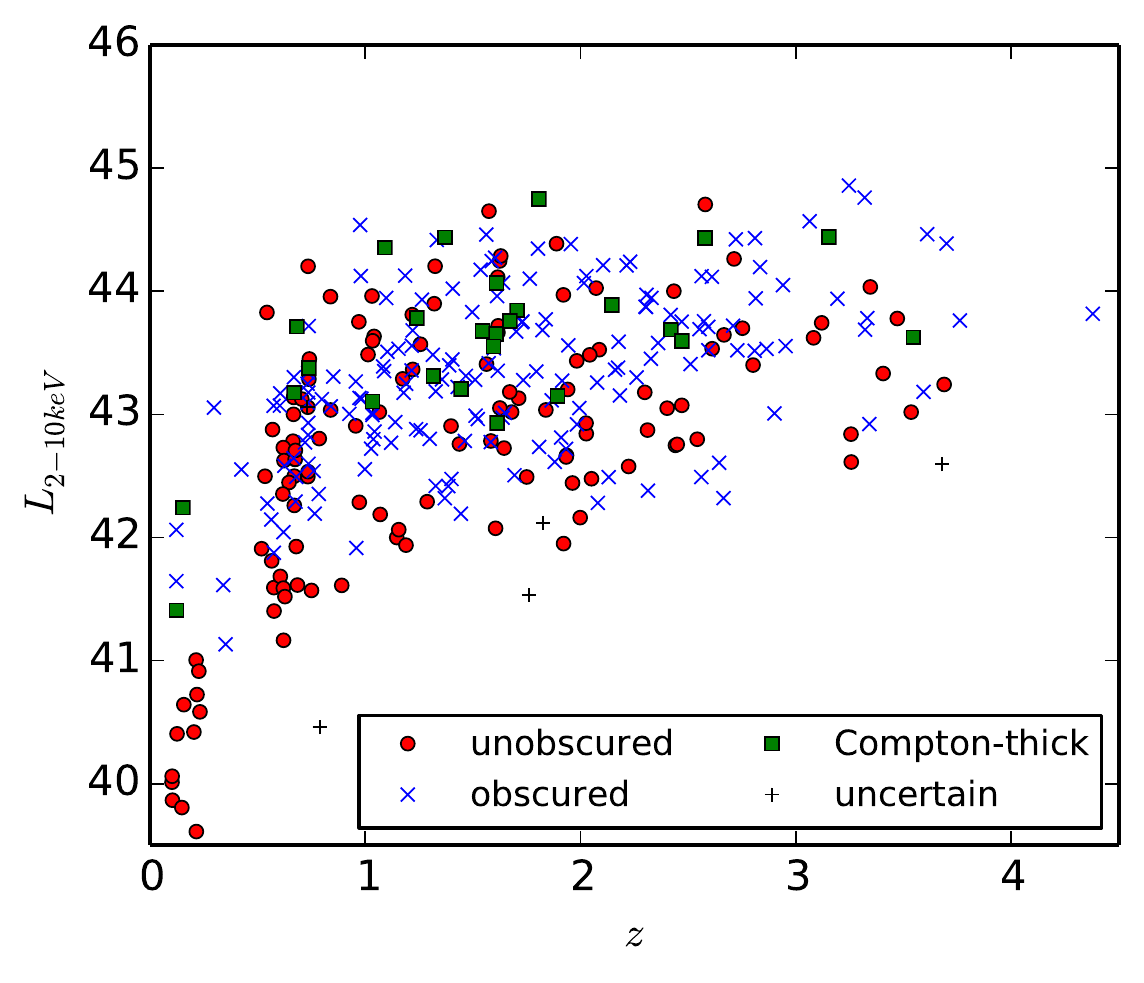}}
\par\end{centering}

\caption[L-z plane]{\label{fig:Lzplane}Luminosity-redshift plot of the sample. The median
of the intrinsic luminosity (logarithmic, in $\text{erg}/\text{s}$)
and redshift posterior probabilities have been used from the \mo{torus+pexmon+scattering}
model. Sources are classified as Compton-thick ($N_{H}>10^{24}\text{cm}^{-2}$),
obscured ($10^{22}\text{cm}^{-2}<N_{H}<10^{24}\text{cm}^{-2}$) or
unobscured ($N_{H}<10^{22}\text{cm}^{-2}$) when the majority of the
probability posterior of $N_{H}$ lies in the respective range. \replace{}{Because
of their heavy absorption, the detection of Compton-Thick AGN is biased
towards higher luminosities, compared to Compton-thin AGN.}}
\end{figure}

\begin{figure*}[t]
\begin{centering}
\includegraphics[bb=50bp 0bp 670bp 313bp,clip,width=17cm]{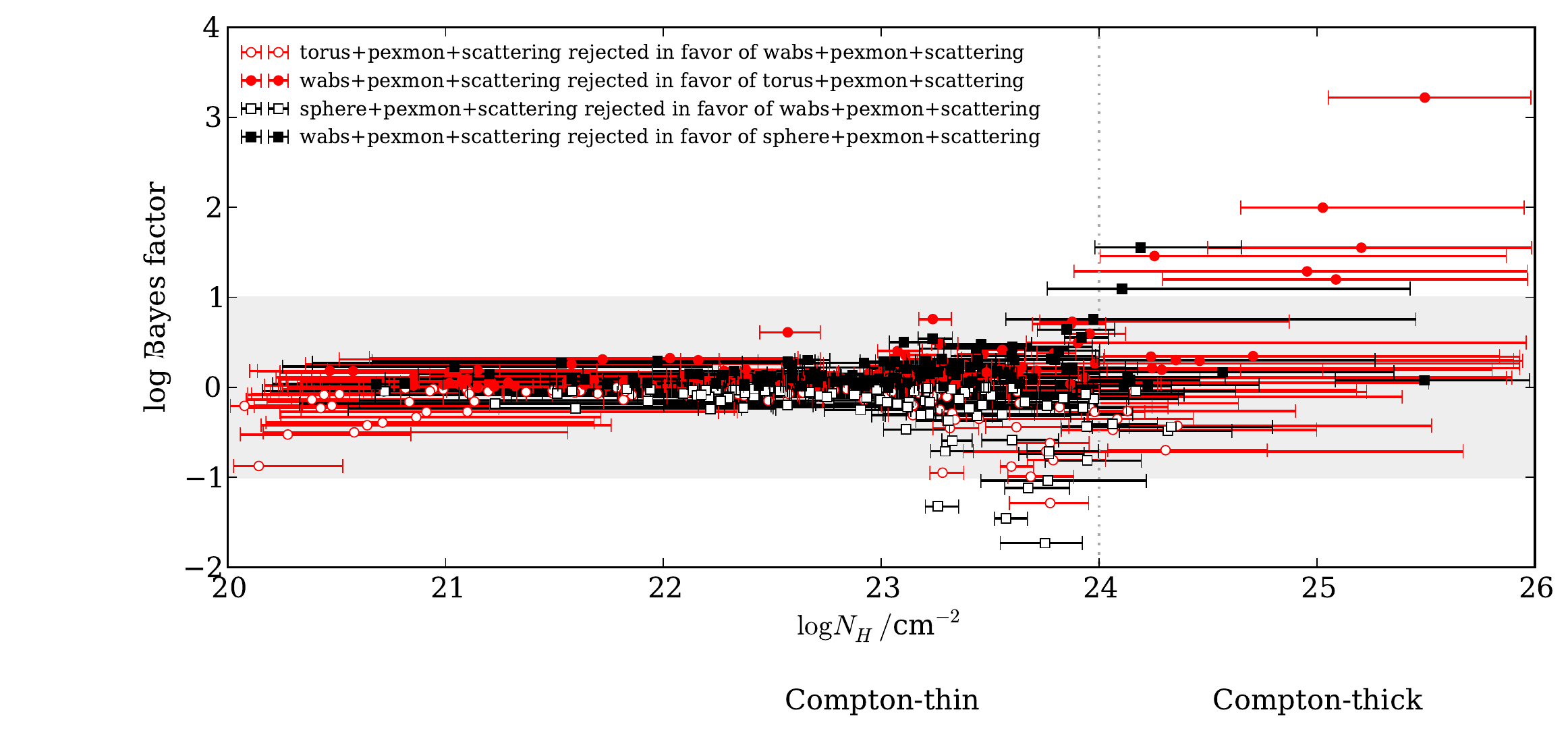}
\par\end{centering}

\caption{\label{fig:Individual-evidence}Evidence contribution from each source
with secure spectroscopic redshift. The vertical axis shows the Bayes
factor between \mo{torus+pexmon+scattering} and \mo{wabs+pexmon+scattering}
(red circles), where strong preference for the torus is above $\log\,10=1$.
The same is shown for \mo{sphere+pexmon+scattering} and \mo{wabs+pexmon+scattering}
(black squares). In both model comparisons, there are obscured objects
showing significant preference for either model.}
\end{figure*}
\begin{figure*}[p]
\begin{centering}
\includegraphics[width=8.5cm]{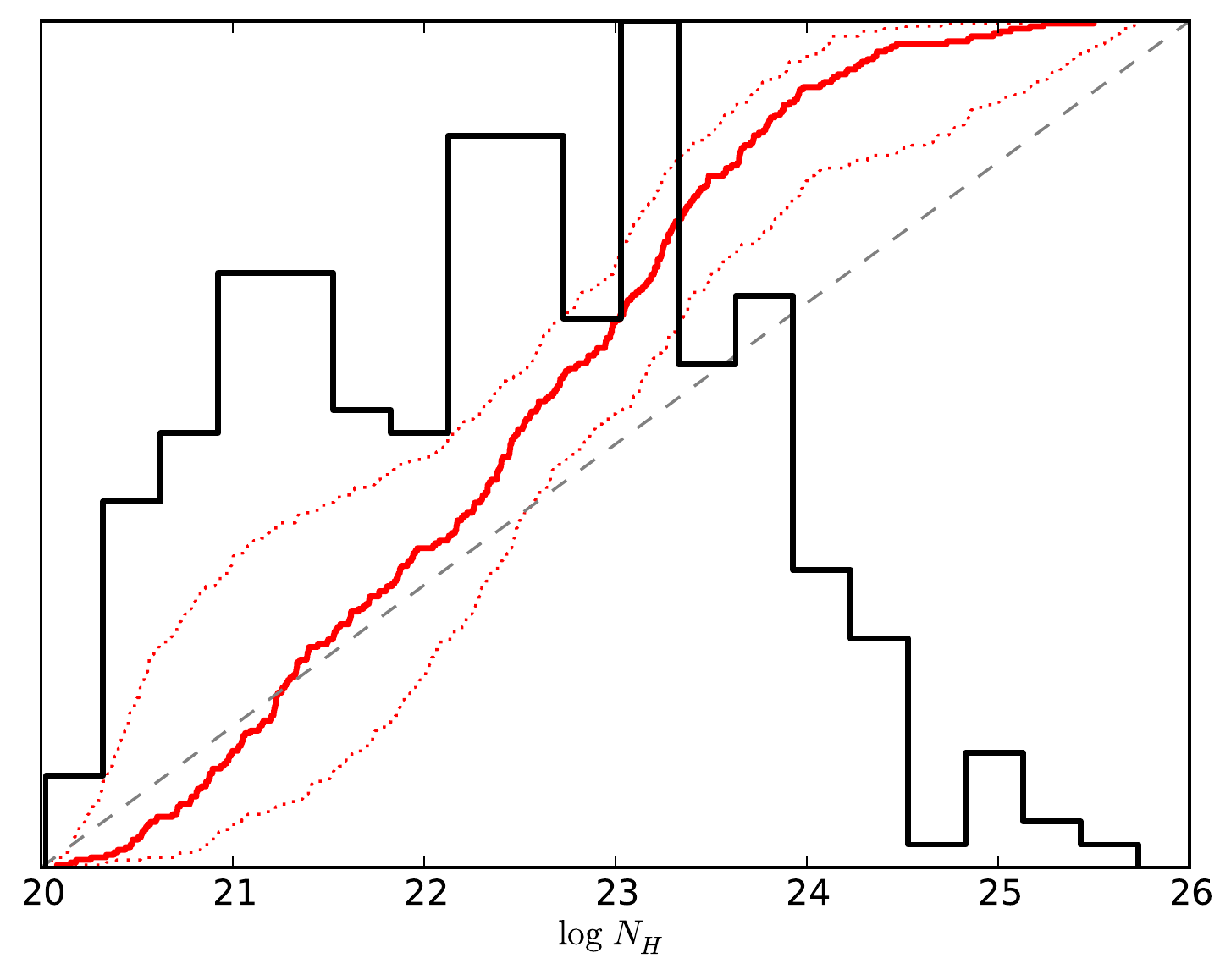}\includegraphics[width=8.5cm]{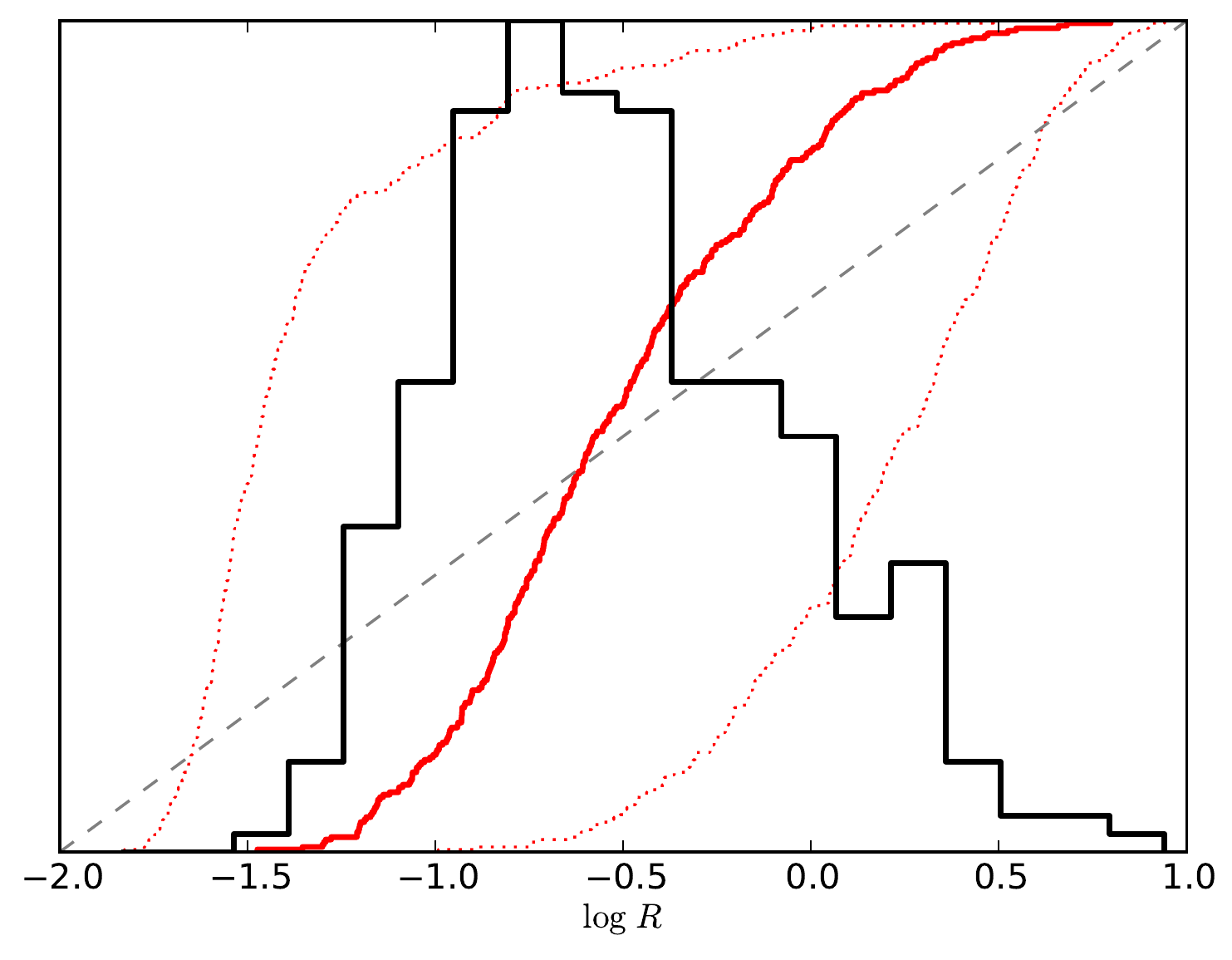}
\par\end{centering}

\begin{centering}
\includegraphics[width=8.5cm]{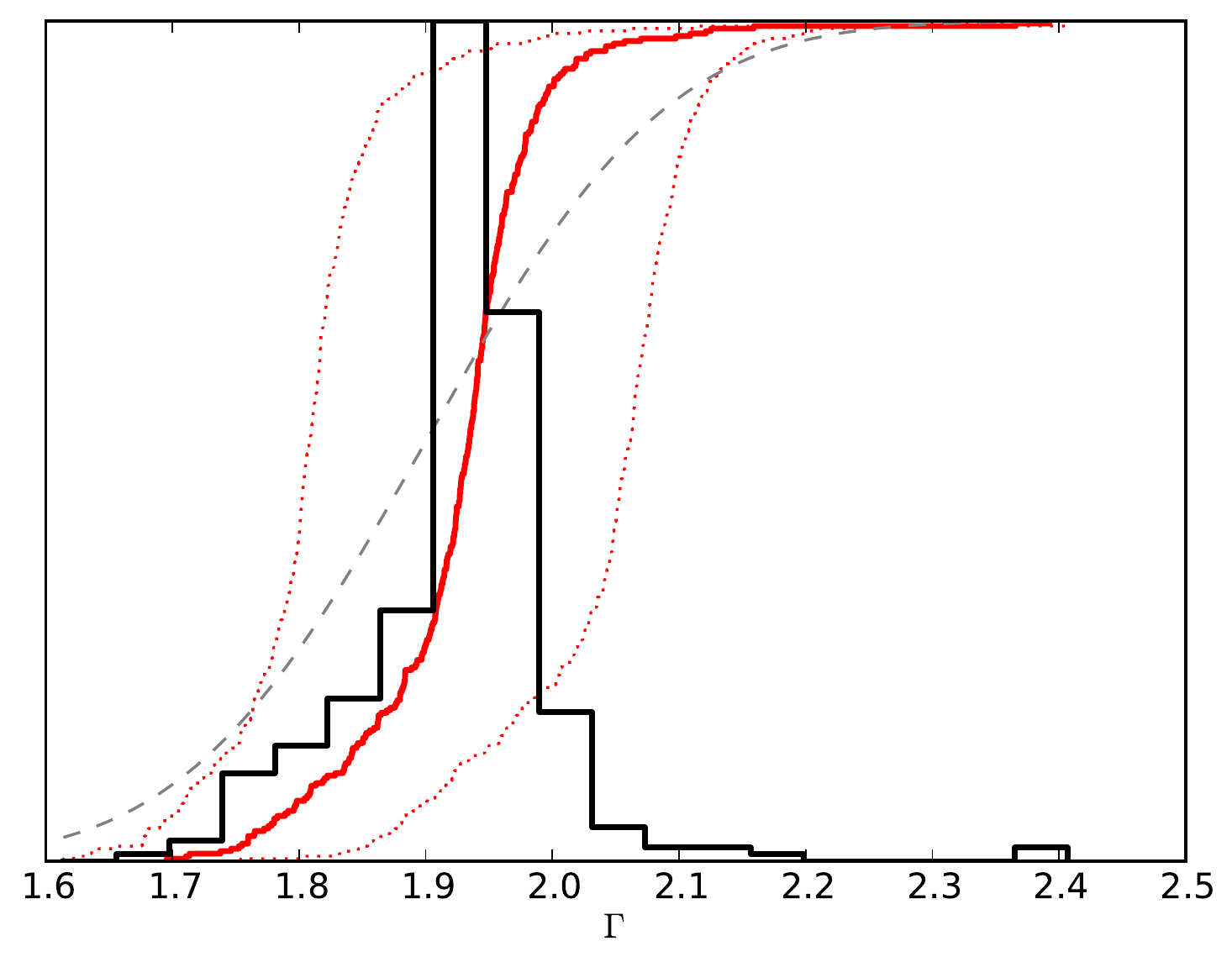}\includegraphics[width=8.5cm]{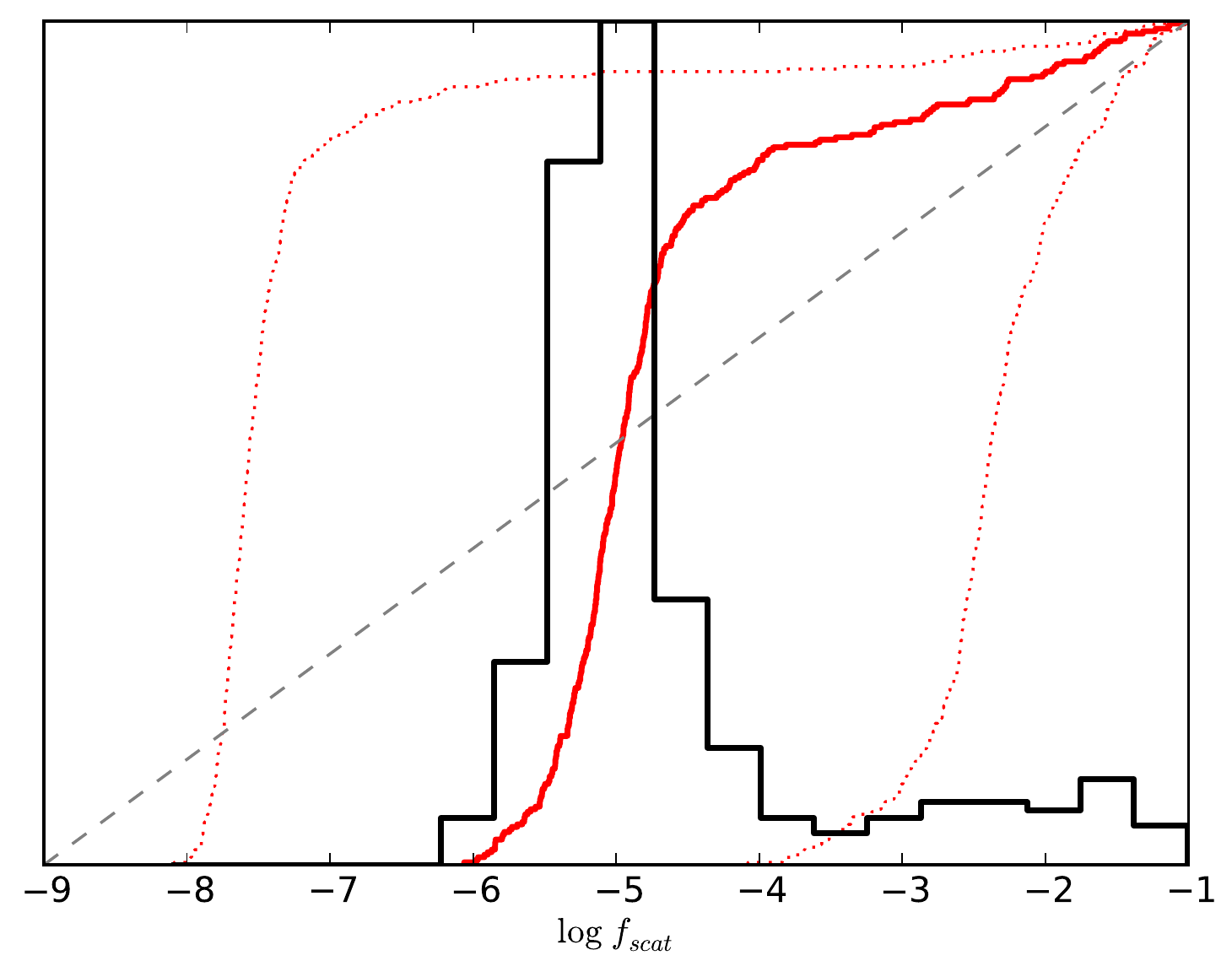}
\par\end{centering}

\caption[Parameter distributions]{\label{fig:distributions}Histograms of the best parameter values
derived using the \mo{torus+pexmon+scattering} model. The median
of the marginal posterior distribution for each object is histogrammed
in black. The thick red line shows the same as a cumulative distribution.
To illustrate the uncertainty in the parameters, the dotted red lines
show the cumulative distribution of the 10\% and 90\% quantiles instead
of the median. The dashed gray line shows the used prior.}
\end{figure*}

Overall however, \mo{torus+pexmon+scattering} can be considered the
best model. We release a catalogue of the derived quantities for each
source in the CDFS (Table \ref{table:catalogueinteresting} shows
an excerpt, Table \ref{table:catalogueCT} lists all Compton-thick
sources; the complete catalogue is available electronically). In particular,
we list column densities, intrinsic power law index, intrinsic luminosity
as well as the relative normalisations of the additional scattering
and reflection components. Full probability distributions are available
on request. The most important parameters for e.g. luminosity function
studies are $L_{2-10\text{keV}}$, $z$ and $N_{H}$, which are visualised
in Figure \ref{fig:Lzplane}. 

\begin{figure*}[h]
\begin{centering}
\includegraphics[width=8.5cm]{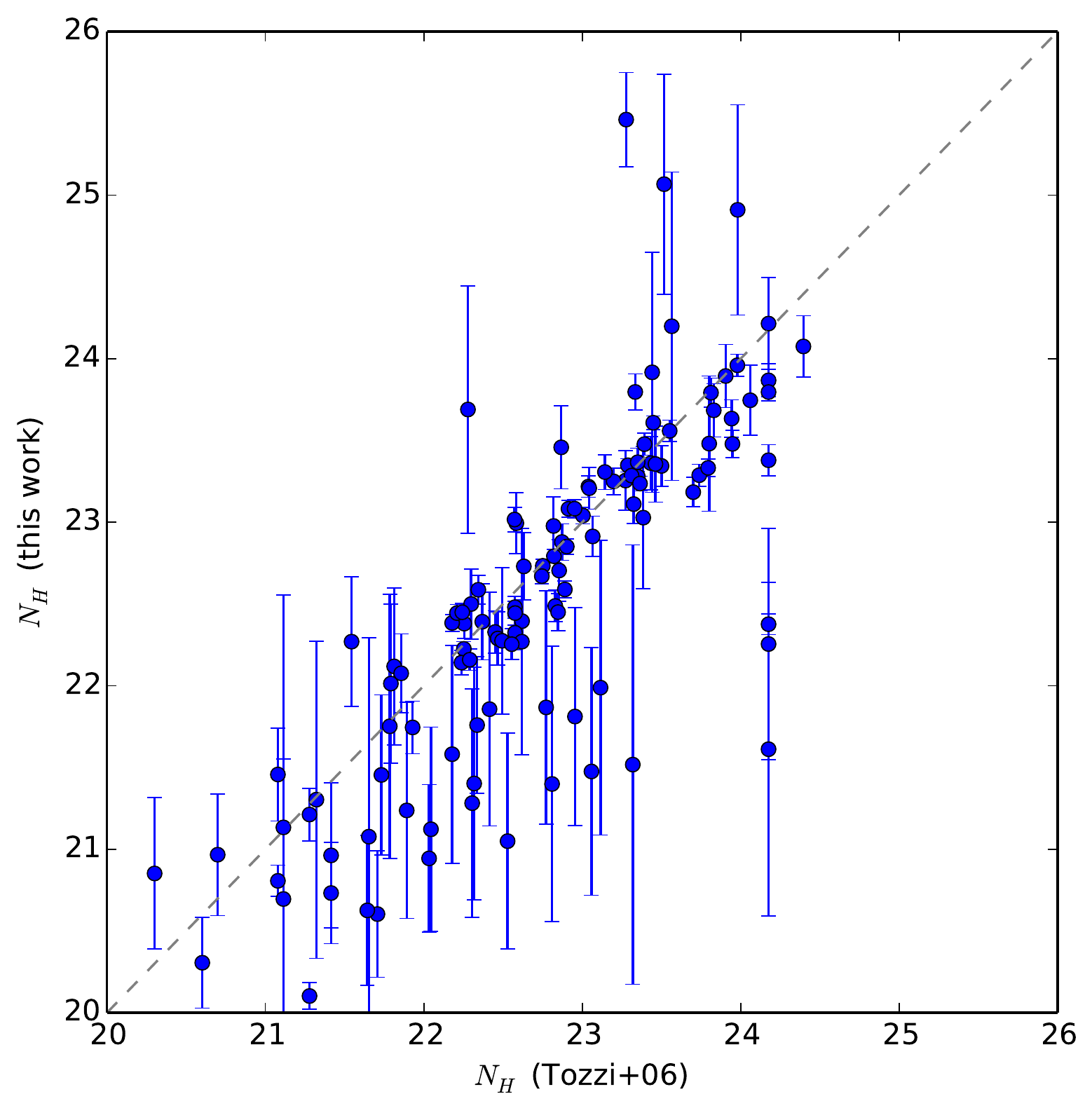}\includegraphics[width=8.5cm]{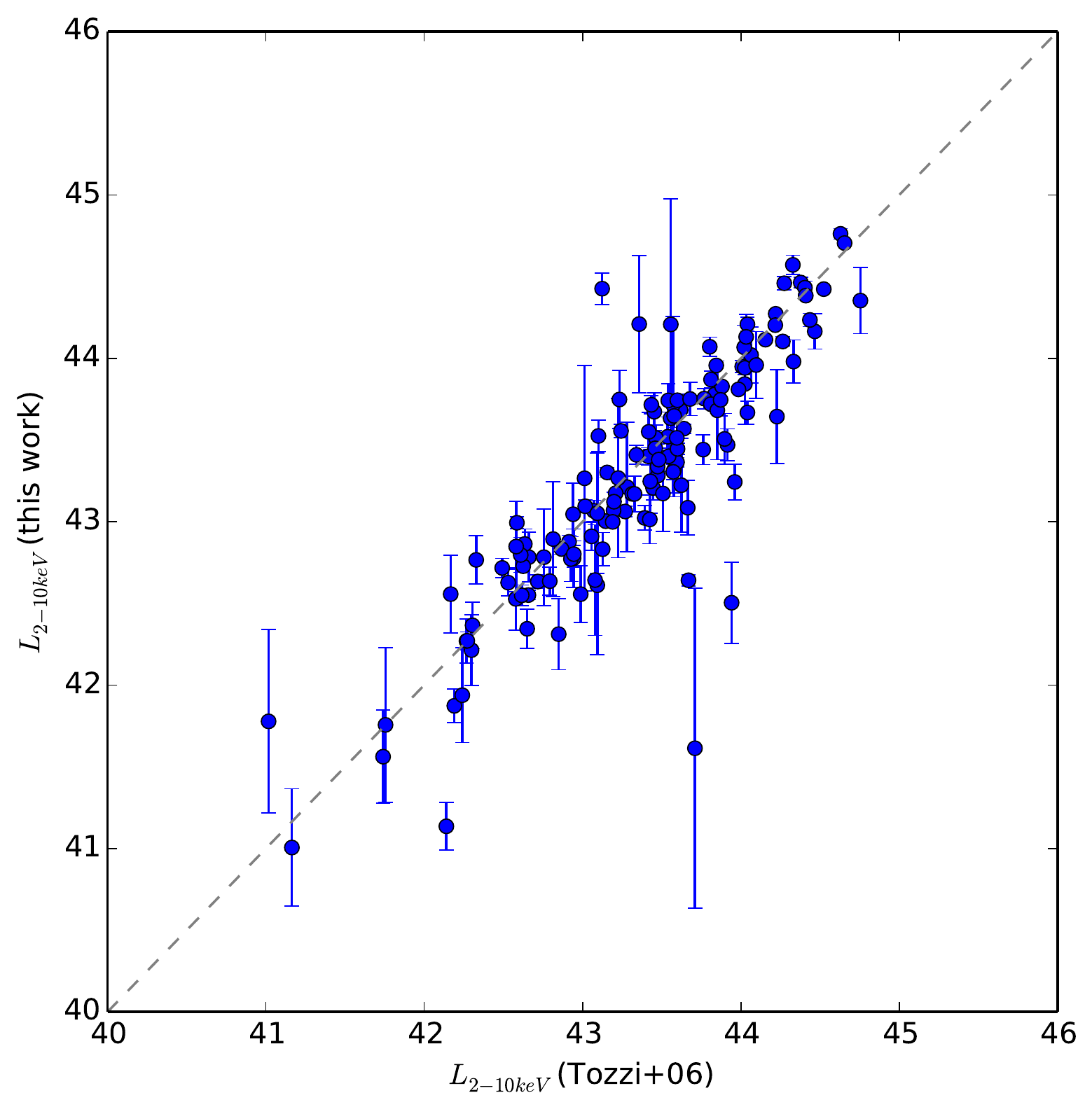}
\par\end{centering}

\caption{\change{}{\label{fig:Comparison-previous-works}Comparison of the
derived column density (left panels, $N_{H}$, here in logarithmic)
and intrinsic luminosity (right panel, logarithmic, in $\text{erg}/\text{s}$
for the $2-10\text{\,\ keV}$ rest frame band) with the analysis of
\cite{Tozzi2006}. We selected only objects from our sample which
have the same redshift in \cite{Tozzi2006} and this work. \protect \\
We plot the median and 1-sigma equivalent quantiles of the posterior
in our analysis against the best fit found in \cite{Tozzi2006}. There
are important differences between the works. The \cite{Tozzi2006}
analysis is based on only the first 1Ms data, and thus has much fewer
counts. Furthermore, only simple absorption models have been considered
in their maximum likelihood fitting.}}
\end{figure*}

\change{}{Figure \ref{fig:Comparison-previous-works} shows a comparison
to previously published works in the CDFS. }Without going into detail
here (see figure caption), the found Compton-thick AGN are in agreement
with the sample found by \cite{Brightman2012}, except that our selection
criterion removes a number of sources whose soft photons are dominated
by stellar processes. One source (ID 186 in their paper), is not found
to be a Compton-thick AGN, as a different redshift from the improved
catalogue (Hsu et al., in prep) was used.

\begin{center}\small

\begin{sidewaystable*}[p]
\caption{Catalogue (excerpt). The parameters are derived using the \mo{torus+pexmon+scattering} model. (1) XID, (2,3) Right ascension and Declination (JD2000) in degrees, (4) Photon counts obtained in the analysed $0.5-8 \text{keV}$ range. (5) Redshift with (posterior) uncertainties if photometric (shortened to 2 significant digits). (6) Intrinsic luminosity (2-10 keV) in erg/s. (7) logarithm of the column density $N_H$ in ${\text{cm}}^{-2}$. (8) Photon index; The prior was $1.95\pm0.15$, so if no information was gained this value remains. (9) Scattering normalization relative to the intrinsic power law. (10) Reflection normalization of the \mo{pexmon} component relative to the intrinsic power law. (11) Information gain measured from the $N_H$ posterior in bans. As a reference, the narrowing of a Gaussian from prior to posterior by a factor of 2 corresponds to $0.13 \text{ban}$, and thus values higher than that correspond to significant discriminatory information in the data. (12) Annotations; S when $f_{scat}>3\%$, s when $f_{scat}>0.5\%$ with $\geq90\%$ probability; R when $R > 0.3$ with $\geq 90\%$ probability, i.e. strong additional \mo{pexmon} reflection; Compton-thick (CT) if $N_H > {10}^{24} {\text{cm}}^{-2}$, Compton-thin (O) if ${10}^{22} {\text{cm}}^{-2} < N_H < {10}^{24} {\text{cm}}^{-2}$, Unobscured (U, $N_H < {10}^{22} {\text{cm}}^{-2}$, each with $\geq 50\%$ probability. Only sources with R, s or S are shown here.}
\label{table:catalogueinteresting}
\centering
\begin{tabular}{l r r r r r r r r r r r}
\hline\hline
ID & RA & Dec & counts & $z$ & $L_{2-10 \text{keV}}$ & $N_H$ & $\Gamma$ & $f_{scat}$ & $R$ & $KL|_{N_H}$ & notes \\ 
 (1) & (2) & (3) & (4) & (5) & (6) & (7) & (8) & (9) & (10) & (11) & (12)\\
\hline
49 & 3:32:14.01 & -27:51:00.53 & 1997 & 0.122 & $41.65_{-0.01}^{+0.01}$ & $22.17_{-0.02}^{+0.02}$ & $1.80_{-0.05}^{+0.05}$ & $2.2_{-0.4}^{+0.4}\%$ & $0.06_{-0.03}^{+0.06}$ & $1.49$ & O,s\\

242 & 3:31:52.35 & -27:47:52.79 & 762 & $1.84_{-0.03}^{+0.02}$ & $43.04_{-0.03}^{+0.03}$ & $21.62_{-0.44}^{+0.31}$ & $1.87_{-0.05}^{+0.05}$ & $0.0_{-0.0}^{+0.0}\%$ & $6.31_{-1.33}^{+1.30}$ & $0.36$ & U,R\\

21 & 3:32:12.94 & -27:52:36.57 & 957 & 2.562 & $44.12_{-0.03}^{+0.03}$ & $23.21_{-0.03}^{+0.03}$ & $1.76_{-0.04}^{+0.06}$ & $0.0_{-0.0}^{+0.0}\%$ & $1.08_{-0.31}^{+0.37}$ & $1.34$ & O,R\\

131 & 3:32:22.54 & -27:46:03.84 & 597 & 1.727 & $43.75_{-0.05}^{+0.05}$ & $23.37_{-0.05}^{+0.04}$ & $1.85_{-0.05}^{+0.05}$ & $5.9_{-0.8}^{+0.8}\%$ & $1.25_{-0.68}^{+0.66}$ & $1.25$ & O,S\\

118 & 3:31:52.53 & -27:46:42.30 & 1295 & 0.673 & $42.63_{-0.02}^{+0.02}$ & $21.77_{-0.06}^{+0.06}$ & $1.90_{-0.06}^{+0.07}$ & $0.0_{-0.0}^{+0.0}\%$ & $2.18_{-0.61}^{+0.64}$ & $1.02$ & U,R\\

338 & 3:32:49.20 & -27:40:50.01 & 667 & $1.26_{-0.00}^{+0.00}$ & $43.93_{-0.08}^{+0.08}$ & $23.88_{-0.06}^{+0.06}$ & $1.99_{-0.06}^{+0.06}$ & $3.6_{-0.7}^{+0.9}\%$ & $0.14_{-0.08}^{+0.25}$ & $1.05$ & O,s\\

144 & 3:32:29.98 & -27:45:29.96 & 3961 & 1.218 & $43.81_{-0.01}^{+0.01}$ & $20.52_{-0.17}^{+0.19}$ & $1.75_{-0.02}^{+0.02}$ & $0.0_{-0.0}^{+0.0}\%$ & $1.14_{-0.21}^{+0.24}$ & $0.71$ & U,R\\

104 & 3:32:08.66 & -27:47:34.34 & 28019 & 0.543 & $43.83_{-0.00}^{+0.00}$ & $20.82_{-0.05}^{+0.04}$ & $1.95_{-0.01}^{+0.01}$ & $0.0_{-0.0}^{+0.0}\%$ & $0.79_{-0.09}^{+0.09}$ & $1.21$ & U,R\\

190 & 3:32:27.00 & -27:41:05.11 & 32099 & 0.734 & $44.20_{-0.00}^{+0.00}$ & $20.08_{-0.03}^{+0.04}$ & $1.97_{-0.01}^{+0.01}$ & $0.0_{-0.0}^{+0.0}\%$ & $0.78_{-0.08}^{+0.08}$ & $1.36$ & U,R\\

92 & 3:31:58.11 & -27:48:33.97 & 915 & 0.734 & $42.53_{-0.02}^{+0.02}$ & $21.26_{-0.18}^{+0.14}$ & $1.84_{-0.06}^{+0.07}$ & $0.0_{-0.0}^{+0.0}\%$ & $3.39_{-0.72}^{+0.84}$ & $0.61$ & U,R\\

179 & 3:32:13.23 & -27:42:41.02 & 2485 & 0.605 & $43.17_{-0.01}^{+0.01}$ & $22.45_{-0.03}^{+0.02}$ & $1.90_{-0.04}^{+0.05}$ & $7.4_{-0.9}^{+0.8}\%$ & $1.61_{-0.31}^{+0.29}$ & $1.45$ & O,S,R\\

287 & 3:31:55.40 & -27:54:47.21 & 618 & 0.737 & $43.72_{-0.03}^{+0.03}$ & $23.02_{-0.04}^{+0.04}$ & $1.89_{-0.06}^{+0.07}$ & $3.4_{-0.6}^{+0.6}\%$ & $0.13_{-0.08}^{+0.19}$ & $1.31$ & O,s\\

4 & 3:32:38.90 & -27:57:00.17 & 4687 & 0.297 & $43.06_{-0.01}^{+0.01}$ & $22.59_{-0.01}^{+0.01}$ & $1.91_{-0.04}^{+0.04}$ & $2.5_{-0.2}^{+0.2}\%$ & $1.16_{-0.18}^{+0.18}$ & $1.76$ & O,s,R\\

133 & 3:32:39.08 & -27:46:01.79 & 1355 & 1.216 & $43.56_{-0.02}^{+0.02}$ & $22.60_{-0.04}^{+0.03}$ & $2.11_{-0.05}^{+0.05}$ & $9.4_{-0.4}^{+0.3}\%$ & $0.87_{-0.43}^{+0.43}$ & $1.26$ & O,S\\

50 & 3:32:18.34 & -27:50:55.13 & 647 & 1.536 & $44.18_{-0.05}^{+0.04}$ & $23.80_{-0.03}^{+0.03}$ & $1.89_{-0.06}^{+0.06}$ & $1.0_{-0.2}^{+0.2}\%$ & $0.17_{-0.10}^{+0.24}$ & $1.46$ & O,s\\

181 & 3:32:47.88 & -27:42:32.78 & 8718 & 0.979 & $44.12_{-0.00}^{+0.00}$ & $22.30_{-0.02}^{+0.02}$ & $1.86_{-0.02}^{+0.02}$ & $7.3_{-1.1}^{+1.0}\%$ & $0.05_{-0.02}^{+0.04}$ & $1.56$ & O,S\\

132 & 3:32:03.66 & -27:46:03.74 & 1380 & 0.574 & $43.07_{-0.02}^{+0.02}$ & $22.79_{-0.02}^{+0.02}$ & $1.88_{-0.05}^{+0.06}$ & $0.5_{-0.5}^{+0.4}\%$ & $1.26_{-0.30}^{+0.32}$ & $1.56$ & O,R\\

7 & 3:32:40.82 & -27:55:46.76 & 1557 & $3.25_{-0.01}^{+0.01}$ & $44.86_{-0.04}^{+0.04}$ & $23.91_{-0.02}^{+0.02}$ & $1.86_{-0.05}^{+0.06}$ & $3.6_{-0.6}^{+0.6}\%$ & $0.06_{-0.03}^{+0.06}$ & $1.49$ & O,s\\

200 & 3:32:34.39 & -27:39:13.55 & 1848 & $1.56_{-0.01}^{+0.01}$ & $44.46_{-0.02}^{+0.02}$ & $23.09_{-0.03}^{+0.02}$ & $2.01_{-0.05}^{+0.04}$ & $4.8_{-0.7}^{+0.7}\%$ & $0.20_{-0.13}^{+0.26}$ & $1.45$ & O,s\\

\hline
\end{tabular}
\end{sidewaystable*}

\end{center}

\begin{center}\small

\begin{sidewaystable*}[p]
\caption{Catalogue (Compton Thicks only). The parameters are derived using the \mo{torus+pexmon+scattering} model. (1) XID, (2,3) Right ascension and Declination (JD2000) in degrees, (4) Photon counts obtained in the analysed $0.5-8 \text{keV}$ range. (5) Redshift with (posterior) uncertainties if photometric (shortened to 2 significant digits). (6) Intrinsic luminosity (2-10 keV) in erg/s. (7) logarithm of the column density $N_H$ in ${\text{cm}}^{-2}$. (8) Photon index; The prior was $1.95\pm0.15$, so if no information was gained this value remains. (9) Scattering normalization relative to the intrinsic power law. (10) Reflection normalization of the \mo{pexmon} component relative to the intrinsic power law. (11) Information gain measured from the $N_H$ posterior in bans. As a reference, the narrowing of a Gaussian from prior to posterior by a factor of 2 corresponds to $0.13 \text{ban}$, and thus values higher than that correspond to significant discriminatory information in the data. (12) Annotations; S when $f_{scat}>3\%$, s when $f_{scat}>0.5\%$ with $\geq90\%$ probability; R when $R > 0.3$ with $\geq 90\%$ probability, i.e. strong additional \mo{pexmon} reflection; Compton-thick (CT) if $N_H > {10}^{24} {\text{cm}}^{-2}$, Compton-thin (O) if ${10}^{22} {\text{cm}}^{-2} < N_H < {10}^{24} {\text{cm}}^{-2}$, Unobscured (U, $N_H < {10}^{22} {\text{cm}}^{-2}$, each with $\geq 50\%$ probability. Only Compton-thick sources are shown here.}
\label{table:catalogueCT}
\centering
\begin{tabular}{l r r r r r r r r r r r}
\hline\hline
ID & RA & Dec & counts & $z$ & $L_{2-10 \text{keV}}$ & $N_H$ & $\Gamma$ & $f_{scat}$ & $R$ & $KL|_{N_H}$ & notes \\ 
 (1) & (2) & (3) & (4) & (5) & (6) & (7) & (8) & (9) & (10) & (11) & (12)\\
\hline
460 & 3:32:23.41 & -27:42:55.89 & 194 & 2.145 & $43.89_{-0.19}^{+0.16}$ & $24.37_{-0.13}^{+0.15}$ & $1.94_{-0.06}^{+0.07}$ & $0.0_{-0.0}^{+0.0}\%$ & $0.32_{-0.23}^{+0.56}$ & $0.61$ & CT\\

420 & 3:32:23.60 & -27:46:01.23 & 48 & 1.033 & $43.11_{-0.15}^{+0.12}$ & $24.32_{-0.09}^{+0.07}$ & $1.93_{-0.07}^{+0.08}$ & $0.0_{-0.0}^{+0.0}\%$ & $3.45_{-1.81}^{+2.03}$ & $0.91$ & CT\\

412 & 3:32:42.87 & -27:48:09.44 & 67 & $2.42_{-0.05}^{+0.04}$ & $43.69_{-0.19}^{+0.16}$ & $24.42_{-0.13}^{+0.14}$ & $1.94_{-0.06}^{+0.06}$ & $0.0_{-0.0}^{+0.0}\%$ & $0.28_{-0.20}^{+0.51}$ & $0.59$ & CT\\

448 & 3:32:19.41 & -27:40:51.76 & 528 & 0.682 & $43.71_{-0.09}^{+0.07}$ & $25.07_{-0.32}^{+0.33}$ & $1.99_{-0.07}^{+0.08}$ & $0.3_{-0.1}^{+0.1}\%$ & $0.55_{-0.40}^{+1.09}$ & $0.37$ & CT\\

474 & 3:32:15.81 & -27:42:06.78 & 339 & $1.71_{-0.05}^{+0.21}$ & $43.85_{-0.22}^{+0.21}$ & $24.21_{-0.15}^{+0.15}$ & $1.95_{-0.06}^{+0.06}$ & $0.0_{-0.0}^{+0.0}\%$ & $0.19_{-0.12}^{+0.38}$ & $0.55$ & CT\\

271 & 3:32:33.86 & -27:42:04.30 & 306 & $1.45_{-0.01}^{+0.01}$ & $43.20_{-0.99}^{+0.42}$ & $24.16_{-2.29}^{+0.34}$ & $1.95_{-0.06}^{+0.05}$ & $0.4_{-0.4}^{+1.0}\%$ & $1.03_{-0.69}^{+1.71}$ & $0.25$ & CT\\

478 & 3:32:14.60 & -27:52:57.06 & 164 & $1.24_{-0.01}^{+0.01}$ & $43.78_{-0.13}^{+0.09}$ & $24.84_{-0.24}^{+0.29}$ & $1.95_{-0.06}^{+0.06}$ & $0.0_{-0.0}^{+0.0}\%$ & $0.41_{-0.27}^{+0.91}$ & $0.36$ & CT\\

467 & 3:32:00.80 & -27:53:33.84 & 541 & $1.60_{-0.03}^{+0.02}$ & $43.55_{-0.17}^{+0.17}$ & $24.14_{-0.08}^{+0.09}$ & $1.91_{-0.06}^{+0.06}$ & $0.0_{-0.0}^{+0.0}\%$ & $2.05_{-1.40}^{+2.01}$ & $0.85$ & CT\\

437 & 3:32:39.15 & -27:48:32.26 & 54 & 2.470 & $43.60_{-0.15}^{+0.13}$ & $24.27_{-0.09}^{+0.10}$ & $1.94_{-0.06}^{+0.06}$ & $0.0_{-0.0}^{+0.0}\%$ & $0.36_{-0.26}^{+0.72}$ & $0.84$ & CT\\

345 & 3:32:24.67 & -27:54:43.24 & 255 & 0.123 & $41.41_{-1.21}^{+0.41}$ & $24.13_{-2.02}^{+0.74}$ & $1.95_{-0.07}^{+0.06}$ & $0.5_{-0.2}^{+0.6}\%$ & $0.37_{-0.28}^{+0.96}$ & $0.16$ & CT\\

451 & 3:32:31.49 & -27:50:28.64 & 38 & 1.613 & $42.93_{-0.96}^{+0.44}$ & $24.34_{-0.55}^{+0.43}$ & $1.95_{-0.07}^{+0.07}$ & $0.0_{-0.0}^{+0.1}\%$ & $0.40_{-0.30}^{+0.88}$ & $0.11$ & CT\\

266 & 3:32:14.94 & -27:42:24.83 & 976 & $1.81_{-0.02}^{+0.02}$ & $44.75_{-1.38}^{+0.14}$ & $24.95_{-3.94}^{+0.46}$ & $1.94_{-0.05}^{+0.05}$ & $2.0_{-2.0}^{+0.8}\%$ & $1.94_{-1.65}^{+3.60}$ & $0.35$ & CT\\

416 & 3:32:21.77 & -27:46:56.99 & 64 & $1.32_{-0.02}^{+0.03}$ & $43.31_{-0.12}^{+0.10}$ & $24.09_{-0.05}^{+0.05}$ & $1.94_{-0.07}^{+0.06}$ & $0.0_{-0.0}^{+0.0}\%$ & $0.28_{-0.18}^{+0.59}$ & $1.08$ & CT\\

406 & 3:31:45.16 & -27:49:48.72 & 571 & 3.153 & $44.44_{-0.24}^{+0.14}$ & $24.73_{-0.18}^{+0.18}$ & $1.91_{-0.05}^{+0.07}$ & $0.0_{-0.0}^{+0.0}\%$ & $1.09_{-0.75}^{+1.29}$ & $0.45$ & CT\\

400 & 3:32:25.18 & -27:54:49.72 & 346 & 1.090 & $44.36_{-0.04}^{+0.03}$ & $25.23_{-0.37}^{+0.26}$ & $1.88_{-0.04}^{+0.05}$ & $0.0_{-0.0}^{+0.0}\%$ & $0.94_{-0.61}^{+1.39}$ & $0.46$ & CT\\

404 & 3:32:36.15 & -27:50:36.97 & 108 & 1.608 & $43.65_{-0.12}^{+0.12}$ & $24.07_{-0.08}^{+0.08}$ & $1.96_{-0.07}^{+0.07}$ & $0.2_{-0.2}^{+0.4}\%$ & $0.16_{-0.10}^{+0.32}$ & $0.91$ & CT\\

401 & 3:31:50.45 & -27:52:11.49 & 592 & 1.370 & $44.44_{-0.25}^{+0.08}$ & $25.17_{-0.45}^{+0.27}$ & $1.95_{-0.07}^{+0.07}$ & $0.5_{-0.3}^{+0.2}\%$ & $1.02_{-0.81}^{+2.52}$ & $0.27$ & CT\\

158 & 3:32:22.59 & -27:44:25.97 & 129 & 0.738 & $43.38_{-0.20}^{+0.17}$ & $24.25_{-0.10}^{+0.11}$ & $1.95_{-0.06}^{+0.06}$ & $0.4_{-0.2}^{+0.3}\%$ & $0.75_{-0.54}^{+1.48}$ & $0.73$ & CT\\

77 & 3:32:35.71 & -27:49:16.18 & 95 & 2.578 & $44.43_{-0.04}^{+0.04}$ & $25.50_{-0.14}^{+0.12}$ & $1.93_{-0.06}^{+0.06}$ & $0.0_{-0.0}^{+0.0}\%$ & $0.88_{-0.68}^{+1.41}$ & $0.75$ & CT\\

417 & 3:32:21.97 & -27:46:56.00 & 35 & 0.670 & $43.18_{-0.09}^{+0.07}$ & $24.99_{-0.40}^{+0.30}$ & $1.92_{-0.07}^{+0.07}$ & $0.0_{-0.0}^{+0.0}\%$ & $0.26_{-0.19}^{+0.50}$ & $0.41$ & CT\\

570 & 3:33:00.76 & -27:48:57.46 & 438 & $1.67_{-0.08}^{+0.04}$ & $43.76_{-0.18}^{+0.19}$ & $24.36_{-0.12}^{+0.11}$ & $1.93_{-0.06}^{+0.07}$ & $1.0_{-0.9}^{+1.3}\%$ & $1.24_{-0.89}^{+1.49}$ & $0.64$ & CT\\

226 & 3:32:14.42 & -27:51:10.43 & 141 & 1.544 & $43.68_{-0.13}^{+0.14}$ & $24.16_{-0.06}^{+0.07}$ & $1.96_{-0.07}^{+0.07}$ & $1.2_{-0.5}^{+0.7}\%$ & $0.23_{-0.14}^{+0.55}$ & $0.94$ & CT\\

441 & 3:32:22.79 & -27:45:28.54 & 46 & $1.89_{-0.04}^{+0.04}$ & $43.15_{-0.47}^{+0.27}$ & $24.44_{-0.24}^{+0.23}$ & $1.93_{-0.06}^{+0.06}$ & $0.0_{-0.0}^{+0.0}\%$ & $0.45_{-0.33}^{+1.00}$ & $0.26$ & CT\\

408 & 3:32:33.22 & -27:49:15.92 & 38 & $3.55_{-0.03}^{+0.03}$ & $43.63_{-0.24}^{+0.30}$ & $24.29_{-0.22}^{+0.54}$ & $1.95_{-0.06}^{+0.07}$ & $0.0_{-0.0}^{+0.0}\%$ & $0.26_{-0.16}^{+0.40}$ & $0.40$ & CT\\

459 & 3:32:29.48 & -27:43:22.27 & 143 & 1.609 & $44.07_{-0.06}^{+0.05}$ & $25.01_{-0.13}^{+0.14}$ & $1.90_{-0.05}^{+0.06}$ & $0.0_{-0.0}^{+0.0}\%$ & $1.77_{-1.21}^{+1.81}$ & $0.60$ & CT\\

384 & 3:32:42.02 & -27:39:49.95 & 507 & 0.152 & $42.24_{-1.70}^{+0.42}$ & $24.46_{-2.62}^{+0.52}$ & $1.92_{-0.06}^{+0.06}$ & $0.1_{-0.1}^{+0.2}\%$ & $0.81_{-0.61}^{+1.56}$ & $0.22$ & CT\\

\hline
\end{tabular}
\end{sidewaystable*}

\end{center}

\section{Discussion\label{sec:Discussion}}

Before putting the results into context, we review our methodology.

\subsection{X-ray spectral analysis methodology}

We have presented a new framework and method for analysing X-ray spectra,
relying on Bayesian inference using nested sampling. In particular,
parameter estimation and model comparison are easily possible and
overcome considerable limitations of current methods (see Section
\ref{sub:Statistical-analysis-methods} for a detail discussion of
various methods):
\begin{enumerate}
\item \textbf{No binning of data}. Low-count and high-count sources are
treated the same way using Poisson statistics, as with C-stat in the
well-established maximum likelihood estimation methods. No information
loss by binning needs to be introduced.
\item \textbf{Background modelling}. The background is modelled with a continuous
non-physical model (Gaussian mixture). Unlike other options (background
subtractions, bin-wise background estimation), this method remains
consistent with the used Poisson statistics.
\item \textbf{Bayesian parameter estimation}. The presented Bayesian framework
allows the estimation of parameters where full probability distributions
for each parameter are a natural outcome. Constraints such as unphysical
regions in parameter space, knowledge from local samples, and information
from other studies can be incorporated. For instance, with the $\Gamma$-prior
we include the assumption that high-redshift AGN behave like local
AGN in some regards, and we propagate the uncertainty of redshifts
estimates for each source.
\item \textbf{Model comparison}. The comparison of models used here overcomes
the limitations of current methods. Likelihood-ratio based methods
are approximate results in the limit, which can not compare non-nested
models. Unlike approximations like information criteria, the approach
is general so that it is unproblematic for model comparison at boundaries.
\end{enumerate}
The implementation overcomes the weaknesses of standard MCMC, namely
unknown convergence and multi-modal parameter spaces (see the discussion
of methodology in Section \ref{sub:Statistical-analysis-methods},
and also the Appendix \ref{sec:pdz-test} for a specific case). The
computational cost is not higher than classical fitting with error
estimation or MCMC.

Taking the small step from the MLE-based approach (``C-stat'') to
a Bayesian methodology, one might be concerned that the priors influence
the result too much. Similarly, one may ask why parametric models
are used when no physical model is available? Non-parametric methods
would remove the a-priori assumption of a specific model. Often however,
physically motivated models \textit{are} available, and the same is
true for priors. Similarly to comparing multiple competing models,
multiple priors can be tried to test the robustness of the results.
To address the concern that prior distributions may dominate the posterior
distribution, the Kullback-Leibler divergence (``information gain'')
is a useful characterisation of how strongly the posterior was influenced
by the prior (see Section \ref{sub:Statistical-analysis-methods}).

The presented method performs better in model selection than likelihood-ratios,
which are also problematic for the models considered here (see Appendix
\ref{sec:False-discovery-rate}, and the extensive discussion of methods
in Section \ref{sub:Statistical-analysis-methods}). For goodness-of-fit,
we use the Kolmogorov-Smirnov measure to detect potential problems
in background model fits, and visually inspect the deviations using
Q\textendash{}Q plots (see Appendix \ref{sec:Appendix-I:-Goodness-1}).

\subsection{Implications for the geometry of the obscurer of AGN}

\para{Absorption exists}{\stmt{AGN are confirmed to have line-of-sight
obscuration using our methodology by ruling out a simple power law
in favour of any of the obscuration models (see Table \ref{table:refl-modelselection}
for decisive evidence on a single source, and Table \ref{table:totalmodel}
for the full sample). }\just{The improvement of the fit by adding
photo-electric absorption (\mo{wabs}) is shown in Figure \ref{fig:refl-example-spec}
for a single source (ID 179), where the simple power law clearly does
not fit the data.} If a significant portion of AGN consisted of powerlaw
spectra with a low photon index, this simpler model would have been
preferred in the bootstrapped results of Table \ref{table:totalmodel}.
This hypothesis by \cite{Hopkins2009} can clearly be rejected.}

\para{Scattering exists}{\stmt{Additionally, a fraction of energy
which has not seen any obscuration is apparent in the soft energies
(shown in Figure \ref{fig:refl-example-spec}, upper right panel).
This component can be attributed to Thomson-scattering by ionised
material within the opening angle of the torus, scattering the intrinsic
spectrum into the line-of-sight. Models without this component are
ruled out, as can be seen from the large differences in evidence values
in Table \ref{table:refl-modelselection} for Source 179, and in Table
\ref{table:totalmodel} for the full sample. This soft component may
be confused with other processes such as thermal disk emission or
stellar processes. To remedy this, we removed host-dominated sources,
and further only considered the subsample of $z>1$, where only the
$>1\text{keV}$ photons enter the observed band. As the lower segments
of Table \ref{table:totalmodel} show, this component is still strongly
preferred. The detection of the soft scattering component is in agreement
with \cite{Brightman2012}, who used an ad-hoc method for model selection.}}

\para{Absorption models}{We considered three different absorption
models, which differ mainly in the amount of Compton scattering produced
outside the line-of-sight due to volume filling. While \mo{wabs}
represents a bullet-like blob in the line-of-sight with no Compton
scattering, \mo{sphere} and \mo{torus} model a fully and partially
open toroidal absorber respectively. The latter two models, computed
using Monte Carlo simulations on a constant-density geometry, differ
from \mo{wabs} as they consider Compton scattering and K-shell fluorescence.
For the full sample, \mo{wabs+scattering} is ruled out by \mo{torus+scattering}
(and also \mo{sphere+scattering}), indicating that these differences
are important, i.e. that forward-scattered, low-energy radiation and
the additional reflection are observed. This is a relevant finding
because it means that high-redshift data is significantly better modelled
by a more complex model than commonly used. }

\para{Absorption models need reflection}{Next to absorption and
the scattering component, we find that additional Compton reflection
is needed. In Figure \ref{fig:refl-example-spec}, where the spectrum
of \mo{torus+scattering} is shown in the lower left panel, this component
is clearly visible in the data through its most prominent feature,
the $Fe-K\alpha$ line. As \mo{torus+scattering} already models
the Compton scattering and line emission within the well-constrained
line-of-sight obscurer, this component must be of different origin.
Radiation may be scattered into the line of sight from denser regions
of the torus, if a density gradient is assumed. It is worth re-stating
that we photo-electrically absorbed the \mo{+pexmon} component, requiring
the reflection to occur behind the LOS column density. Alternatively,
the accretion disk may contribute a reflection spectrum as is known
from unobscured objects, which is transmitted through the obscurer.
In principle, a higher iron abundance is another hypothesis to increase
the yield of the line. These options are discussed further in Buchner
et al. (in prep).}

\para{Absorption model comparison}{Overall, comparing the absorption
models for obscured sources, \mo{torus+pexmon+scattering} is the
best model for obscured AGN, especially when considering Compton-thick
AGN. However, the combination of photo-electric absorption with a
Compton scattering and reflection component (\mo{wabs+pexmon+scattering})
can emulate the observed spectra almost equally well, especially in
the Compton-thin regime where the diversity does not seem to suggest
one common geometry (see Figure \ref{fig:Individual-evidence}). In
this phenomenological model, the three interaction processes -- photo-electric
absorption, Compton scattering and Thomson scattering -- are independent
and not physically connected. Figure \ref{fig:model-geometry}f,g
illustrates possible geometries. But the \mo{wabs+pexmon+scattering}
model is ruled out for Compton-thick AGN (see lower segment of Table
\ref{table:totalmodel}), in favour of \mo{torus+scattering}.

The considered obscurer geometries differ in their covering factor,
and thus in the strength of the Compton reflection hump. The \mo{sphere}
geometry has the largest reflective area, while \mo{wabs} does not
have any Compton scattering; \mo{torus} constitutes an intermediate
case. The \mo{sphere+pexmon+scattering} model is ruled out in favour
of \mo{torus+pexmon+scattering}. As the geometry is the only difference
between the models, we conclude that the opening angle of AGN must
not be vanishing. The spherical geometry can thus be excluded not
only by the amount of scattering needed, but independent of that due
to the shape of the reflection hump (obscured sources in Table \ref{table:totalmodel}
rule out the \mo{sphere+scattering} and \mo{wabs+scattering} model).
This result also holds when only the Compton-thick AGN are considered.}

\para{Reflection in unobscured}{\stmt{Unobscured objects, in contrast,
are often well-described by a simple power law. However, they may
show Fe lines originating from reflection off Compton-thick material
outside the line of sight, either from the accretion disk or the torus.
For this reason, e.g. \mo{wabs+pexmon+scattering} provides a good
fit here. The torus simulation used may be a good fit as well if \mo{torus}
had not been constrained to an edge-on view. In the face-on view,
the Compton scattering off the torus is part of the model spectrum
\citep[see ][]{Brightman2011a}.}}

\para{Other stuff tried}{A number of more complicated variations
of the best model, \mo{torus+pexmon+scattering}, have been tried,
namely (1) linking the opening angle to $\log\, N_{H}$ by decreasing
it linearly from $60\text{\textdegree}$ to $40\text{\textdegree}$
from unobscured to Compton-thick sources, (2) making the opening angle
a free parameter for each source, and (3) freeing both the opening
and viewing angle. These models yield comparable evidence to \mo{torus+pexmon+scattering}
and thus are not justified by the data.}

\section{Conclusions\label{sec:Conclusions}}

We develop a Bayesian framework for analysing X-ray spectra. We apply
this methodology to $\sim350$ faint, low-count spectra of AGN in
the CDFS to infer model parameters. The approach propagates all uncertainties
e.g. the Poisson process of collecting counts, or errors in photometric
redshifts determination. The novelty of this work however is to apply
Bayesian model comparison.

We consider physically motivated models where various geometries --
no obscurer, bullet-like blob in the LOS, toroidal and spherical obscurer
-- are considered. The best model has (1) an intrinsic power law obscured
by (2) a constant-density toroid where photo-electric absorption,
Compton scattering and Fe-K fluorescence are considered. We detect
the presence of (3) an unabsorbed power law associated with Thomson
scattering off ionised clouds. Additional (4) Compton reflection,
most noticeable through a stronger Fe-K$\alpha$ line, is also found.
We find strong evidence against a completely closed, or entirely open,
toroidal obscurer geometry.

\begin{figure}[h]
\begin{centering}
\resizebox{\hsize}{!}{\includegraphics{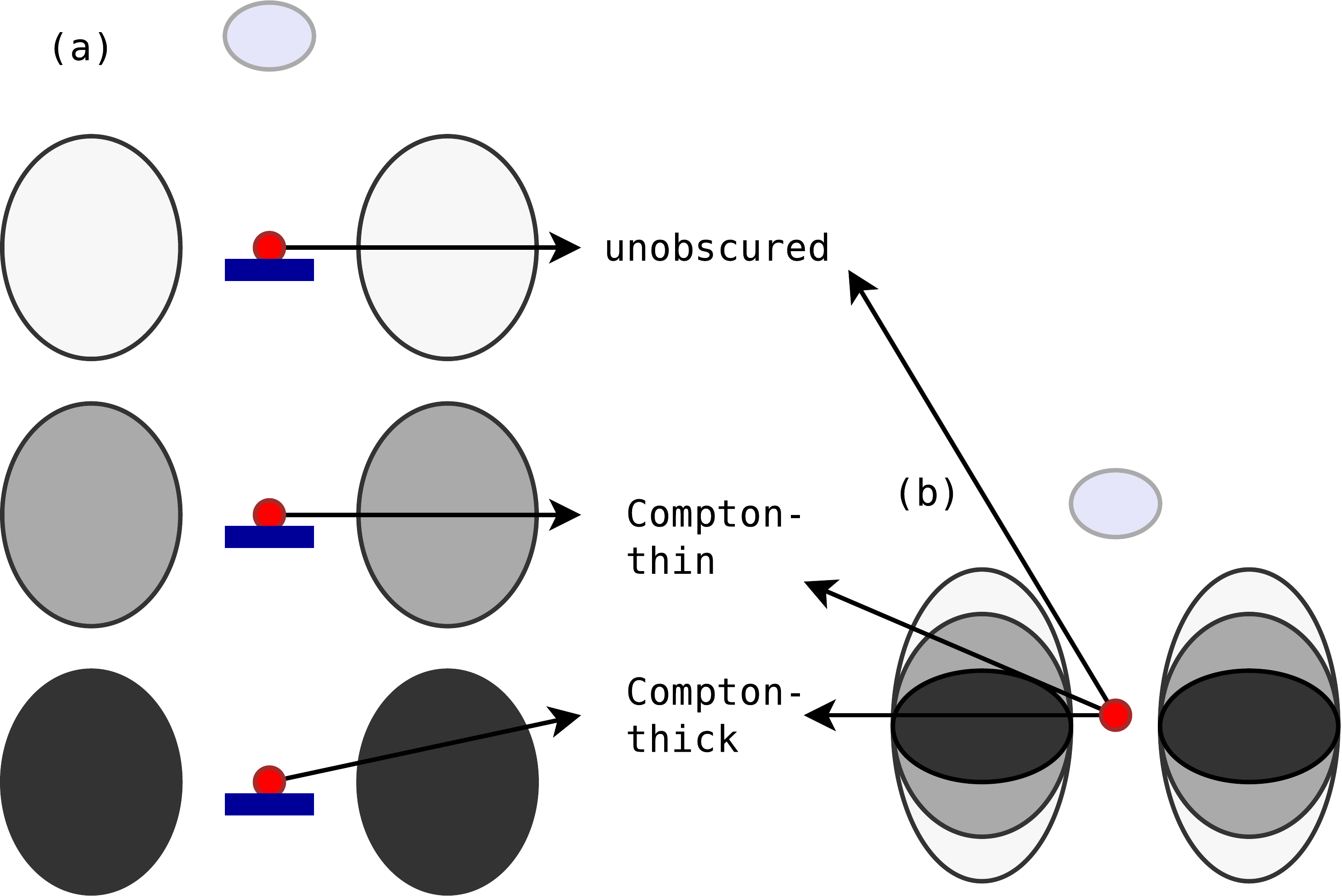}}
\par\end{centering}

\caption[Inferred possible geometries]{\label{fig:posterior-geometry}Cartoon illustrations of a-posteriori
possible geometries (see text).}
\end{figure}

The geometry of the obscurer in the deepest field to date is thus,
from the point of view of X-ray spectra, compatible with two simple
scenarios illustrated in Figure~\ref{fig:posterior-geometry}: (a)
Per-source density variations of a constant-density torus, with an
accretion disk contributing extra reflection in some sources or (b),
following the unification scheme, a torus with a column density gradient
where the LOS obscuration depends on the viewing angle and the observed
additional reflection originates in denser regions of the torus. In
both scenarios, ionised clouds can scatter intrinsic radiation past
the torus.

\begin{acknowledgements}

JB thanks Farhan Feroz for proofreading the methodology section. Remaining
errors are JB's. JB acknowledges financial support through a Max-Planck-Gesellschaft
stipend. We also thank the builders and operators of \textit{Chandra}.
This research has made use of software provided by the Chandra X-ray
Center (CXC) in the application package CIAO.

\end{acknowledgements}


\bibliographystyle{aa}
\bibliography{agn}

\begin{thebibliography}{93}
\expandafter\ifx\csname natexlab\endcsname\relax\def\natexlab#1{#1}\fi

\bibitem[{{Aird} {et~al.}(2010){Aird}, {Nandra}, {Laird}, {Georgakakis},
  {Ashby}, {Barmby}, {Coil}, {Huang}, {Koekemoer}, {Steidel}, \&
  {Willmer}}]{Aird2010}
{Aird}, J., {Nandra}, K., {Laird}, E.~S., {et~al.} 2010, \mnras, 401, 2531

\bibitem[{Akaike(1974)}]{AIC}
Akaike, H. 1974, Automatic Control, IEEE Transactions on, 19, 716

\bibitem[{{Akylas} {et~al.}(2012){Akylas}, {Georgakakis}, {Georgantopoulos},
  {Brightman}, \& {Nandra}}]{Akylas2012}
{Akylas}, A., {Georgakakis}, A., {Georgantopoulos}, I., {Brightman}, M., \&
  {Nandra}, K. 2012, \aap, 546, A98

\bibitem[{{Alexander} \& {Hickox}(2012)}]{Alexander2012}
{Alexander}, D.~M. \& {Hickox}, R.~C. 2012, \nar, 56, 93

\bibitem[{{Antonucci}(1993)}]{Antonucci1993}
{Antonucci}, R. 1993, \araa, 31, 473

\bibitem[{{Antonucci}(1982)}]{Antonucci1982}
{Antonucci}, R.~R.~J. 1982, \nat, 299, 605

\bibitem[{{Antonucci} \& {Miller}(1985)}]{AntonucciUnification1985}
{Antonucci}, R.~R.~J. \& {Miller}, J.~S. 1985, \apj, 297, 621

\bibitem[{{Arnouts} {et~al.}(1999){Arnouts}, {Cristiani}, {Moscardini},
  {Matarrese}, {Lucchin}, {Fontana}, \& {Giallongo}}]{Arnouts1999}
{Arnouts}, S., {Cristiani}, S., {Moscardini}, L., {et~al.} 1999, \mnras, 310,
  540

\bibitem[{{Ashby} {et~al.}(2013){Ashby}, {Willner}, {Fazio}, {Huang}, {Arendt},
  {Barmby}, {Barro}, {Bell}, {Bouwens}, {Cattaneo}, {Croton}, {Dav{\'e}},
  {Dunlop}, {Egami}, {Faber}, {Finlator}, {Grogin}, {Guhathakurta},
  {Hernquist}, {Hora}, {Illingworth}, {Kashlinsky}, {Koekemoer}, {Koo},
  {Labb{\'e}}, {Li}, {Lin}, {Moseley}, {Nandra}, {Newman}, {Noeske}, {Ouchi},
  {Peth}, {Rigopoulou}, {Robertson}, {Sarajedini}, {Simard}, {Smith}, {Wang},
  {Wechsler}, {Weiner}, {Wilson}, {Wuyts}, {Yamada}, \& {Yan}}]{Ashby2013}
{Ashby}, M.~L.~N., {Willner}, S.~P., {Fazio}, G.~G., {et~al.} 2013, \apj, 769,
  80

\bibitem[{Baganoff(1999)}]{ACISMemo162}
Baganoff, F. 1999, ACIS On-orbit Background Rates and Spectra from Chandra OAC
  Phase 1, ACIS Memo 162, Massachusetts Institute of Technology, Centerfor
  Space Research

\bibitem[{{Bayarri} \& {Castellanos}(2008)}]{Bayarri2008}
{Bayarri}, M.~J. \& {Castellanos}, M.~E. 2008, ArXiv e-prints

\bibitem[{{Brightman} \& {Nandra}(2011)}]{Brightman2011a}
{Brightman}, M. \& {Nandra}, K. 2011, \mnras, 413, 1206

\bibitem[{{Brightman} {et~al.}(2013){Brightman}, {Silverman}, {Mainieri},
  {Ueda}, {Schramm}, {Matsuoka}, {Nagao}, {Steinhardt}, {Kartaltepe},
  {Sanders}, {Treister}, {Shemmer}, {Brandt}, {Brusa}, {Comastri}, {Ho},
  {Lanzuisi}, {Lusso}, {Nandra}, {Salvato}, {Zamorani}, {Akiyama}, {Alexander},
  {Bongiorno}, {Capak}, {Civano}, {Del Moro}, {Doi}, {Elvis}, {Hasinger},
  {Laird}, {Masters}, {Mignoli}, {Ohta}, {Schawinski}, \&
  {Taniguchi}}]{Brightman2013}
{Brightman}, M., {Silverman}, J.~D., {Mainieri}, V., {et~al.} 2013, \mnras,
  433, 2485

\bibitem[{{Brightman} \& {Ueda}(2012)}]{Brightman2012}
{Brightman}, M. \& {Ueda}, Y. 2012, \mnras, 423, 702

\bibitem[{{Broos} {et~al.}(2010){Broos}, {Townsley}, {Feigelson}, {Getman},
  {Bauer}, \& {Garmire}}]{BroosACISEXTRACT2010}
{Broos}, P.~S., {Townsley}, L.~K., {Feigelson}, E.~D., {et~al.} 2010, \apj,
  714, 1582

\bibitem[{{Budav{\'a}ri} \& {Szalay}(2008)}]{Budavari2008}
{Budav{\'a}ri}, T. \& {Szalay}, A.~S. 2008, \apj, 679, 301

\bibitem[{{Burtscher} {et~al.}(2013){Burtscher}, {Meisenheimer}, {Tristram},
  {Jaffe}, {H{\"o}nig}, {Davies}, {Kishimoto}, {Pott}, {R{\"o}ttgering},
  {Schartmann}, {Weigelt}, \& {Wolf}}]{Burtscher2013}
{Burtscher}, L., {Meisenheimer}, K., {Tristram}, K.~R.~W., {et~al.} 2013, ArXiv
  e-prints

\bibitem[{{Cardamone} {et~al.}(2010){Cardamone}, {van Dokkum}, {Urry},
  {Taniguchi}, {Gawiser}, {Brammer}, {Taylor}, {Damen}, {Treister}, {Cobb},
  {Bond}, {Schawinski}, {Lira}, {Murayama}, {Saito}, \&
  {Sumikawa}}]{Cardamone2010}
{Cardamone}, C.~N., {van Dokkum}, P.~G., {Urry}, C.~M., {et~al.} 2010, \apjs,
  189, 270

\bibitem[{{Cash}(1979)}]{Cash1979}
{Cash}, W. 1979, \apj, 228, 939

\bibitem[{{Comastri} {et~al.}(2002){Comastri}, {Vignali}, {Brusa}, {Hellas}, \&
  {Hellas2XMM Consortia}}]{Comastri2002}
{Comastri}, A., {Vignali}, C., {Brusa}, M., {Hellas}, \& {Hellas2XMM
  Consortia}. 2002, in Astronomical Society of the Pacific Conference Series,
  Vol. 284, IAU Colloq. 184: AGN Surveys, ed. R.~F. {Green}, E.~Y.
  {Khachikian}, \& D.~B. {Sanders}, 235

\bibitem[{{Damen} {et~al.}(2011){Damen}, {Labb{\'e}}, {van Dokkum}, {Franx},
  {Taylor}, {Brandt}, {Dickinson}, {Gawiser}, {Illingworth}, {Kriek},
  {Marchesini}, {Muzzin}, {Papovich}, \& {Rix}}]{Damen2011}
{Damen}, M., {Labb{\'e}}, I., {van Dokkum}, P.~G., {et~al.} 2011, \apj, 727, 1

\bibitem[{{de Rosa} {et~al.}(2012){de Rosa}, {Panessa}, {Bassani}, {Bazzano},
  {Bird}, {Landi}, {Malizia}, {Molina}, \& {Ubertini}}]{deRosa2012}
{de Rosa}, A., {Panessa}, F., {Bassani}, L., {et~al.} 2012, \mnras, 420, 2087

\bibitem[{Efron {et~al.}(2001)Efron, Gous, Kass, Datta, \&
  Lahiri}]{efron2001scales}
Efron, B., Gous, A., Kass, R., Datta, G., \& Lahiri, P. 2001, Lecture
  Notes-Monograph Series, 208

\bibitem[{{Fabian}(1989)}]{Fabian1989}
{Fabian}, A.~C. 1989, in ESA Special Publication, Vol. 296, Two Topics in X-Ray
  Astronomy, Volume 1: X Ray Binaries. Volume 2: AGN and the X Ray Background,
  ed. J.~{Hunt} \& B.~{Battrick}, 1097--1104

\bibitem[{{Fabian}(1999)}]{Fabian1999}
{Fabian}, A.~C. 1999, \mnras, 308, L39

\bibitem[{{Feroz} {et~al.}(2009){Feroz}, {Hobson}, \& {Bridges}}]{Feroz2009}
{Feroz}, F., {Hobson}, M.~P., \& {Bridges}, M. 2009, \mnras, 398, 1601

\bibitem[{{Ferrarese} {et~al.}(1996){Ferrarese}, {Ford}, \&
  {Jaffe}}]{Ferrarese1996}
{Ferrarese}, L., {Ford}, H.~C., \& {Jaffe}, W. 1996, \apj, 470, 444

\bibitem[{{Freeman} {et~al.}(2001){Freeman}, {Doe}, \&
  {Siemiginowska}}]{Freeman2001}
{Freeman}, P., {Doe}, S., \& {Siemiginowska}, A. 2001, in Society of
  Photo-Optical Instrumentation Engineers (SPIE) Conference Series, Vol. 4477,
  Society of Photo-Optical Instrumentation Engineers (SPIE) Conference Series,
  ed. J.-L. {Starck} \& F.~D. {Murtagh}, 76--87

\bibitem[{{Gaskell} {et~al.}(2008){Gaskell}, {Goosmann}, \&
  {Klimek}}]{Gaskell2008}
{Gaskell}, C.~M., {Goosmann}, R.~W., \& {Klimek}, E.~S. 2008, \memsai, 79, 1090

\bibitem[{{Georgakakis} {et~al.}(2013){Georgakakis}, {Carrera}, {Lanzuisi},
  {Brightman}, {Buchner}, {Aird}, {Page}, {Cappi}, {Afonso}, {Alonso-Herrero},
  {Ballo}, {Barcons}, {Ceballos}, {Comastri}, {Georgantopoulos}, {Mateos},
  {Nandra}, {Rosario}, {Salvato}, {Schawinski}, {Severgnini}, \&
  {Vignali}}]{Georgakakis2013}
{Georgakakis}, A., {Carrera}, F., {Lanzuisi}, G., {et~al.} 2013, ArXiv e-prints

\bibitem[{{Georgantopoulos} \& {Georgakakis}(2005)}]{Georgantopoulos2005}
{Georgantopoulos}, I. \& {Georgakakis}, A. 2005, \mnras, 358, 131

\bibitem[{{George} \& {Fabian}(1991)}]{pexmonlines}
{George}, I.~M. \& {Fabian}, A.~C. 1991, \mnras, 249, 352

\bibitem[{{Gierli{\'n}ski} \& {Done}(2004)}]{Gierlinski2004}
{Gierli{\'n}ski}, M. \& {Done}, C. 2004, \mnras, 349, L7

\bibitem[{{Gondhalekar} {et~al.}(1997){Gondhalekar}, {Rouillon-Foley}, \&
  {Kellett}}]{Gondhalekar1997}
{Gondhalekar}, P.~M., {Rouillon-Foley}, C., \& {Kellett}, B.~J. 1997, \mnras,
  288, 260

\bibitem[{{Grogin} {et~al.}(2012){Grogin}, {Rajan}, {Donley}, {Kartaltepe},
  {Koekemoer}, {Lucas}, {Rosario}, \& {Salvato}}]{Grogin2012}
{Grogin}, N.~A., {Rajan}, A., {Donley}, J.~L., {et~al.} 2012, in American
  Astronomical Society Meeting Abstracts, Vol. 220, American Astronomical
  Society Meeting Abstracts 220, 335.23

\bibitem[{{Guainazzi} \& {Bianchi}(2007)}]{Guainazzi2007}
{Guainazzi}, M. \& {Bianchi}, S. 2007, \mnras, 374, 1290

\bibitem[{{Guo} {et~al.}(2013){Guo}, {Ferguson}, {Giavalisco}, {Barro},
  {Willner}, {Ashby}, {Dahlen}, {Donley}, {Faber}, {Fontana}, {Galametz},
  {Grazian}, {Huang}, {Kocevski}, {Koekemoer}, {Koo}, {McGrath}, {Peth},
  {Salvato}, {Wuyts}, {Castellano}, {Cooray}, {Dickinson}, {Dunlop}, {Fazio},
  {Gardner}, {Gawiser}, {Grogin}, {Hathi}, {Hsu}, {Lee}, {Lucas}, {Mobasher},
  {Nandra}, {Newman}, \& {van der Wel}}]{Guo2013}
{Guo}, Y., {Ferguson}, H.~C., {Giavalisco}, M., {et~al.} 2013, \apjs, 207, 24

\bibitem[{{Hopkins} {et~al.}(2009){Hopkins}, {Hickox}, {Quataert}, \&
  {Hernquist}}]{Hopkins2009}
{Hopkins}, P.~F., {Hickox}, R., {Quataert}, E., \& {Hernquist}, L. 2009,
  \mnras, 398, 333

\bibitem[{{Hsieh} {et~al.}(2012){Hsieh}, {Wang}, {Hsieh}, {Lin}, {Yan}, {Lim},
  \& {Ho}}]{Hsieh2012}
{Hsieh}, B.-C., {Wang}, W.-H., {Hsieh}, C.-C., {et~al.} 2012, \apjs, 203, 23

\bibitem[{Hunter(2007)}]{matplotlib}
Hunter, J.~D. 2007, Computing In Science \& Engineering, 9, 90

\bibitem[{{Ilbert} {et~al.}(2006){Ilbert}, {Arnouts}, {McCracken},
  {Bolzonella}, {Bertin}, {Le F{\`e}vre}, {Mellier}, {Zamorani}, {Pell{\`o}},
  {Iovino}, {Tresse}, {Le Brun}, {Bottini}, {Garilli}, {Maccagni}, {Picat},
  {Scaramella}, {Scodeggio}, {Vettolani}, {Zanichelli}, {Adami}, {Bardelli},
  {Cappi}, {Charlot}, {Ciliegi}, {Contini}, {Cucciati}, {Foucaud}, {Franzetti},
  {Gavignaud}, {Guzzo}, {Marano}, {Marinoni}, {Mazure}, {Meneux}, {Merighi},
  {Paltani}, {Pollo}, {Pozzetti}, {Radovich}, {Zucca}, {Bondi}, {Bongiorno},
  {Busarello}, {de La Torre}, {Gregorini}, {Lamareille}, {Mathez}, {Merluzzi},
  {Ripepi}, {Rizzo}, \& {Vergani}}]{Ilbert2006}
{Ilbert}, O., {Arnouts}, S., {McCracken}, H.~J., {et~al.} 2006, \aap, 457, 841

\bibitem[{{Ilbert} {et~al.}(2009){Ilbert}, {Capak}, {Salvato}, {Aussel},
  {McCracken}, {Sanders}, {Scoville}, {Kartaltepe}, {Arnouts}, {Le Floc'h},
  {Mobasher}, {Taniguchi}, {Lamareille}, {Leauthaud}, {Sasaki}, {Thompson},
  {Zamojski}, {Zamorani}, {Bardelli}, {Bolzonella}, {Bongiorno}, {Brusa},
  {Caputi}, {Carollo}, {Contini}, {Cook}, {Coppa}, {Cucciati}, {de la Torre},
  {de Ravel}, {Franzetti}, {Garilli}, {Hasinger}, {Iovino}, {Kampczyk},
  {Kneib}, {Knobel}, {Kovac}, {Le Borgne}, {Le Brun}, {F{\`e}vre}, {Lilly},
  {Looper}, {Maier}, {Mainieri}, {Mellier}, {Mignoli}, {Murayama}, {Pell{\`o}},
  {Peng}, {P{\'e}rez-Montero}, {Renzini}, {Ricciardelli}, {Schiminovich},
  {Scodeggio}, {Shioya}, {Silverman}, {Surace}, {Tanaka}, {Tasca}, {Tresse},
  {Vergani}, \& {Zucca}}]{Ilbert2009}
{Ilbert}, O., {Capak}, P., {Salvato}, M., {et~al.} 2009, \apj, 690, 1236

\bibitem[{Jeffreys(1961)}]{jeffreys1961theory}
Jeffreys, H. 1961, International series of monographs on physics.

\bibitem[{Jones {et~al.}(2001--)Jones, Oliphant, Peterson, {et~al.}}]{scipy}
Jones, E., Oliphant, T., Peterson, P., {et~al.} 2001--, {SciPy}: Open source
  scientific tools for {Python}

\bibitem[{{King}(2005)}]{King2005}
{King}, A. 2005, \apjl, 635, L121

\bibitem[{{King}(2010)}]{King2010}
{King}, A.~R. 2010, \mnras, 408, L95

\bibitem[{{Koekemoer} {et~al.}(2011){Koekemoer}, {Faber}, {Ferguson}, {Grogin},
  {Kocevski}, {Koo}, {Lai}, {Lotz}, {Lucas}, {McGrath}, {Ogaz}, {Rajan},
  {Riess}, {Rodney}, {Strolger}, {Casertano}, {Castellano}, {Dahlen},
  {Dickinson}, {Dolch}, {Fontana}, {Giavalisco}, {Grazian}, {Guo}, {Hathi},
  {Huang}, {van der Wel}, {Yan}, {Acquaviva}, {Alexander}, {Almaini}, {Ashby},
  {Barden}, {Bell}, {Bournaud}, {Brown}, {Caputi}, {Cassata}, {Challis},
  {Chary}, {Cheung}, {Cirasuolo}, {Conselice}, {Roshan Cooray}, {Croton},
  {Daddi}, {Dav{\'e}}, {de Mello}, {de Ravel}, {Dekel}, {Donley}, {Dunlop},
  {Dutton}, {Elbaz}, {Fazio}, {Filippenko}, {Finkelstein}, {Frazer}, {Gardner},
  {Garnavich}, {Gawiser}, {Gruetzbauch}, {Hartley}, {H{\"a}ussler},
  {Herrington}, {Hopkins}, {Huang}, {Jha}, {Johnson}, {Kartaltepe},
  {Khostovan}, {Kirshner}, {Lani}, {Lee}, {Li}, {Madau}, {McCarthy},
  {McIntosh}, {McLure}, {McPartland}, {Mobasher}, {Moreira}, {Mortlock},
  {Moustakas}, {Mozena}, {Nandra}, {Newman}, {Nielsen}, {Niemi}, {Noeske},
  {Papovich}, {Pentericci}, {Pope}, {Primack}, {Ravindranath}, {Reddy},
  {Renzini}, {Rix}, {Robaina}, {Rosario}, {Rosati}, {Salimbeni}, {Scarlata},
  {Siana}, {Simard}, {Smidt}, {Snyder}, {Somerville}, {Spinrad}, {Straughn},
  {Telford}, {Teplitz}, {Trump}, {Vargas}, {Villforth}, {Wagner}, {Wandro},
  {Wechsler}, {Weiner}, {Wiklind}, {Wild}, {Wilson}, {Wuyts}, \&
  {Yun}}]{Koekemoer2011}
{Koekemoer}, A.~M., {Faber}, S.~M., {Ferguson}, H.~C., {et~al.} 2011, \apjs,
  197, 36

\bibitem[{{Komatsu} {et~al.}(2011){Komatsu}, {Smith}, {Dunkley}, {Bennett},
  {Gold}, {Hinshaw}, {Jarosik}, {Larson}, {Nolta}, {Page}, {Spergel},
  {Halpern}, {Hill}, {Kogut}, {Limon}, {Meyer}, {Odegard}, {Tucker}, {Weiland},
  {Wollack}, \& {Wright}}]{Komatsu2011cosmology}
{Komatsu}, E., {Smith}, K.~M., {Dunkley}, J., {et~al.} 2011, \apjs, 192, 18

\bibitem[{{Kormendy} \& {Ho}(2013)}]{Kormendy2013}
{Kormendy}, J. \& {Ho}, L.~C. 2013, \araa, 51, 511

\bibitem[{{Krolik} \& {Kallman}(1987)}]{Krolik1987}
{Krolik}, J.~H. \& {Kallman}, T.~R. 1987, \apjl, 320, L5

\bibitem[{Kullback \& Leibler(1951)}]{kullback1951information}
Kullback, S. \& Leibler, R.~A. 1951, The Annals of Mathematical Statistics, 22,
  79

\bibitem[{{Laird} {et~al.}(2009){Laird}, {Nandra}, {Georgakakis}, {Aird},
  {Barmby}, {Conselice}, {Coil}, {Davis}, {Faber}, {Fazio}, {Guhathakurta},
  {Koo}, {Sarajedini}, \& {Willmer}}]{Laird2009}
{Laird}, E.~S., {Nandra}, K., {Georgakakis}, A., {et~al.} 2009, \apjs, 180, 102

\bibitem[{Levenberg(1944)}]{levenberg:1944}
Levenberg, K. 1944, Quart. Applied Math., 2, 164

\bibitem[{{Magdziarz} \& {Zdziarski}(1995)}]{pexrav}
{Magdziarz}, P. \& {Zdziarski}, A.~A. 1995, \mnras, 273, 837

\bibitem[{Marquardt(1963)}]{Marquardt1963}
Marquardt, D. 1963, Journal of the Society for Industrial and Applied
  Mathematics, 11, 431

\bibitem[{{Matt}(2002)}]{pexmonshoulder}
{Matt}, G. 2002, \mnras, 337, 147

\bibitem[{{Morrison} \& {McCammon}(1983)}]{wabs}
{Morrison}, R. \& {McCammon}, D. 1983, \apj, 270, 119

\bibitem[{{Murphy} \& {Yaqoob}(2009)}]{MurphyYaqoobMyTorus2009}
{Murphy}, K.~D. \& {Yaqoob}, T. 2009, \mnras, 397, 1549

\bibitem[{{Mushotzky}(2004)}]{Mushotzky2004}
{Mushotzky}, R. 2004, in Astrophysics and Space Science Library, Vol. 308,
  Supermassive Black Holes in the Distant Universe, ed. A.~J. {Barger}, 53

\bibitem[{{Nandra} \& {George}(1994)}]{Nandra1994a}
{Nandra}, K. \& {George}, I.~M. 1994, \mnras, 267, 974

\bibitem[{{Nandra} {et~al.}(2005){Nandra}, {Laird}, {Adelberger}, {Gardner},
  {Mushotzky}, {Rhodes}, {Steidel}, {Teplitz}, \& {Arnaud}}]{NandraSrcDet2005}
{Nandra}, K., {Laird}, E.~S., {Adelberger}, K., {et~al.} 2005, \mnras, 356, 568

\bibitem[{{Nandra} {et~al.}(2007){Nandra}, {O'Neill}, {George}, \&
  {Reeves}}]{pexmon}
{Nandra}, K., {O'Neill}, P.~M., {George}, I.~M., \& {Reeves}, J.~N. 2007,
  \mnras, 382, 194

\bibitem[{{Nandra} \& {Pounds}(1994)}]{Nandra1994}
{Nandra}, K. \& {Pounds}, K.~A. 1994, \mnras, 268, 405

\bibitem[{{Perola} {et~al.}(2002){Perola}, {Matt}, {Cappi}, {Fiore},
  {Guainazzi}, {Maraschi}, {Petrucci}, \& {Piro}}]{Perola2002}
{Perola}, G.~C., {Matt}, G., {Cappi}, M., {et~al.} 2002, \aap, 389, 802

\bibitem[{{Protassov} {et~al.}(2002){Protassov}, {van Dyk}, {Connors},
  {Kashyap}, \& {Siemiginowska}}]{Protassov2002}
{Protassov}, R., {van Dyk}, D.~A., {Connors}, A., {Kashyap}, V.~L., \&
  {Siemiginowska}, A. 2002, \apj, 571, 545

\bibitem[{{Rangel} {et~al.}(2013{\natexlab{a}}){Rangel}, {Nandra}, {Laird}, \&
  {Orange}}]{Rangel2013b}
{Rangel}, C., {Nandra}, K., {Laird}, E.~S., \& {Orange}, P. 2013{\natexlab{a}},
  \mnras

\bibitem[{{Rangel} {et~al.}(2013{\natexlab{b}}){Rangel}, {Nandra}, {Laird}, \&
  {Orange}}]{Rangel2013}
{Rangel}, C., {Nandra}, K., {Laird}, E.~S., \& {Orange}, P. 2013{\natexlab{b}},
  \mnras, 428, 3089

\bibitem[{{Richstone} {et~al.}(1998){Richstone}, {Ajhar}, {Bender}, {Bower},
  {Dressler}, {Faber}, {Filippenko}, {Gebhardt}, {Green}, {Ho}, {Kormendy},
  {Lauer}, {Magorrian}, \& {Tremaine}}]{Richstone1998}
{Richstone}, D., {Ajhar}, E.~A., {Bender}, R., {et~al.} 1998, \nat, 395, A14

\bibitem[{{Risaliti} {et~al.}(1999){Risaliti}, {Maiolino}, \&
  {Salvati}}]{Risaliti1999}
{Risaliti}, G., {Maiolino}, R., \& {Salvati}, M. 1999, \apj, 522, 157

\bibitem[{{Ross} \& {Fabian}(1993)}]{Ross1993}
{Ross}, R.~R. \& {Fabian}, A.~C. 1993, \mnras, 261, 74

\bibitem[{{Salvato} {et~al.}(2009){Salvato}, {Hasinger}, {Ilbert}, {Zamorani},
  {Brusa}, {Scoville}, {Rau}, {Capak}, {Arnouts}, {Aussel}, {Bolzonella},
  {Buongiorno}, {Cappelluti}, {Caputi}, {Civano}, {Cook}, {Elvis}, {Gilli},
  {Jahnke}, {Kartaltepe}, {Impey}, {Lamareille}, {Le Floc'h}, {Lilly},
  {Mainieri}, {McCarthy}, {McCracken}, {Mignoli}, {Mobasher}, {Murayama},
  {Sasaki}, {Sanders}, {Schiminovich}, {Shioya}, {Shopbell}, {Silverman},
  {Smol{\v c}i{\'c}}, {Surace}, {Taniguchi}, {Thompson}, {Trump}, {Urry}, \&
  {Zamojski}}]{Salvato2009}
{Salvato}, M., {Hasinger}, G., {Ilbert}, O., {et~al.} 2009, \apj, 690, 1250

\bibitem[{{Salvato} {et~al.}(2011){Salvato}, {Ilbert}, {Hasinger}, {Rau},
  {Civano}, {Zamorani}, {Brusa}, {Elvis}, {Vignali}, {Aussel}, {Comastri},
  {Fiore}, {Le Floc'h}, {Mainieri}, {Bardelli}, {Bolzonella}, {Bongiorno},
  {Capak}, {Caputi}, {Cappelluti}, {Carollo}, {Contini}, {Garilli}, {Iovino},
  {Fotopoulou}, {Fruscione}, {Gilli}, {Halliday}, {Kneib}, {Kakazu},
  {Kartaltepe}, {Koekemoer}, {Kovac}, {Ideue}, {Ikeda}, {Impey}, {Le Fevre},
  {Lamareille}, {Lanzuisi}, {Le Borgne}, {Le Brun}, {Lilly}, {Maier},
  {Manohar}, {Masters}, {McCracken}, {Messias}, {Mignoli}, {Mobasher}, {Nagao},
  {Pello}, {Puccetti}, {Perez-Montero}, {Renzini}, {Sargent}, {Sanders},
  {Scodeggio}, {Scoville}, {Shopbell}, {Silvermann}, {Taniguchi}, {Tasca},
  {Tresse}, {Trump}, \& {Zucca}}]{Salvato2011}
{Salvato}, M., {Ilbert}, O., {Hasinger}, G., {et~al.} 2011, \apj, 742, 61

\bibitem[{Schwarz(1978)}]{Schwarz1978}
Schwarz, G. 1978, Ann. Stat., 6, 461

\bibitem[{{Severgnini} {et~al.}(2003){Severgnini}, {Caccianiga}, {Braito},
  {Della Ceca}, {Maccacaro}, {Wolter}, {Sekiguchi}, {Sasaki}, {Yoshida},
  {Akiyama}, {Watson}, {Barcons}, {Carrera}, {Pietsch}, \&
  {Webb}}]{Severgnini2003}
{Severgnini}, P., {Caccianiga}, A., {Braito}, V., {et~al.} 2003, \aap, 406, 483

\bibitem[{{Shankar} {et~al.}(2004){Shankar}, {Salucci}, {Granato}, {De Zotti},
  \& {Danese}}]{Shankar2004}
{Shankar}, F., {Salucci}, P., {Granato}, G.~L., {De Zotti}, G., \& {Danese}, L.
  2004, \mnras, 354, 1020

\bibitem[{{Shankar} {et~al.}(2009){Shankar}, {Weinberg}, \&
  {Miralda-Escud{\'e}}}]{Shankar2009}
{Shankar}, F., {Weinberg}, D.~H., \& {Miralda-Escud{\'e}}, J. 2009, \apj, 690,
  20

\bibitem[{{Silk} \& {Rees}(1998)}]{Silk1998}
{Silk}, J. \& {Rees}, M.~J. 1998, \aap, 331, L1

\bibitem[{Sinharay \& Stern(2003)}]{Sinharay2003209}
Sinharay, S. \& Stern, H.~S. 2003, Journal of Statistical Planning and
  Inference, 111, 209 , special issue I: Model Selection, Model Diagnostics,
  Empirical Bayes and Hierarchical Bayes

\bibitem[{Skilling(2004)}]{Skilling2004}
Skilling, J. 2004, in AIP Conference Proceedings, Vol. 735, 395

\bibitem[{{Stark} {et~al.}(1992){Stark}, {Gammie}, {Wilson}, {Bally}, {Linke},
  {Heiles}, \& {Hurwitz}}]{Stark1992GalHImap}
{Stark}, A.~A., {Gammie}, C.~F., {Wilson}, R.~W., {et~al.} 1992, \apjs, 79, 77

\bibitem[{{Sunyaev} \& {Titarchuk}(1980)}]{Sunyaev1980}
{Sunyaev}, R.~A. \& {Titarchuk}, L.~G. 1980, \aap, 86, 121

\bibitem[{Thompson {et~al.}(2001)Thompson, Vaughan, {et~al.}}]{thompson2001x}
Thompson, A., Vaughan, D., {et~al.} 2001, X-ray data booklet (Lawrence Berkeley
  National Laboratory, University of California Berkeley, CA)

\bibitem[{{Tozzi} {et~al.}(2006){Tozzi}, {Gilli}, {Mainieri}, {Norman},
  {Risaliti}, {Rosati}, {Bergeron}, {Borgani}, {Giacconi}, {Hasinger},
  {Nonino}, {Streblyanska}, {Szokoly}, {Wang}, \& {Zheng}}]{Tozzi2006}
{Tozzi}, P., {Gilli}, R., {Mainieri}, V., {et~al.} 2006, \aap, 451, 457

\bibitem[{{Treister} {et~al.}(2004){Treister}, {Urry}, {Chatzichristou},
  {Bauer}, {Alexander}, {Koekemoer}, {Van Duyne}, {Brandt}, {Bergeron},
  {Stern}, {Moustakas}, {Chary}, {Conselice}, {Cristiani}, \&
  {Grogin}}]{Treister2004}
{Treister}, E., {Urry}, C.~M., {Chatzichristou}, E., {et~al.} 2004, \apj, 616,
  123

\bibitem[{{Turner} {et~al.}(1997){Turner}, {George}, {Nandra}, \&
  {Mushotzky}}]{Turner1997}
{Turner}, T.~J., {George}, I.~M., {Nandra}, K., \& {Mushotzky}, R.~F. 1997,
  \apjs, 113, 23

\bibitem[{{Turner} \& {Pounds}(1989)}]{Turner1989}
{Turner}, T.~J. \& {Pounds}, K.~A. 1989, \mnras, 240, 833

\bibitem[{{Ueda} {et~al.}(2007){Ueda}, {Eguchi}, {Terashima}, {Mushotzky},
  {Tueller}, {Markwardt}, {Gehrels}, {Hashimoto}, \& {Potter}}]{Ueda2007}
{Ueda}, Y., {Eguchi}, S., {Terashima}, Y., {et~al.} 2007, \apjl, 664, L79

\bibitem[{{van der Marel} \& {van den Bosch}(1998)}]{vanderMarel1998}
{van der Marel}, R.~P. \& {van den Bosch}, F.~C. 1998, \aj, 116, 2220

\bibitem[{{van Dyk} {et~al.}(2001){van Dyk}, {Connors}, {Kashyap}, \&
  {Siemiginowska}}]{Dyk2001}
{van Dyk}, D.~A., {Connors}, A., {Kashyap}, V.~L., \& {Siemiginowska}, A. 2001,
  \apj, 548, 224

\bibitem[{Wilk \& Gnanadesikan(1968)}]{1968}
Wilk, M.~B. \& Gnanadesikan, R. 1968, Biometrika, 55, pp. 1

\bibitem[{Wilks(1938)}]{wilks1938large}
Wilks, S. 1938, The Annals of Mathematical Statistics, 60

\bibitem[{{Xue} {et~al.}(2011){Xue}, {Luo}, {Brandt}, {Bauer}, {Lehmer},
  {Broos}, {Schneider}, {Alexander}, {Brusa}, {Comastri}, {Fabian}, {Gilli},
  {Hasinger}, {Hornschemeier}, {Koekemoer}, {Liu}, {Mainieri}, {Paolillo},
  {Rafferty}, {Rosati}, {Shemmer}, {Silverman}, {Smail}, {Tozzi}, \&
  {Vignali}}]{Xue2011}
{Xue}, Y.~Q., {Luo}, B., {Brandt}, W.~N., {et~al.} 2011, \apjs, 195, 10

\bibitem[{{Zdziarski} {et~al.}(2000){Zdziarski}, {Poutanen}, \&
  {Johnson}}]{Zdziarski2000}
{Zdziarski}, A.~A., {Poutanen}, J., \& {Johnson}, W.~N. 2000, \apj, 542, 703

\end{thebibliography}


\begin{appendix}

\section{Goodness-of-Fit of the Background model\label{sec:Appendix-I:-Goodness-1}}

\begin{figure*}[h]
\begin{centering}
\includegraphics[width=17cm]{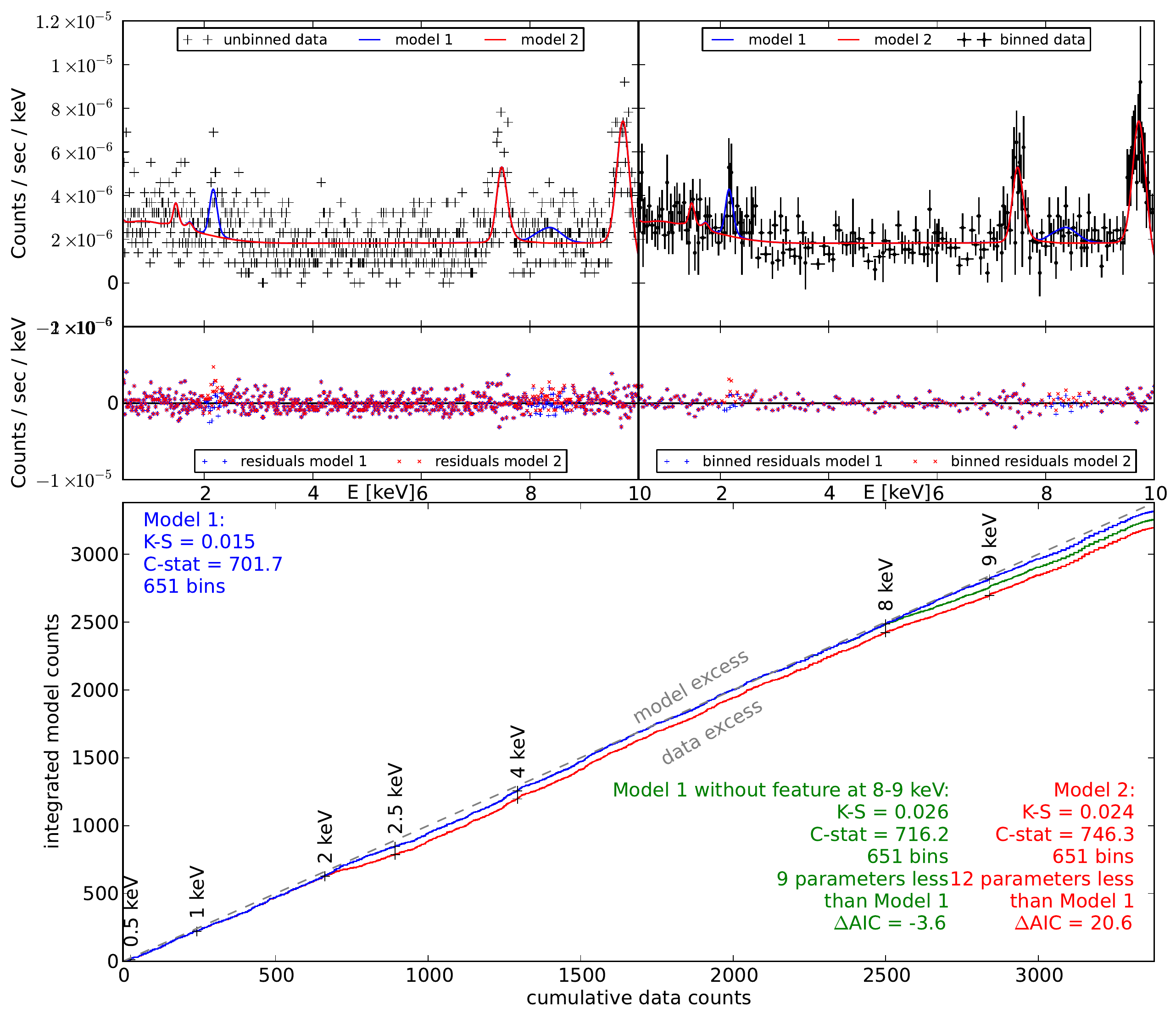}
\par\end{centering}

\caption{\label{fig:gof-318}\change{See text in Section \ref{sec:Appendix-I:-Goodness-1}.}{Comparison
of the background data from Source 318 with background models. The
best model is shown in blue, while the red model has several features
removed (see text in Appendix \ref{sec:Appendix-I:-Goodness-1}).
In the top two panels, the usual count spectrum is shown with residuals
(left unbinned, right binned to at least 20 counts per bin).\protect \\
In the large, bottom panel we present the corresponding Q\textendash{}Q
(quantile-quantile) plot. For each energy $E$, the model counts predicted
and the counts observed below $E$ are recorded on the plot. The grey
dashed line is where data and model would perfectly agree. Model 1
(blue solid top line) follows this line very closely, and thus can
be considered a good model. Model 2 (red solid bottom line) deviates
from the grey dashed line at $2\,\text{keV}$, indicating that a feature
in the data may be missing in the model. The shape and size of the
deviation also indicates the shape of the needed feature. The significance
of the feature can be tested using model selection. Here, the AIC
shows that the feature at $2-2.5\,\text{keV}$ is necessary $\Delta AIC>0$,
but adding the more complicated feature at $8-9\,\text{keV}$ is not
(see text in Appendix \ref{sec:Appendix-I:-Goodness-1} for details).}}
\end{figure*}

We intend to demonstrate that the used background description is a
good model. To this end, we present a goodness of fit (GoF) methodology
for X-ray spectra. We use Q\-Q plots for model discovery and the
AIC model comparison method to test for the significance of model
improvements\change{}{, although any of the model comparison methods
introduced in Section \ref{sub:Model-selection} could be chosen}.
We demonstrate the method using a background source spectrum and our
best fit, comparing it to a simplified model. In our analysis, every
spectrum is fitted individually to accommodate the diversity of background
spectra.

\subsection{Background model definition}

We present ``Model 1'', which is our final Gaussian mixture model
with a constant base continuum:

\[
M_{1}(E)=C\times\left(1+\sum_{i}A_{i}\times\exp\left\{ -\frac{1}{2}\left(\frac{E-E_{i}}{FWHM_{i}}\right)^{2}\right\} \right)
\]

To model the rise in the soft energies two Gaussians (called ``softend''
and ``softsoftend'') are used of widths $\sim0.5\,\text{keV}$ and
$\sim2\,\text{keV}$ at centred at $\sim0\,\text{keV}$. Another five
Gaussian lines describe the spectral bumps due to e.g. transition
lines of the detector material. These are centred around 1.486 ($Al\, K\alpha$),
1.739 ($Si\, K\alpha$), 2.142 ($Au\, M\alpha,\beta$), 7.478 ($Ni\, K\alpha$),
9.713 ($Au\, L\alpha$; all in keV). A feature at $\sim8.3\,\text{keV}$
possibly composed of $Ni$/$Au$ lines is modelled by three further
Gaussians at $8.012$, $8.265$, $8.494\,\text{keV}$. We allow all
centres to vary within $0.1\,\text{keV}$. The parameters start from
reasonable guesses and are optimised as long as the fit statistic
(C-Stat) improves.

Figure \ref{fig:gof-318} shows the comparison between data and model
for a source with the sequential number 318 in the catalogue. This
source has 3380 counts and was chosen because it constitutes an intermediate
case between the most high-count spectra with many peculiarities and
low-count spectra with almost no visible features. The final parameter
values after fitting are shown in Table \ref{tab:Background-model-parameters}
(middle column).
\begin{table}[h]
\begin{centering}
\begin{tabular}{lcc}
\hline 
\hline
Source & 318 & 179\tabularnewline
\hline 
$C$ & $7.278\times10^{-5}$ & $23.28\times10^{-5}$\tabularnewline
line1.ampl & 413 & 511\tabularnewline
line1.fwhm & 0.0329 & 0.0162\tabularnewline
line1.pos & 1.475 & 1.49\tabularnewline
line2.ampl & 0 & 74.7\tabularnewline
line2.fwhm & 0.1 & 0.002\tabularnewline
line2.pos & 1.84 & 1.82\tabularnewline
line3.ampl & 420 & 581\tabularnewline
line3.fwhm & 0.1 & 0.0827\tabularnewline
line3.pos & 2.16 & 2.16\tabularnewline
line4.ampl  & 3769 & 1159\tabularnewline
line4.fwhm  & 0.0257 & 0.0768\tabularnewline
line4.pos  & 7.48 & 7.49\tabularnewline
line5.ampl  & 1446 & 42.9\tabularnewline
line5.fwhm  & 0.002 & 0.102\tabularnewline
line5.pos  & 8.10 & 8.07\tabularnewline
line6.ampl & 97.1 & 280\tabularnewline
line6.fwhm & 0.4 & 0.0793\tabularnewline
line6.pos  & 8.37 & 8.25\tabularnewline
line7.ampl  & 111.5 & 6656\tabularnewline
line7.fwhm  & 0.002 & 0.002\tabularnewline
line7.pos  & 8.47 & 8.47\tabularnewline
line8.ampl  & 1993 & 2225\tabularnewline
line8.fwhm  & 0.1 & 0.0886\tabularnewline
line8.pos  & 9.72 & 9.71\tabularnewline
softend.ampl & 21092 & 21824\tabularnewline
softend.fwhm &  2.0 & 2.6\tabularnewline
softend.pos &  1.0 & 1.0\tabularnewline
softsoftend.ampl & 72642 & 78711\tabularnewline
softsoftend.fwhm &  0.325 & 0.638\tabularnewline
softsoftend.pos  & 0.167 & 0.243\tabularnewline
\hline 
\end{tabular}
\par\end{centering}

\caption{\label{tab:Background-model-parameters}Background model parameters
for two sources. For the left source, the model is shown in Figure
\ref{fig:gof-318}. Positions and FWHM are in keV. Amplitudes are
unitless, except for the normalisation $C$, which is in $cts/s/keV/cm^{-2}$.}
\end{table}

\subsection{Goodness of Fit and Model discovery}

For comparison, we present Model 2 which has several Gaussians disabled,
namely the one centred at $2.1\,\text{keV}$ (line 3) and the three
between $8-9\text{\,\ keV}$ (line 5-7). The intent is to compare
methods evaluating whether Model 2 is a good model, where it deviates
from the data, and whether the deviation is significant.

The classic method is to plot the data, model and the residuals. This
is shown in the upper left panel of Figure \ref{fig:gof-318} for
the un-binned data and model, with the residuals below. The upper
right panel shows the same, but with adaptive binning requiring 20
counts in each bin. The feature at $2.1\,\text{keV}$ is visible immediately,
while the feature at $8-9\text{\,\ keV}$ is less striking.

We present an alternative method of analysing the quality of a model:
the \textbf{Q\textendash{}Q plot}, shown in the large lower panel.
For each energy $E$, the model counts predicted and the counts observed
below $E$ are recorded on the plot. Here the counts are shown, while
statisticians typically use quantiles, i.e. the fraction of observations
that lie below a quantity. This does not influence the main point,
namely the shape of the curve. The grey dashed line is where data
and model would perfectly agree. A steeper curve means the model predicts
more counts than observed, while a shallower curve indicates an excess
of observed counts.

Model 1 (blue solid top line) follows this line very closely, and
thus can be considered a good model. Model 2 (red solid bottom line)
deviates from the grey dashed line at $2\,\text{keV}$, indicating
that a feature in the data is not modelled. Above $2.5\,\text{keV}$,
the line is parallel to Model 1, indicating no further difference.
This means the feature is confined to this energy range. With a bit
of practise one can also see that the difference required to bring
the lines into agreement looks like the cumulative distribution of
a Gaussian (a S-shape rather than e.g. a straight line for a flat
distribution). Another, but more subtle, deviation is visible between
$8-9\,\text{keV}$. This is highlighted using the green solid middle
line which does model the $2.1\,\text{keV}$ feature.

\begin{figure*}[h]
\begin{centering}
\includegraphics[width=17cm]{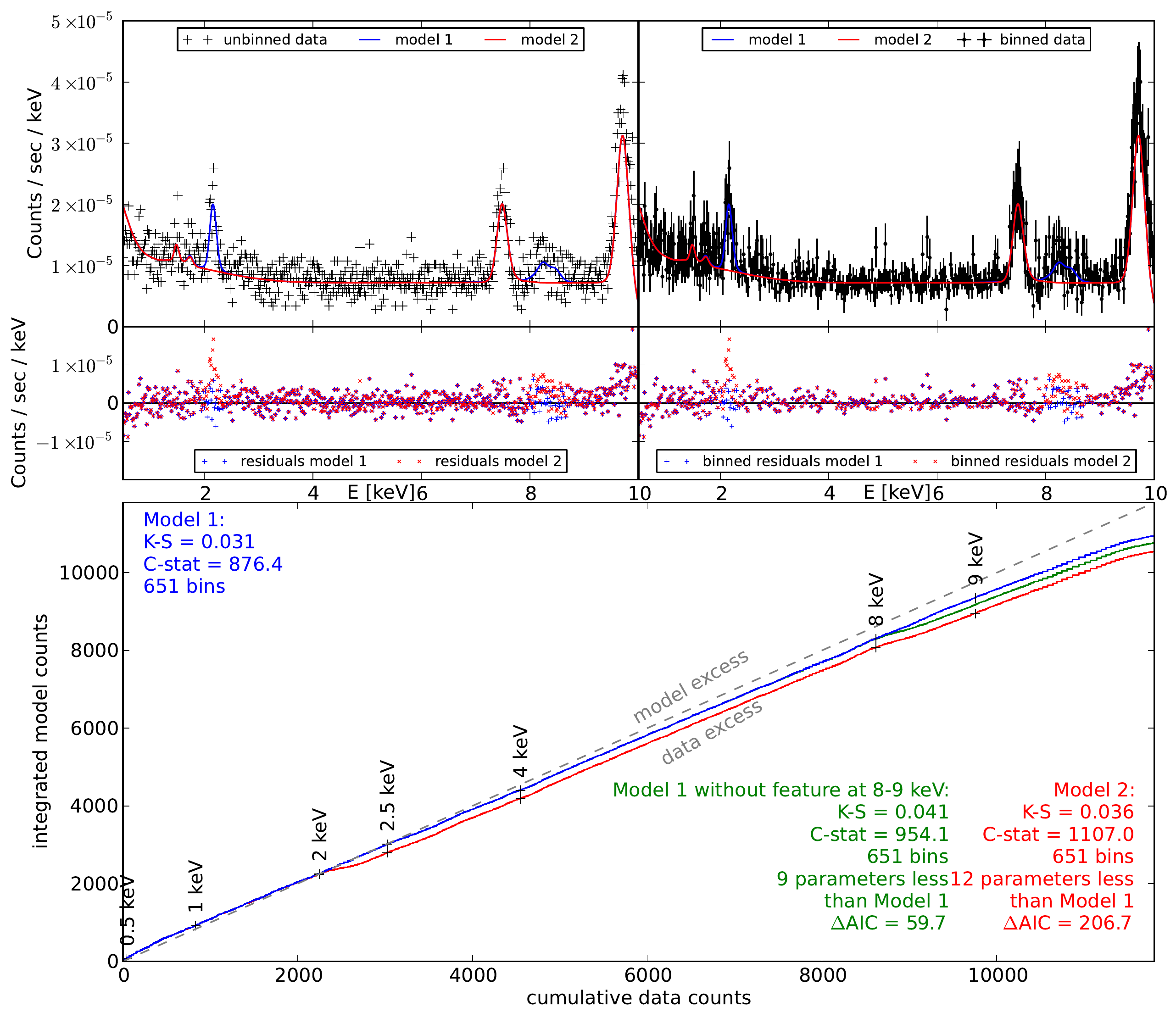}
\par\end{centering}

\caption{\label{fig:gof-179}Same as Figure \ref{fig:gof-318}, but for Source
179 (11802 counts). Contrary to Figure \ref{fig:gof-318}, the feature
between $8-9\text{\,\ keV}$ is required ($\Delta AIC>0$). In the
lower panel, there is a mild, continuous deviation from the grey dashed
line indicating that a mild increase in the higher energy counts.
This hints that the model could be improved further. Although perfection
in background modelling is a noble quest, the source spectra have
1 order of magnitude fewer counts, allowing us to be satisfied with
this model.}
\end{figure*}

Having found a good model, and slightly worse, simpler models, we
can now test whether the improvement is significant. For instance,
it seems doubtful that the minute feature at $8-9\text{keV}$ justifies
modelling with 3 Gaussian components (9 parameters). For this, we
employ the AIC, which punishes the difference in the likelihood (C-stat)
by adding twice the number of parameters. As $\Delta AIC=20.6>0$
(Model 2 vs Model 1, red text), the worsening is significant. But
if we only remove the feature at $8-9\text{\,\ keV}$, the AIC decreases
due to the simplification of the model (green text). Thus, in this
source, the $8-9\text{\,\ keV}$ feature can be ignored. However,
in sources with more counts, it is required (see Figure \ref{fig:gof-179}
for one example).

The deviation between model and data can be summarised using GoF measures.
The Kolmogorov-Smirnov measure, $K-S=\sup_{E}|observed(<E)-predicted(<E)|$,
records the largest deviation between the empirical cumulative distribution
of observed counts (0 at the lowest energy, 1 at the highest), and
the model cumulative distribution. Other measures, such as the ones
used for the Cramér\textendash{}von Mises test or the Anderson\textendash{}Darling
test take all deviations into account, not merely the largest. When
many spectra are analysed in an automated fashion, the highest values
can hint to problematic cases which require further, visual analysis.

\section{Propagation of redshift uncertainty\label{sec:pdz-test}}

It is common practise to use a redshift point estimator (best fit
redshift) from photometric redshifts and analyse spectra with this
value. In this work, in contrast, the redshift uncertainty from photometric
redshift is propagated into the analysis of X-ray spectra in the form
of a probability distribution on the redshift parameter. In this section,
we discuss the differences between the approaches.

\begin{figure}[h]
\begin{centering}
\includegraphics[width=1\columnwidth]{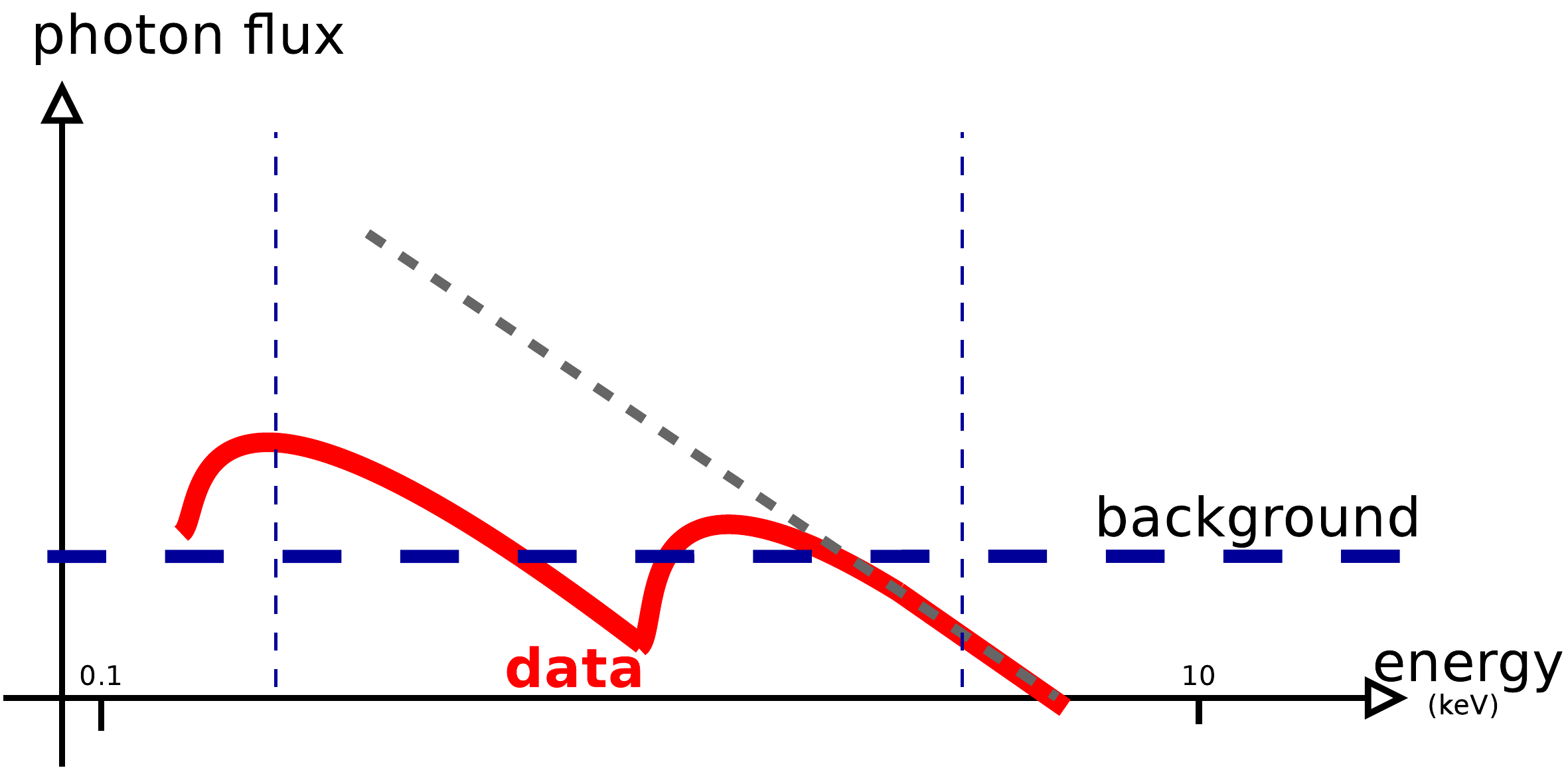}
\par\end{centering}

\begin{centering}
\includegraphics[width=1\columnwidth]{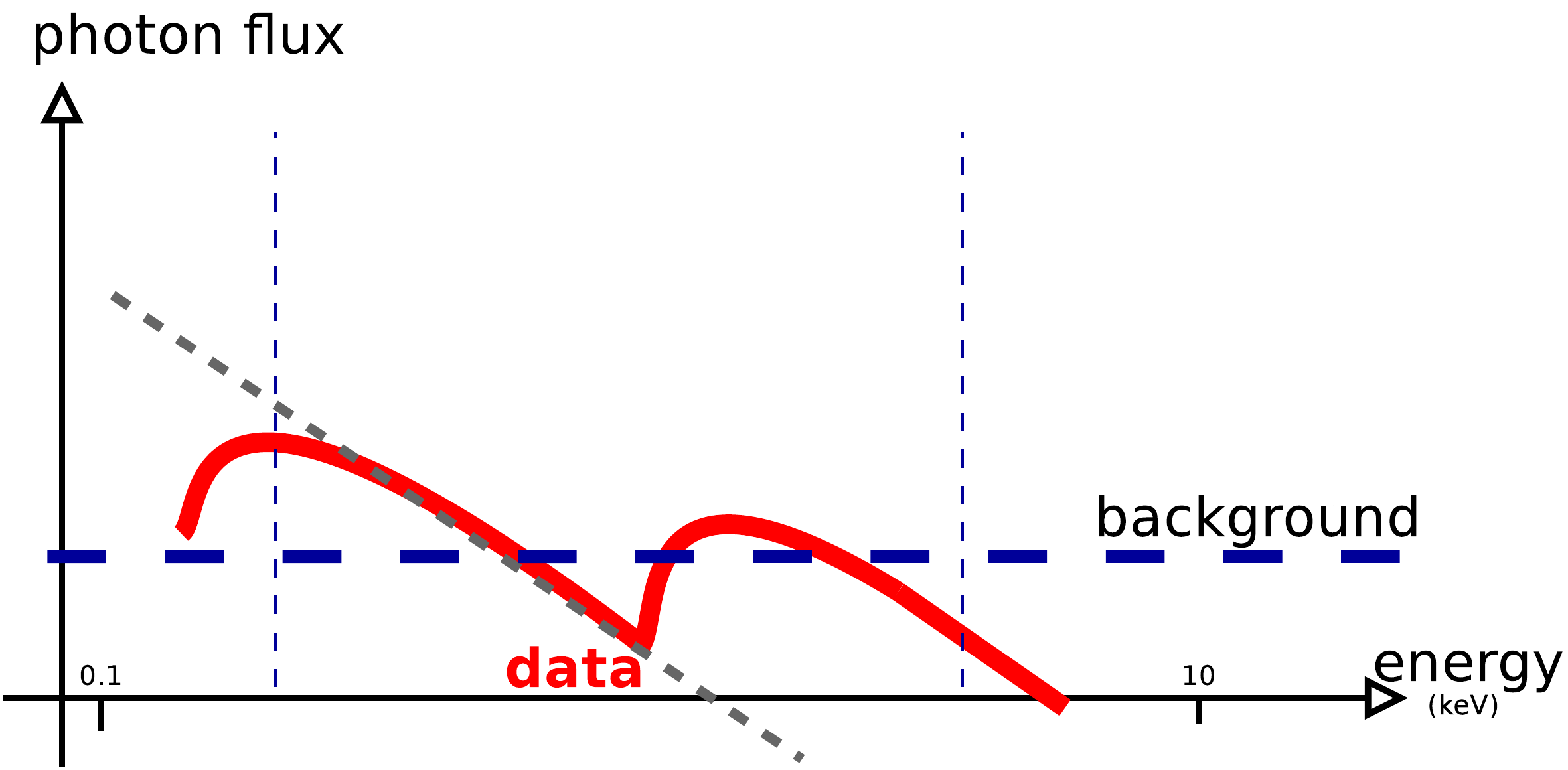}
\par\end{centering}

\caption{\label{fig:dualsol-cartoon}The origin of the two distinct solutions
in Figure \ref{fig:gof-179-1} is highlighted in these two cartoons.
Given the data shown in the red thick line, one can either A) consider
a bright, highly obscured source (top panel), or B) a low-luminosity,
low-obscuration solution where the hard counts are due to the background.
An intermediate solution is ruled out however. Depending on the background
level and redshift, the two solutions will have different likelihoods.}
\end{figure}

\begin{figure*}[t]
\begin{centering}
\includegraphics[width=17cm]{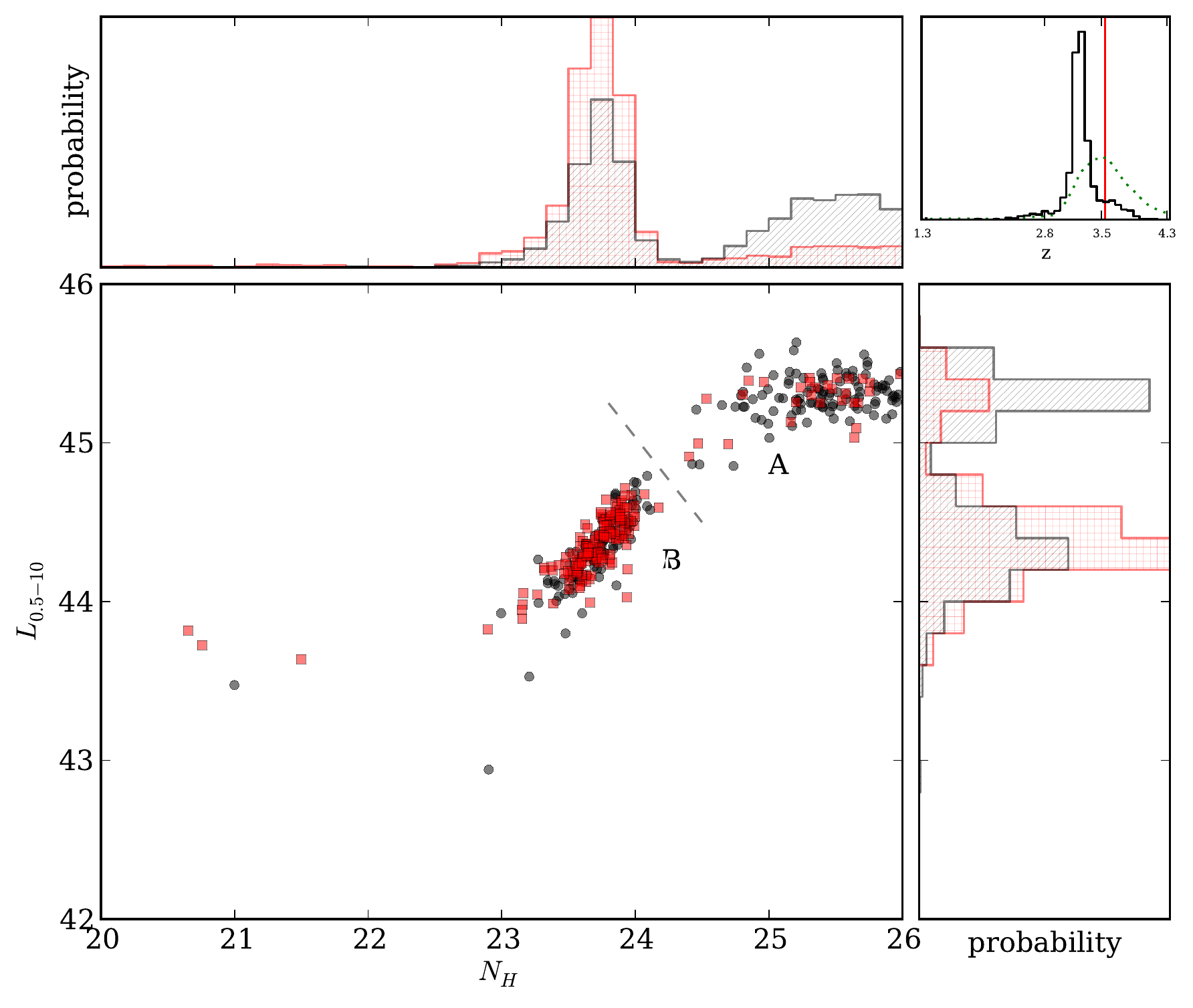}
\par\end{centering}

\caption{\label{fig:gof-179-1}Demonstration of parameter estimation results
under different redshift inputs. Source 551 in the catalogue is analysed
twice with the methodology laid out in this paper using the \mo{torus}
model. We consider once the case of using only the best fit photometric
redshift and once the case of using the photometric redshift probability
distribution (PDZ). Both inputs are shown in the upper right panel
as a vertical red line at $z=3.5444$ and with a dotted green line
respectively. For brevity, only two resulting parameters are presented,
namely the column density $N_{H}$ (logarithmic, in $\text{cm}^{-2}$)
and the derived intrinsic luminosity $L$ in the $0.5-10\,\text{keV}$
band (logarithmic, in $\text{erg}/\text{s}$). The large, lower left
panel shows the derived intrinsic luminosity and column density parameters
by equally probable points, similar to a Markov Chain. As each point
on the plane is equally likely to be the true value, denser regions
represent more probable parameter values. Here, the red squares represent
the fixed redshift analysis while the black circles show the analysis
results using the PDZ. The marginal distributions are shown in the
upper left and lower right panels as probability histograms (red crossed
hatching and black striped hatching). \protect \\
Two separated solutions are clearly visible and highlighted using
the labels ``A'' for the highly obscured solution and ``B'' for
the less obscured solution. The associated spectra are illustrated
in Figure \ref{fig:dualsol-cartoon}. The fixed redshift analysis
has the highest likelihood at the less obscured solution. When given
the freedom to vary the redshift parameter, the X-ray data have the
highest likelihood at the highly obscured solution. Maximum Likelihood
analysis methods thus may fail to account for the uncertainty (see
text). The difference in results come from the freedom to move to
a lower redshift, as can be seen in the upper right panel, where the
black line shows the redshift posterior probability distribution.
These results thus also show that redshift information can be improved
using X-ray data (see Buchner et al. in prep).}
\end{figure*}

We consider the source 551 in our catalogue and analyse its X-ray
spectrum using the methodology laid out in this paper with the \mo{torus}
model. We perform the analysis twice, a) with the probability distribution
from photometric redshifts and b) with the best fit photo-metric redshift.
Figure \ref{fig:gof-179-1} demonstrates how the different input redshifts
(upper right panel) influence the results in the derived column density
and intrinsic luminosity parameters (large panel). The Bayesian analysis
shows that the parameter space is broad, and split to two distinct
solutions (see lower left panel in Figure \ref{fig:gof-179-1}): a
highly obscured solution and a less obscured solution. The maximum
likelihood is in the less obscured solution for the fixed redshift
value, but in the highly-obscured solution if the redshift distribution
is used, because the likelihood improves when using a slightly lower
value than the best fit redshift. The two solutions are strictly separated,
i.e. an intermediate solution is ruled out. Common Maximum Likelihood
methods, like fitting and error estimation by Fisher matrix or contour
search, will fail to estimate the uncertainty correctly and hide the
respectively other solution. Methods building on these results can
therefore make false conclusions about e.g. the number of Compton-thick
sources. The Bayesian inference method presented in this work can
handle the separated solutions well and re-weighs them based on the
redshift information given. The various methods are discussed and
compared in Section \ref{sub:Statistical-analysis-methods}.

\section{False discovery rate of model comparison methods\label{sec:False-discovery-rate}}

In this section, we compare the efficiency and error rate of model
comparison methods. 

On the one hand we consider methods based on the ratio of the highest
likelihood, where, if the likelihood improves beyond a certain log-likelihood
ratio threshold, the more complex model is accepted. The threshold
may be dependent on the difference in the number of model parameters.
In this general form, Wilks' theorem, $\chi^{2}$-test, $F$-test,
AIC and BIC methods are covered.

On the other hand, the Bayesian inference based on computing the evidence
$Z$ is considered, where, assuming a-priori equality, the Bayes factor
$B_{12}=Z_{1}/Z_{2}$ is used to decide which model to use. In particular,
when $B_{12}$ is greater than the threshold, Model 1 is accepted,
while if $B_{12}^{-1}$ is greater than the threshold, Model 2 is
accepted. An important difference is that when the Bayesian inference
can conclude that not enough information to make a decision.

\subsection{Nested Problems}

\begin{figure}[h]
\begin{centering}
\resizebox{\hsize}{!}{\includegraphics{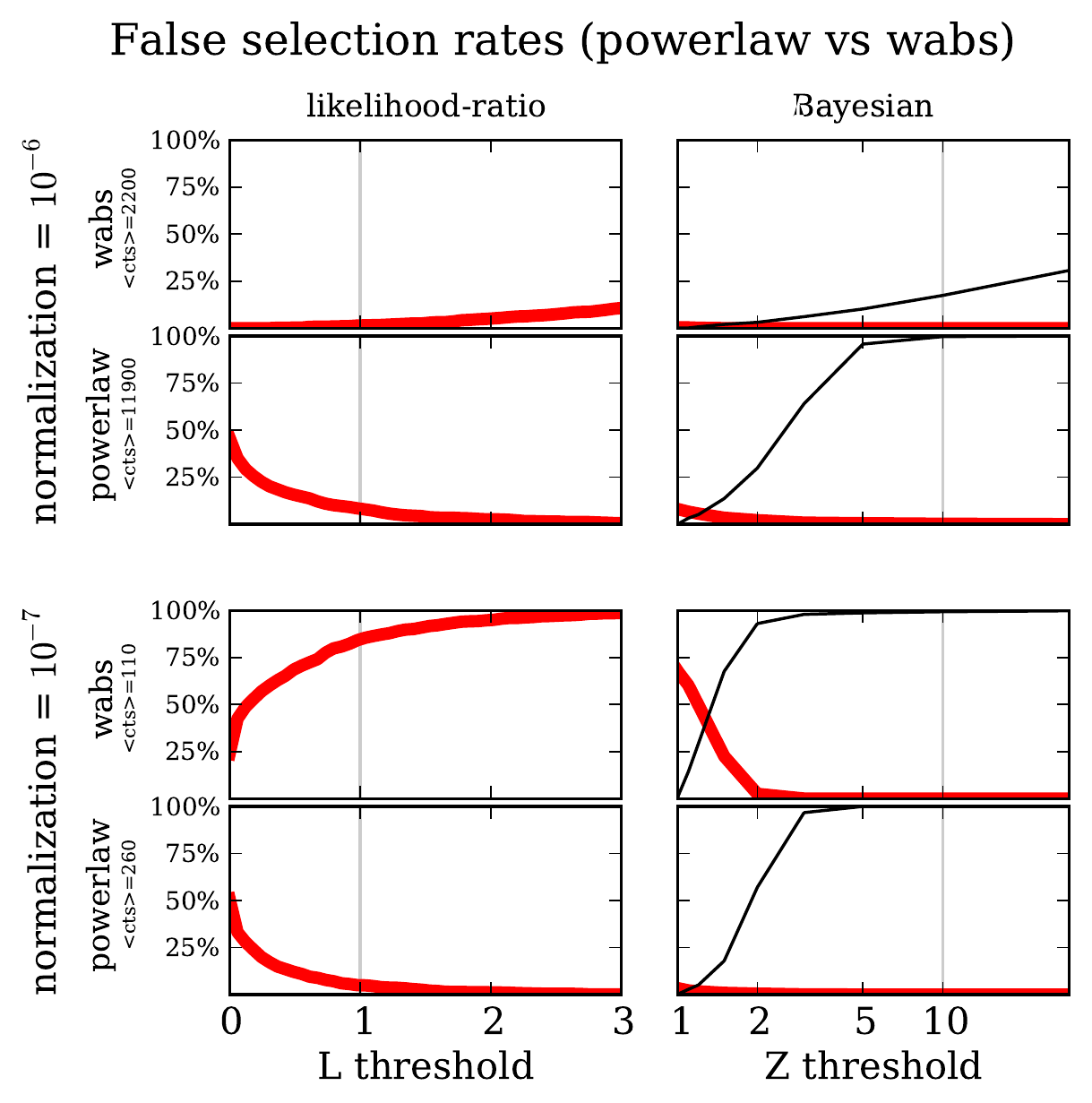}}
\par\end{centering}

\caption{\label{fig:nested-efficiency}Model selection results between \mo{powerlaw}
and \mo{wabs}. Likelihood-ratio based methods (left panels) and Bayesian
model selection (right panels) show different fractions of false choices
decisions (thick red line). The top left plot for example shows the
fraction of generated spectra where the likelihood ratio method selected
\mo{powerlaw} instead of the model used for generating the spectrum,
\mo{wabs}. The plot below should be considered simultaneously, as
it shows the fraction where \mo{powerlaw} was selected in spectra
generated with \mo{wabs}. Results are shown in dependence of the
threshold applied. The threshold may be optimised, but typical choices
are $\Delta L=1$ from AIC and $\log\,10=1$ for the Bayes factor
if the model priors are equal (gray vertical lines).\protect \\
Considering the sum of thick red lines in panel pairs, the Bayesian
model selection has a false selection rate below $1\%$ at the marked
threshold, which can not be achieved using likelihood-ratios regardless
of the threshold chosen. The Bayesian model selection may decline
to decide due to insufficient discriminatory power of the data (thin
black line). This is the case for the faint normalisations (bottom
four panels), where the likelihood-ratio methods remain with the simpler
model (yielding an error of $\sim50\%$). In the brighter normalisations
(upper right panels), the Bayesian method beyond a threshold of $5$
declines to distinguish when the \mo{powerlaw} model was used to
generate the data. This is a consequence of the degeneracy between
the nested \mo{powerlaw} and \mo{wabs} models. Thus for nested models,
a lower threshold, such as $\Delta\log\, Z=\log\,3=0.5$ can be appropriate.
When considering the efficiency only, the likelihood ratio based method
yields more correct decisions, as the Bayesian method declines to
choose very often. However, one may also remain with the simpler model,
in which case the results would be comparable.}
\end{figure}

We generate 2000 spectra each based on the \mo{powerlaw} and \mo{wabs}
input model. We assume a photon index of $\Gamma=1.9$, redshift $z=1.5$
and column density $N_{H}=10^{23}/\text{cm}^{2}$ (for \mo{wabs}).
The normalisation of the source power-law at 1 keV in $\text{photons}/\text{keV}/\text{cm}^{2}/\text{s}$
is set to either $10^{-6}$ or $10^{-7}$ (1000 simulated spectra
each). A fixed, flat background with $10^{-8}\,\text{photons}/\text{keV}/\text{cm}^{2}/\text{s}$
is used. We re-use the exposure time, ARF and RMF of Source 179 to
produce realistic, low-count spectra.

We analyse all simulated X-ray spectra in the $0.5-7\text{\,\ keV}$
band using the methodology laid out in this paper, once with the \mo{powerlaw}
and once with the \mo{wabs} model. We also store the maximum likelihood
(best fit). We apply model selection (Bayesian and likelihood-ratio-based)
between \mo{powerlaw} and the more complicated model \mo{wabs},
and record the number of false choices (e.g. \mo{wabs} was preferred,
although \mo{powerlaw} was used for generating the spectrum). These
results shown in Figure \ref{fig:nested-efficiency} using thick red
lines. For the Bayesian model selection, we also record the number
of cases where no significant preference was found (black lines).

The Bayesian model selection has a false selection rate below $1\%$
at the marked threshold, which can not be achieved by likelihood-ratio
based methods regardless of the threshold chosen. The Bayesian model
selection may decline to decide due to insufficient discriminatory
power of the data (thin black line). When considering the efficiency
only, the likelihood ratio based method yields more correct decisions,
as the Bayesian method declines to choose very often. However, one
may also remain with the simpler model, in which case the results
would be comparable.

\subsection{Non-nested problems}

\begin{figure}[h]
\begin{centering}
\resizebox{\hsize}{!}{\includegraphics{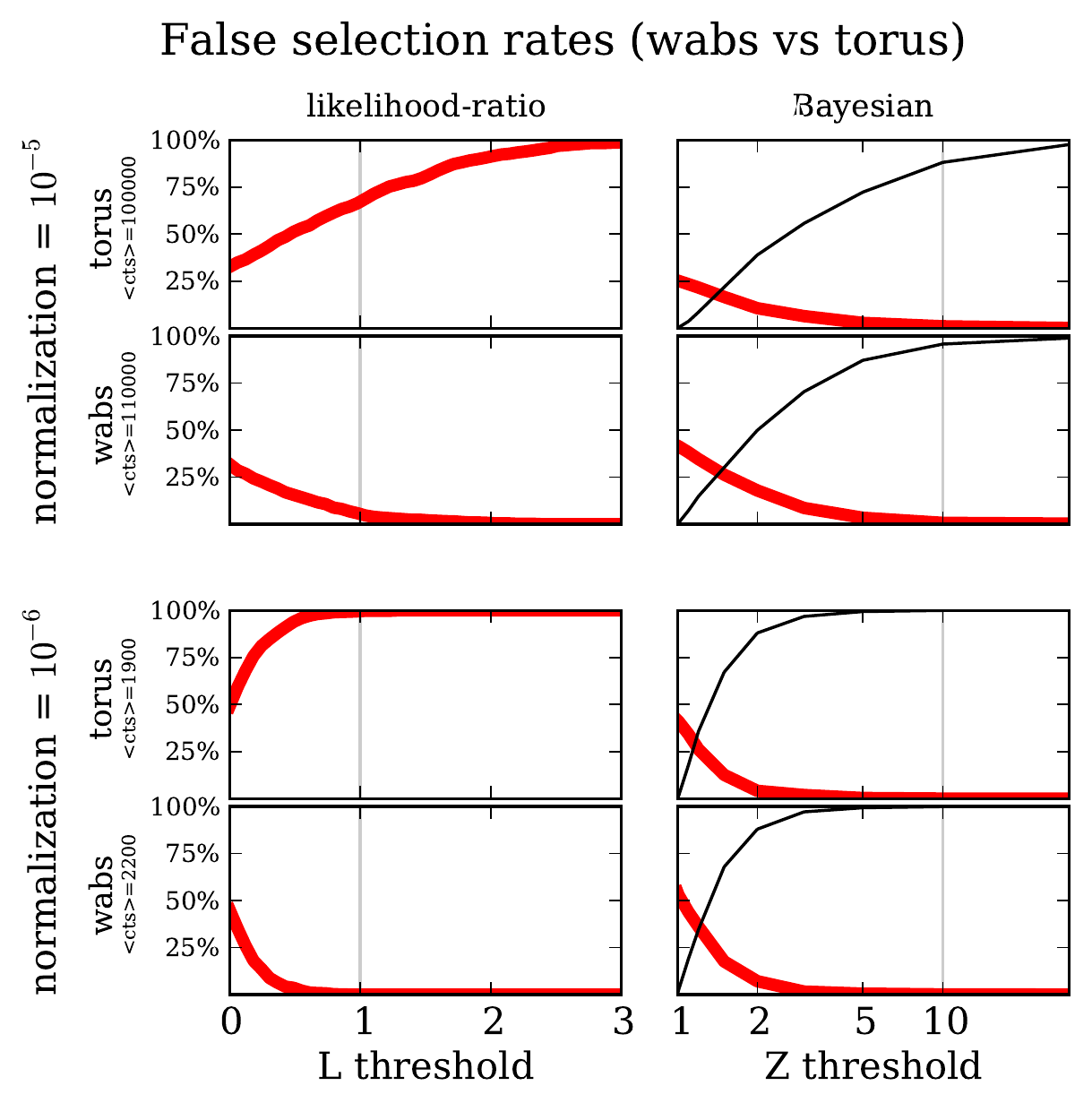}}
\par\end{centering}

\caption{\label{fig:nested-efficiency-1}Same as Figure \ref{fig:nested-efficiency}
for the non-nested model selection between \mo{wabs} and \mo{torus}
(instead of \mo{powerlaw} and \mo{wabs}). As the differences between
the models are subtle, the likelihood-ratio method often remains with
the null model (\mo{wabs}). This yields an overall error rate (red
line) of $\sim30-50\%$. The Bayesian model selection declares the
data insufficient for a distinction in such cases. In the upper right
panel, the Bayesian method is able to effectively distinguish between
some of the simulated instances, and makes few mistakes ($<1\%$ false
selection rate, considering the total of panel pairs).}
\end{figure}
We now consider a non-nested model selection problem, applying the
same methodology. We select between the models \mo{wabs} and \mo{torus}
-- neither is a special case of the other. We apply likelihood-ratio
based methods using \mo{wabs} as the null hypothesis, and only select
\mo{torus} when the likelihood-ratio threshold is exceeded. These
results are presented in a similar fashion as above in Figure \ref{fig:nested-efficiency-1}.
The panels show higher normalisations, as the models are more subtle
in their differences.

\subsection{Conclusion}

To summarise the Figures \ref{fig:nested-efficiency} and \ref{fig:nested-efficiency-1},
error rates below $1\%$ can be achieved using the in the Bayesian
model selection, with a threshold of e.g. $\log\, Z>1=\log\,10$.
It should be clear that a Bayes factor of $10$ is actually a quite
conservative choice, and does not necessarily refer to a false positive
rate of 1 in 10, but a much lower value. This point has been appreciated
in the literature before: \cite{efron2001scales} show that the Bayes
factor scale by \cite{jeffreys1961theory} is probably overly conservative.

In contrast, methods based on the likelihood-ratio always have higher
error rates than $1\%$ regardless of the chosen threshold. This is
due to the procedure remaining with the simpler model if the data
has insufficient discriminatory power. The Bayesian method has the
benefit of recognising these cases and can decline to decide. It should
be stressed again that several likelihood ratio based methods are
not valid for non-nested problems (Wilks' theorem, F-test, $\chi^{2}$-test,
likelihood-ratio test), although e.g. the AIC is. Furthermore, all
likelihood ratio based methods are not valid at testing against the
border of the parameter space (e.g. \mo{powerlaw} is a special case
of \mo{wabs} with minimal $N_{H}$). This is discussed further in
Section \ref{sub:Model-selection}. These results demonstrate that
even if a valid method based on the likelihood ratio was introduced,
it would perform poorly in the shown cases.

\end{appendix}


\end{document}